%% file: main.tex
\input{_constants}
\arxiv

\pdfoutput=1
\documentclass[10pt,twocolumn,letterpaper]{article}
\input{cvpr_header}

\unless\ifarxiv \myexternaldocument{_supplementary} \fi

\begin{document}
\title{\paperTitle}
\author{\authorBlock}
\maketitle

\input{00_abstract}
\input{01_intro}
\input{02_related}
\input{03_mle}
\input{04_regularization}
\input{05_lab}
\input{06_conclusion}

{\small
\bibliographystyle{ieeenat_fullname}
\bibliography{main}
}

\ifarxiv \clearpage \appendix \input{appendix} \fi

\end{document}

%% file: _constants.tex
\def\paperID{10375}
\def\confName{ICCV}
\def\confYear{2025}

\def\paperTitle{Free-running vs.\ Synchronous: Single-Photon Lidar for High-flux 3D Imaging
}

\def\authorBlock{
    Ruangrawee~Kitichotkul$^\dagger$
    \and
    Shashwath~Bharadwaj$^\dagger$
    \and
    Joshua~Rapp$^*$
    \and
    Yanting~Ma$^*$
    \and
    Alexander~Mehta$^\mathsection$
    \and
    Vivek~K~Goyal$^\dagger$
    \and
    {$^\dagger$Boston~University} \quad
    {$^*$Mitsubishi Electric Research Laboratories} \quad
    {$^\mathsection$University of California, Berkeley}
}

\newif\ifreview 
\newif\ifarxiv \newcommand{\arxiv}{\arxivtrue}
\newif\ifcamera 
\newif\ifrebuttal 

%% file: cvpr_header.tex
\ifreview \usepackage[review]{cvpr} \fi
\ifarxiv \usepackage[pagenumbers]{cvpr} \fi
\ifrebuttal \usepackage[rebuttal]{cvpr} \fi
\ifcamera \usepackage{cvpr} \fi

\input{_macros}  

\usepackage{xr-hyper}

\makeatletter
\newcommand*{\addFileDependency}[1]{
  \typeout{(#1)}
  \@addtofilelist{#1}
  \IfFileExists{#1}{}{\typeout{No file #1.}}
}

\makeatother
\newcommand*{\myexternaldocument}[1]{
    \externaldocument{#1}
    \addFileDependency{#1.tex}
    \addFileDependency{#1.aux}
}

\definecolor{cvprblue}{rgb}{0.21,0.49,0.74}
\usepackage[pagebackref,breaklinks,colorlinks,allcolors=cvprblue]{hyperref}
\usepackage[capitalize]{cleveref}
\usepackage[normalem]{ulem}
\crefname{section}{Sec.}{Secs.}
\crefname{table}{Table}{Tables}
\crefname{figure}{Fig.}{Figs.}

\ifarxiv \crefname{appendix}{App.}{Apps.}
\else \crefname{appendix}{Suppl.}{Suppls.} \fi

%% file: _macros.tex
\usepackage{graphicx}	
\usepackage{amsmath}	
\usepackage{amssymb}	
\usepackage{booktabs}
\usepackage{times}
\usepackage{microtype}
\usepackage{epsfig}
\usepackage{caption}
\usepackage{float}
\usepackage{placeins}
\usepackage{color, colortbl}
\usepackage{stfloats}
\usepackage{enumitem}
\usepackage{tabularx}
\usepackage{xstring}
\usepackage{multirow}
\usepackage{xspace}
\usepackage{url}
\usepackage{subcaption}
\usepackage{xcolor}
\usepackage[hang,flushmargin]{footmisc}

\usepackage{bm}
\usepackage{algorithm,algorithmic}
\usepackage{multirow,rotating}
\usepackage{subcaption}
\usepackage[per-mode=symbol]{siunitx}

\ifcamera \usepackage[accsupp]{axessibility} \fi

\ifarxiv  \fi

\newcommand{\Rev}[1]{{%
    \textbf{%
        \ifstrequal{#1}{1}{\textcolor{red}{R#1}}{%
        \ifstrequal{#1}{2}{\textcolor{blue}{R#1}}{%
        \ifstrequal{#1}{3}{\textcolor{teal}{R#1}}{%
        \ifstrequal{#1}{4}{\textcolor{magenta}{R#1}}{%
                           \textcolor{cyan}{R#1}%
        }}}}%
    }%
}}

\newcommand{\RV}[1]{{\color{black}{#1}}}

\def\lamtil{\widetilde{\lambda}}

\def\Phitil{\widetilde{\Phi}}

\def\tauhat{\widehat{\tau}}

\def\Xset{\{X_i\}_{i = 1}^N}

\def\Tseq{(T_i)_{i = 1}^N}

\def\Bin{\text{Binomial}}
\def\id{\rm{ideal}}
\def\sc{\rm{sync}}
\def\fr{\rm{free}}
\def\rmse{\text{RMSE}}
\def\mae{\text{MAE}}
\def\nrmse{\text{NRMSE}}

\def\sinit{\what{S}_{\rm{init}}}
\def\binit{\what{B}_{\rm{init}}}
\def\tauinit{\what{\tau}_{\rm{init}}}

\def\tr{t_r}
\def\ton{t_{\rm{on}}}
\def\nr{n_r}
\def\td{t_d}

\def\nrp{N_r'}
\def\Ncl{N_{\rm{cl}}}
\def\twin{t_{\rm{win}}}
\def\sbr{\text{SBR}}
\def\median{\text{median}}

\def\KNN{K\text{NN}}
\def\pw{\text{pw}}
\def\sigbar{\overline{\sigma}}

\def\hdel{h_{\Delta}}
\def\bxtil{\widetilde{\bx}}
\def\pitil{\widetilde{\pi}}

\renewcommand{\P}{\mathbb{P}}

\newcommand{\cN}{\mathcal{N}}

\newcommand{\R}{\mathbb{R}}

\newcommand{\norm}[1]{\left\Vert#1\right\Vert}

\newcommand{\ones}{\bm{1}}
\newcommand{\what}{\widehat}

\def\bzero{{\boldsymbol 0}}

\def\de{{\rm d}}

\def\bc{{\boldsymbol c}}

\def\br{{\boldsymbol r}}
\def\bs{{\boldsymbol s}}

\def\bx{{\boldsymbol x}}

\def\cA{{\mathcal A}}

\def\cK{{\mathcal K}}
\def\cL{{\mathcal L}}

\def\cN{{\mathcal N}}
\def\cO{{\mathcal O}}

\def\cS{{\mathcal S}}

\def\cX{{\mathcal X}}

\newcommand{\eps}{\varepsilon}

\DeclareMathOperator*{\argmax}{arg\,max}

\def\ie{{\em{i.e.}}}

%% file: 00_abstract.tex
\begin{abstract}
Conventional wisdom suggests that single-photon lidar (SPL) should operate in low-light conditions to minimize dead-time effects.
Many methods have been developed to mitigate these effects in synchronous SPL systems. 
However, solutions for free-running SPL remain limited despite the advantage of reduced histogram distortion from dead times.
To improve the accuracy of free-running SPL, we propose a computationally efficient joint maximum likelihood estimator of the signal flux, the background flux, and the depth using only histograms, along with a complementary regularization framework that incorporates a learned point cloud score model as a prior.
Simulations and experiments demonstrate that free-running SPL yields lower estimation errors than its synchronous counterpart under identical conditions, with our regularization further improving accuracy.
\end{abstract}

%% file: 01_intro.tex
\section{Introduction}
\label{sec:intro}

Single-photon lidar (SPL) is a direct time-of-flight lidar approach that combines a picosecond laser, single-photon detectors
and timing electronics with picosecond resolution, enabling high-resolution ranging at longer distances than traditional time-of-flight and \RV{frequency-modulated continuous-wave (FMCW)} lidar systems. 
These features make SPL particularly suitable for applications such as terrain mapping~\cite{yu2020comparing,swatantran2016rapid}, underwater exploration~\cite{auroramaccaroneSubmergedSinglephotonLiDAR2023,shangguan2023compact,halimi2021robust}, and autonomous vehicles~\cite{rappAdvancesSinglePhotonLidar2020} 
where precision and efficiency are paramount. SPL's unique advantages have positioned it as a cornerstone technology for next-generation sensing and precision mapping.

Single-photon avalanche diodes (SPADs) are widely used in SPL.
One often overlooked aspect of SPAD modeling is \emph{dead time} --- the period after each photon detection during which the SPAD remains unarmed and cannot register additional photons.
A single photon arriving at an armed SPAD can trigger
a detectable avalanche of charge carriers, which must be quenched before the detector can be reactivated. 
The dead time facilitates quenching and mitigates afterpulsing effects~\cite{cova2013semiconductor}.
Many SPL studies assume a low-flux regime, where long photon inter-arrival times render dead-time effects negligible.
The \emph{5\% rule} suggests keeping flux low to ensure fewer than five detections per 100 laser pulse repetitions~\cite{oconnorTimecorrelatedSinglePhoton1984}.
However, applying these methods to high-flux settings can lead to suboptimal performance due to unmodeled dead-time effects. In this paper, we define ``dead time'' as the minimum duration a SPAD must remain unarmed after detection. Many SPADs, however, can be configured with a ``hold-off time,'' a longer unarmed duration determined by the detector’s reactivation policy.

There are two major approaches to reactivating the SPAD, as illustrated in \Cref{fig:overview}.
In the \emph{synchronous} (clock-driven) mode, the detector remains unarmed until the next laser pulse emission, even after a dead time has passed~\cite{pellegrini2000laser,pediredlaSignalProcessingBased2018,pawlikowskaSinglephotonThreedimensionalImaging2017}.
This approach results in statistically independent detections, simplifying the estimation of depth and reflectivity. However, synchronous measurements suffer from pile-up: a distortion of the histogram due to recording only the first detection in each laser cycle. The pile-up can bury the histogram's peak, which informs the target's distance.
Alternatively, in the \emph{free-running} (event-driven or asynchronous) mode, the detector is reactivated immediately after the dead time ends, maximizing its armed duration~\cite{antolovic2015nonuniformity,antolovic2018dynamic,rappDeadTimeCompensation2019}.
While the free-running mode 
has statistically dependent detection times and complex histogram statistics,
the distortion due to dead times is typically less severe than the pile-up effect in the synchronous mode.

In this paper, we directly compare these two modes and definitively show 
that the free-running mode achieves better performance.
Crucially, we introduce a computationally efficient maximum likelihood (ML) estimator that lowers the barrier to using the free-running mode.
Since log-likelihoods are commonly used as objective functions in regularization frameworks, we further propose a depth regularization algorithm that integrates a trained point cloud score model to exploit spatial structures in 3D reconstruction.

In summary, our main contributions are:
\begin{enumerate}
    \item Novel, computationally efficient joint ML estimators of signal flux, background flux, and depth from free-running and synchronous histograms.
    \item The \emph{Score-based SPL Depth Regularization} (SSDR) algorithm, which uses a point cloud score model as a prior.
\end{enumerate}
Simulation and experimental results reveal two key insights.
First, the free-running mode outperforms the synchronous mode by providing more accurate estimates of signal flux, background flux, and depth across all flux levels, maintaining accuracy at any depth, and performing better even at long dead times.
Second, the optimal flux for SPL is much higher than the recommended 5\% rule. Controlled attenuation for flux higher than the optimal level can enhance free-running accuracy, similar to synchronous mode~\cite{guptaPhotonFloodedSinglePhoton3D2019}.

\input{figs/fig-overview}

%% file: figs/fig-overview.tex
\begin{figure*}[ht]
    \centering
    \begin{tabular}{@{}c|c|c@{}}
         Ideal & Synchronous & Free-running 
        \\
        \begin{subfigure}{0.32\linewidth}
            \centering
            \includegraphics[width=\linewidth]{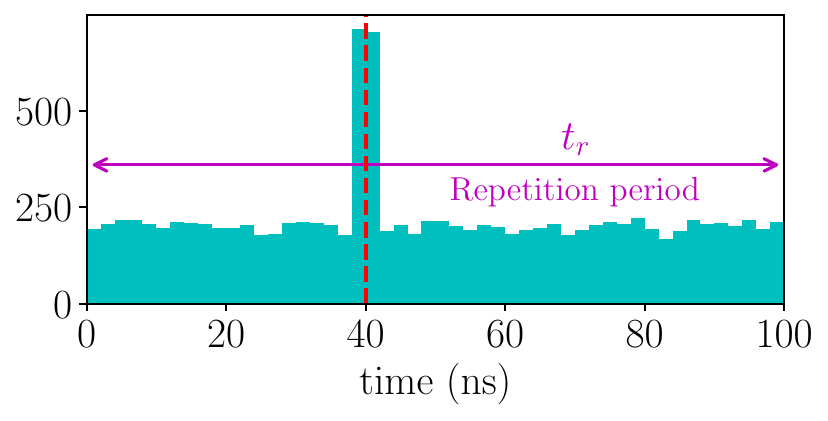}
        \end{subfigure}
        &
        \begin{subfigure}{0.32\linewidth}
            \centering
            \includegraphics[width=\linewidth]{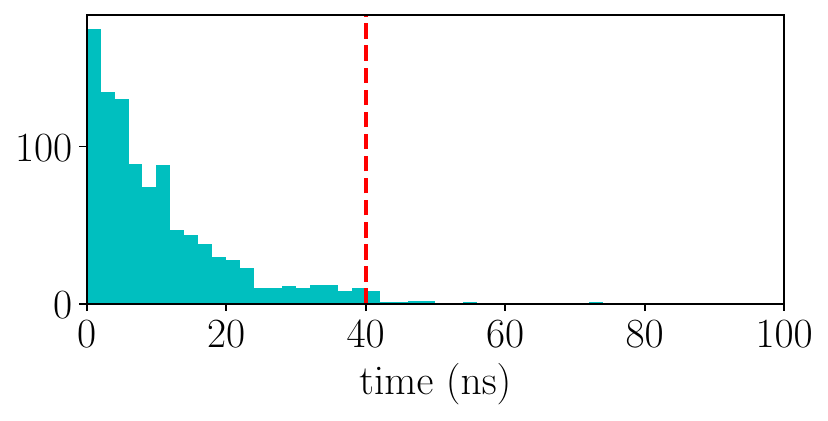}
        \end{subfigure}
        &
        \begin{subfigure}{0.32\linewidth}
            \centering
            \includegraphics[width=\linewidth]{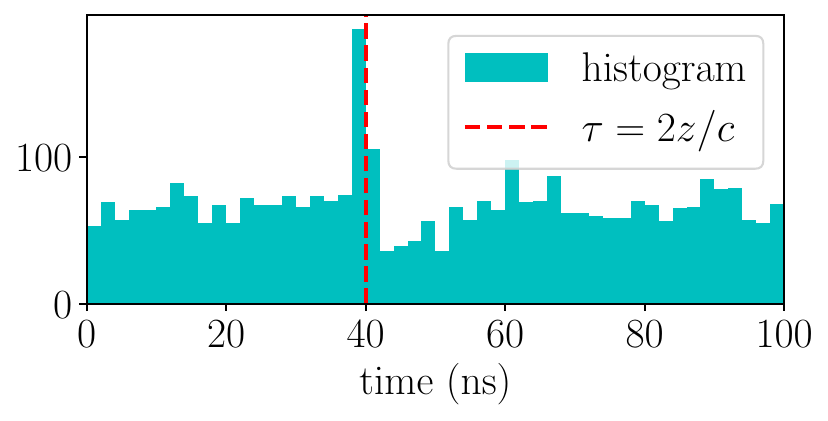}
        \end{subfigure}
        \\
        \begin{subfigure}{0.28\linewidth}
            \centering
            \includegraphics[width=\linewidth]{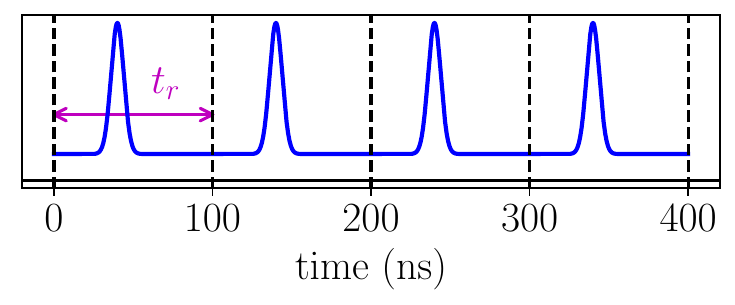}
        \end{subfigure}
         & 
         \begin{subfigure}{0.28\linewidth}
            \centering
            \includegraphics[width=\linewidth]{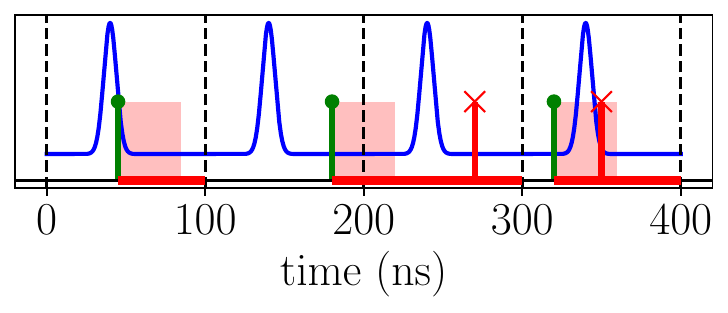}
        \end{subfigure}
         & 
         \begin{subfigure}{0.28\linewidth}
            \centering
            \includegraphics[width=\linewidth]{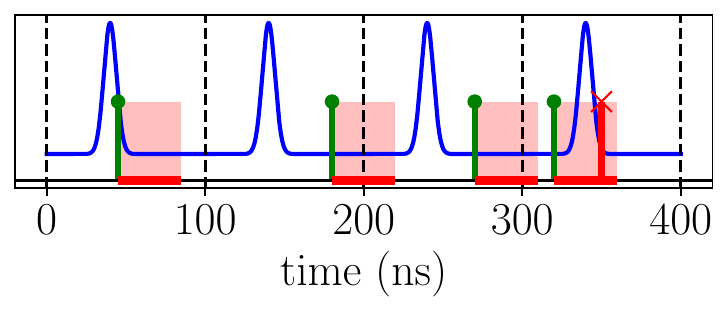}
        \end{subfigure}
    \end{tabular}
    \\
    \vspace{-0.2cm}
    \includegraphics[width=0.8\linewidth]{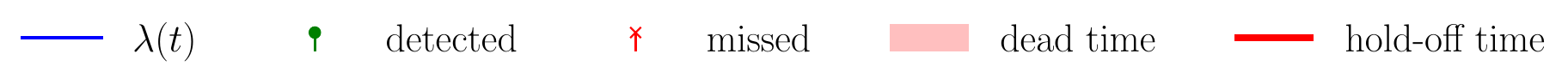}
    \\
    \vspace{0.5em}
    \begin{subfigure}{0.38\linewidth}
        \centering
        \includegraphics[width=\linewidth,trim={1.5cm 4cm 1.5cm 1cm},clip]{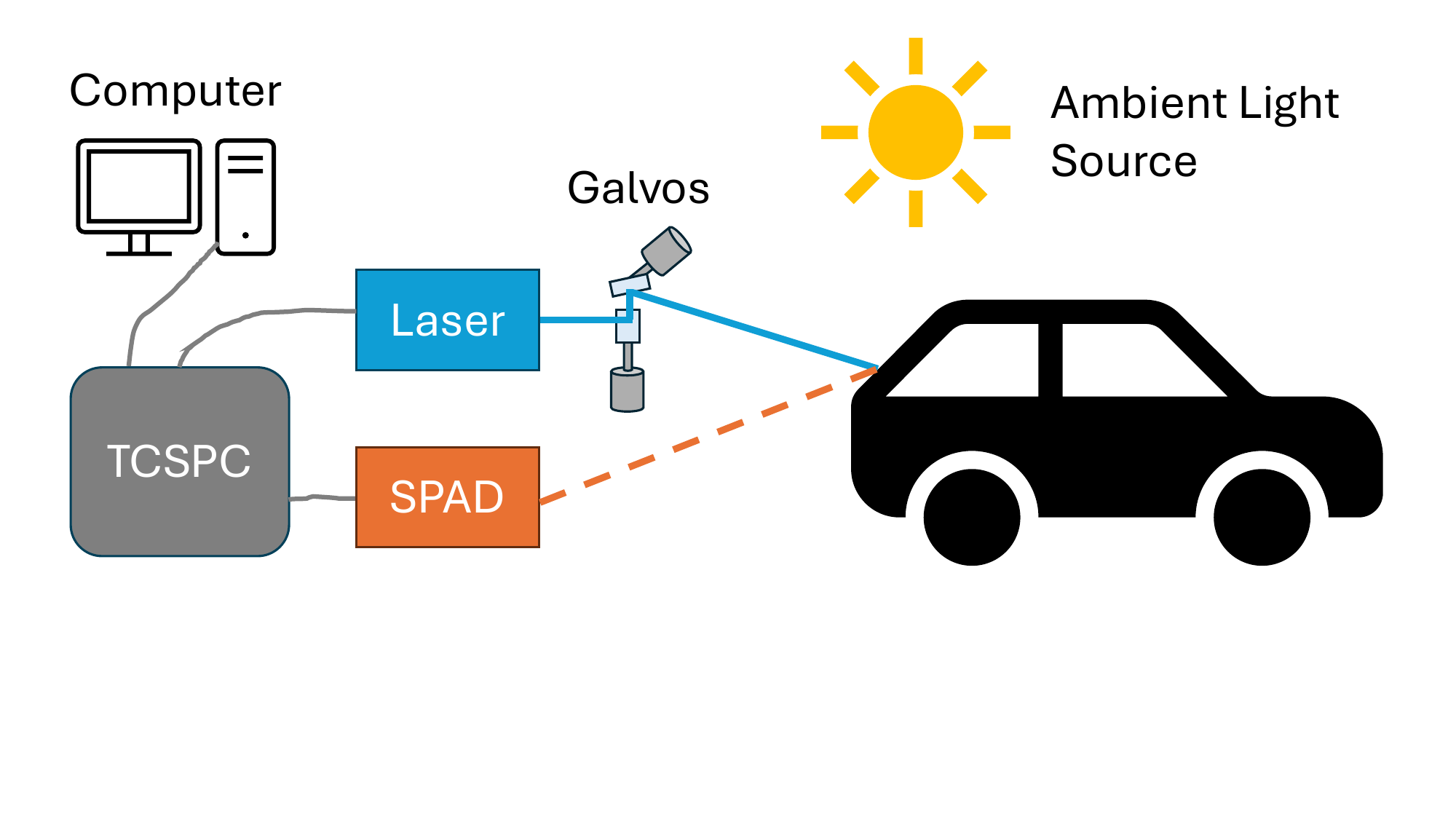}
    \end{subfigure}
    \hspace{0.5em}
    \begin{subfigure}{0.18\linewidth}
        \centering
        \caption*{\small{Sync ML}}
        \includegraphics[width=\linewidth,trim={2.2cm 1cm 5.5cm 2.7cm},clip]{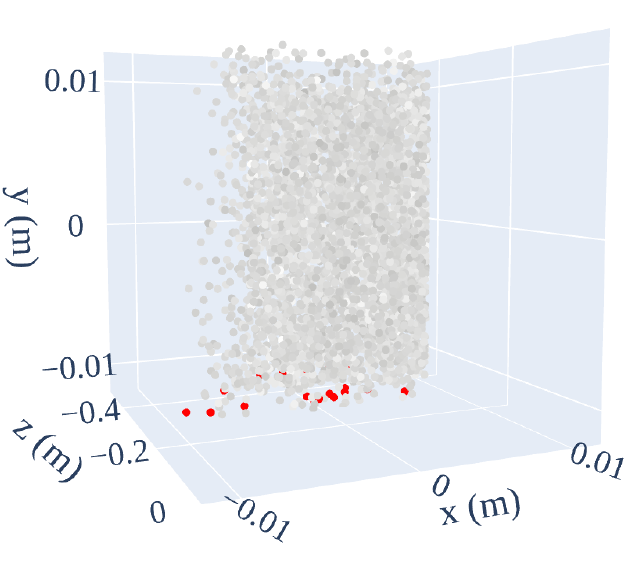}
        \caption*{{$\mae(\what{z}, z) = \SI{2.137}{\meter}$}}
    \end{subfigure}
    \begin{subfigure}{0.18\linewidth}
        \centering
        \caption*{\small{Free ML}}
        \includegraphics[width=\linewidth,trim={2.2cm 1cm 5.5cm 2.7cm},clip]{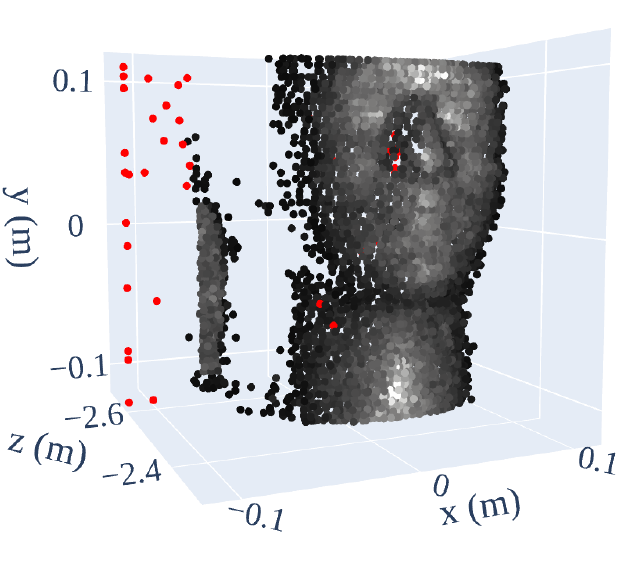}
        \caption*{{$\mae(\what{z}, z) = \SI{0.136}{\meter}$}}
    \end{subfigure}
    \begin{subfigure}{0.18\linewidth}
        \centering
        \caption*{\small{Free ML + SSDR}}
        \includegraphics[width=\linewidth,trim={2.2cm 1cm 5.5cm 2.7cm},clip]{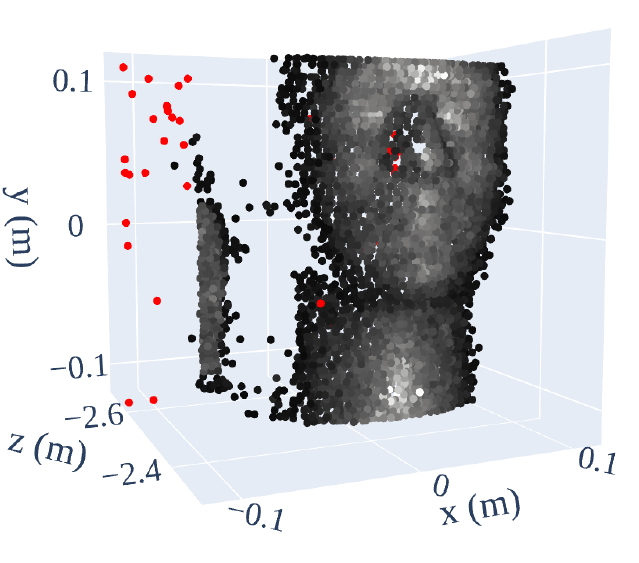}
        \caption*{{$\mae(\what{z}, z) = \SI{0.098}{\meter}$}}
    \end{subfigure}
    \caption{Overview and result highlights.
    \textbf{Top:} The histograms of detection times with illustrations of detector's reactivation protocol. 
    \emph{Top left:} when there is no dead time, the histogram's shape resembles one period of the photon arrival intensity $\lambda(t)$.
    \emph{Top center:} a synchronous detector is reactivated only at the beginning of a repetition period, resulting in pile-up that can bury the histogram's peak.
    \emph{Top right:} a free-running detector is reactivated whenever the dead time has passed, resulting in maximizing the detector's active duration. 
    The histogram's shape is distorted from $\lambda(t)$, but there is still a clear peak.
    \textbf{Bottom left:} 
    Illustration of a SPL system.
    A \emph{Time-Correlated Single Photon Counting} (TCSPC) system is the timing elctronics.
    \textbf{Bottom right:} Pixelwise ML estimation using free-running measurements achieve lower error than that of synchronous measurements. Score-based depth regularization further improves the 3D reconstruction.
    }
    \label{fig:overview}
\end{figure*}

%% file: 02_related.tex
\section{Related Work}
\label{sec:related}

\subsection{High-flux Single-Photon Lidar}

\textbf{Synchronous mode:}
Fixed time gating is applicable when the target's depth is within a known range.
For example, in long-range imaging, it is used to suppress the background and extend the unambiguous depth range~\cite{mccarthy2009long,mccarthy2013kilometer,pawlikowskaSinglephotonThreedimensionalImaging2017}.
If prior knowledge of the depth is not available, uniform gating~\cite{guptaAsynchronousSinglePhoton3D2019} and adaptive gating~\cite{po2022adaptive} can also mitigate the pile-up.
Coates's correction~\cite{coatesCorrectionPhotonPileup1968}, which is ML estimation of the discretized photon arrival intensity, \ie, the \emph{transient}~\cite{pediredlaSignalProcessingBased2018,guptaPhotonFloodedSinglePhoton3D2019,heideSubpicosecondPhotonefficient3D2018}, is a purely computational approach to pile-up mitigation.
Some hardware approaches include~\cite{beer2018background,acconcia2018fast}.

\noindent\textbf{Free-running mode:} 
Rapp et al. showed that the statistical dependence of detection times in free-running mode results in a Markov chain, so ranging and transient estimation can be performed by computing the stationary distribution~\cite{rappDeadTimeCompensation2019,rappHighfluxSinglephotonLidar2020}.
Alternatively, \citet{zhangParametricModelingEstimation2024} proposed a parametric model of the histogram, circumventing the computationally expensive stationary distribution computation~\cite{zhangParametricModelingEstimation2024}.
Free-running detectors are also used in passive imaging, where dead times constrain dynamic range \cite{ingleHighFluxPassive2019,a.inglePassiveInterPhotonImaging2021} and complicate frequency estimation \cite{weiPassiveUltraWidebandSinglePhoton2023}.

\vspace{-1mm}
\subsection{Regularization for Single-Photon Lidar}

\textbf{Transient Estimation:}
in transient estimation, the parameters to be estimated are the discretized photon arrival intensity.
Since the log likelihood is concave, many works proposed to use variational regularization strategies often used in inverse problems~\cite{heideSubpicosecondPhotonefficient3D2018,otooleReconstructingTransientImages2017}, some specializing their methods for scenarios with complex transients, such as multidepth estimation~\cite{shinComputationalMultidepthSinglephoton2016,halimiRestorationMultilayeredSinglephoton2017}, through obscurants~\cite{halimi2021robust,tobinRobustRealtime3D2021}, and non-line-of-sight imaging~\cite{heide2019non,liu2021non,huang2023non}.

\noindent\textbf{Scene parameter estimation}
In a single-depth setting, the estimation parameters are the signal flux, background flux, and the depth, or their subsets.
The log likelihood is nonconcave, so proposed methods are often sampling-based in the Bayesian framework, such as~\cite{altmannRobustBayesianTarget2016a,tachellaBayesian3DReconstruction2019}.
For dynamic scenes, spatiotemporal priors have been proposed~\cite{altmannFastOnline3D2020,legrosRobust3DReconstruction2021}.
Some of the other approaches include nonlocal correlation~\cite{chen2019learning} and plug-and-play prior~\cite{tachellaRealtime3DReconstruction2019}.

Some regularization methods assume a uniform grid of scan locations, enabling priors like total variation that rely on neighboring pixels~\cite{heideSubpicosecondPhotonefficient3D2018,otooleReconstructingTransientImages2017}. However, in scanning systems, transverse locations depend on depth, and scan angles can be arbitrarily chosen. As a result, neighboring pixels may have varying transverse distances, requiring spatially adaptive regularization. 
\RV{While \citet{tachellaRealtime3DReconstruction2019} was first to use a point cloud denoiser for regularizing SPL reconstructions, 
t}o our knowledge, SSDR is the first SPL regularization method to apply a learned model directly to 3D point clouds, allowing it to handle arbitrary scan locations.

%% file: 03_mle.tex
\section{Pixel-wise Maximum Likelihood Estimation} \label{sec:mle}

We first describe a photon arrival model 
and the log likelihood function for each detector mode. Then, we propose computationally efficient ML estimators.
The relevant derivations are deferred to the Supplement.

\subsection{Photon Arrival Model}

At each scan location (pixel), the laser emits $n_r$ pulses with repetition period $t_r$.
The per-pixel acquisition time is therefore $\nr \tr$.
The pulse's temporal profile is $f(t)$, which is normalized such that $\int_{-\infty}^\infty f(t) \de{t} = 1$.
For simulation and experiments, we assume that the pulse profile is Gaussian: $f(t) = \exp(-t^2 / (2w^2)) / \sqrt{2 \pi} w$, where $w$ is the pulse width.
The photon arrivals at the detector follows an inhomogeneous Poisson process with an intensity function
\begin{equation}
    \lambda(t) := \RV{\sum_{k = 0}^{n_r - 1}} \lamtil(t - k t_r), \quad t \in [0, n_r t_r) \label{eq:arrivalIntensity}
\end{equation}
where the single-period intensity is
\begin{equation} \label{eq:singleIntensity}
    \lamtil(t) = S f(t - 2z / c) + b, \quad t \in [0, t_r).
\end{equation}
The signal flux $S$ is the mean number of photons from the laser detected in one repetition period, absorbing the effects of laser power, reflectivity, range falloff, and detector efficiency.
The background intensity is $b = B / \tr$, where the background flux $B$ is the mean number of detections due to ambient light and dark counts per repetition period.
The signal-to-background ratio (SBR) is defined as $S / B$.
We denote the total flux by $\Lambda := S + B$.
The time-of-flight $\tau$ is the round-trip travel time of a laser pulse and is related to the depth $z$ by $\tau := 2z / c$, where $c$ is the speed of light.
We assume that $z \in [0, z_{\max})$, where $z_{\max} = c \tr / 2$, ensuring $\tau \in [0, \tr)$. Exceeding this range causes distance aliasing.
Let $\td$ denote the detector's dead time.
We assume that $\td$ is longer than the electronics dead time, which could otherwise further complicate the measurement model~\cite{rappHighfluxSinglephotonLidar2020}.

\subsection{Data Retention}

Suppose there are $N$ detections during the acquisition interval $[0, \nr\tr)$.
Let $\Tseq$ denote the sequence of \emph{absolute detection times} relative to $t = 0$.
Each $T_i$ has a corresponding \emph{relative detection time} $X_i = T_i \bmod \tr$, which is the time elapsed between $T_i$ and the most recent pulse emission.
Though the timing electronics could be configured to record $\Tseq$, typical SPL systems retain only the histogram of quantized relative detection times.
The theory in this paper neglects quantization effects. 
The SPL system is assumed to retain an unordered set of relative detection times $\Xset$.

\subsection{Ideal Detector}

A hypothetical detector with no dead time serves as our gold standard.
If the detector is always armed, the detection process is identical to the photon arrival process (which has included the detector's efficiency).
The log likelihood of an ideal measurement $\Xset$ follows from the sample function density of an inhomogeneous Poisson process~\cite{snyderRandomPointProcesses2012}:
\begin{equation}
    \cL^{\id} = - n_r \Lambda + \sum_{i = 1}^N \log \lamtil(X_i). \label{eq:lfLogLikelihood}
\end{equation}
We emphasize that $\cL^{\id}$ and other log likelihoods in this section are functions of the scene parameters: $S$, $B$, and $z$.
Practical detectors, whether synchronous or free-running, behave similarly to an ideal detector when the total flux $\Lambda$ is low, so the 5\% rule, \ie, $\Lambda < 0.05$, is traditionally recommended~\cite{oconnorTimecorrelatedSinglePhoton1984}.
Although this photon detection model is simple, it fails to capture real SPL systems in high-flux regimes where dead-time effects are not negligible.

\subsection{Synchronous Detector} \label{sec:syncDetector}

In the synchronous mode, the detector is reactivated when the next laser pulse is emitted, following a hold-off time, which subsumes the dead time after each photon detection.
As demonstrated in \Cref{fig:overview}, a synchronous detector may be active for fewer repetition periods than $\nr$, because some hold-off times span across two (or more) periods.
Let $\nrp$ denote the number of armed repetition periods. Given a measurement $\Xset$, the number of armed periods is $\nrp = \nr - \sum_{i = 1}^N \ones_{[0, \tr - \td)}(X_i)$, where $\ones_{\cA}(x)$ is an indicator function taking the value 1 when $x \in \cA$ and 0 otherwise.

In the synchronous method, only the first photon detection after each laser emission is recorded, resulting in statistically independent detection times.
The log likelihood of a synchronous measurement conditioned on $\nrp$ is
\begin{equation}
    \cL^{\sc} = -(\nrp - N) \Lambda + \sum_{i = 1}^N \log (\lamtil(X_i)) - \Phitil(X_i), \label{eq:syncLogLikelihood}
\end{equation}
where $\Phitil(x) := \int_0^t \lamtil(t') \de{t'} = S F(t - 2z / c) + B t$ is the single-period cumulative flux,
and $F(t) := \int_0^t f(t') \de{t'}$ is the cumulative pulse profile.
We will take this conditional log likelihood $\cL^{\sc}$ as the objective for ML estimation.
We remark that $\cL^{\sc}$ is the (unconditional) log likelihood if the number of active repetition periods is fixed.

\subsection{Free-running Detector}

In the free-running mode, the detector is reactivated immediately after a dead time. The absolute detection times follow a \emph{self-exciting point process}, whose intensity is either $\lambda(t)$ when the detector is armed or $0$ when it is unarmed~\cite{snyderRandomPointProcesses2012}.
An approximation of the log likelihood of \emph{absolute} detection times can be expressed in terms of \emph{relative} detection times:
\begin{equation}
    \cL^{\fr} \approx - \nr \Lambda + \sum_{i = 1}^N \log \lambda(T_i) + \Phi(X_i + \td) - \Phi(X_i), \label{eq:freeLogLikelihood}
\end{equation}
where $\Phi(t) = \int_0^t \lambda(t') \de{t'}$ is the cumulative flux.
The approximation error is relatively small for a large $N$.

\begin{algorithm}[t]
\caption{ML Estimation for SPL} \label{alg:mlEst}
\begin{algorithmic}[1]
\STATE \textbf{Input:} Initialization $(\what{S}_0, \what{B}_0, \what{z}_0)$, Relative detection times $\Xset$, Log likelihood function $\cL$, Number of iterations $K$
\FOR{$k = 1, \ldots, K$}
    \STATE $\what{S}_k, \what{B}_k = \argmax_{S, B} \cL(S, B, \what{z}_k; \Xset)$ \label{eq:fluxUpdate}
    \STATE $\what{z}_k = \argmax_z \cL(\what{S}_k, \what{B}_k, z; \Xset)$ \label{eq:zUpdate}
\ENDFOR
\STATE \textbf{Return:} Estimate $(\what{S}_k, \what{B}_k, \what{z}_k)$
\end{algorithmic}
\end{algorithm}

\subsection{Pixel-wise Maximum Likelihood Estimation}
\label{sec:estimation}

The joint ML estimator of the scene parameters $S$, $B$, and $z$ from a measurement $\Xset$ is
\begin{equation}
    \hat{S}, \hat{B}, \hat{z} = \argmax_{S, B, z} \mathcal{L} (S, B, z; \Xset), \label{eq:mlEst}
\end{equation}
where $\mathcal{L}$ is the log likelihood function, which is one of~\eqref{eq:lfLogLikelihood}, \eqref{eq:syncLogLikelihood}, or~\eqref{eq:freeLogLikelihood}, depending on the detector's mode.
If some of the parameters are known or estimated a priori, then the other parameters can be estimated by optimizing the same objective while fixing the known parameters.

While joint ML estimation of $S$, $B$, and $z$ has been proposed for an ideal detector \cite{kitichotkulRoleDetectionTimes2023}, its extension to synchronous and free-running systems is novel. Computing the ML estimator in \eqref{eq:mlEst} is challenging because all three log likelihoods are nonconcave in $z$, despite being concave in $S$ and $B$. Efficiently maximizing the log-likelihood with respect to $z$ is thus crucial for practical joint ML estimation. 
For ideal systems, ML depth estimation can be framed as matched filtering \cite{bar-davidCommunicationPoissonRegime1969}. Here, we propose analogous matched filtering algorithms for free-running and synchronous measurements.

Following~\cite{kitichotkulRoleDetectionTimes2023}, we implement \eqref{eq:mlEst} using alternating maximization, as detailed in \Cref{alg:mlEst}. We solve the maximization over $S$ and $B$ in line~\ref{eq:fluxUpdate} using the L-BFGS-B algorithm~\cite{zhu1997algorithm}. For the maximization over $z$ in line~\ref{eq:zUpdate}, we apply matched filtering, as discussed in \Cref{sec:mf}. Initial estimates are obtained using censoring-based estimators~\cite{rappFewPhotonsMany2017}.

The proposed ML estimators for the three detector types offer two key advantages over previous methods. First, they enable joint estimation of $S$, $B$, and $z$, whereas prior approaches require a separate calibration for $B$ \cite{rappDeadTimeCompensation2019} or focus solely on ranging \cite{guptaPhotonFloodedSinglePhoton3D2019}. 
Second, the proposed estimators' ranging accuracy is independent of bin size. Although a practical matched filtering implementation requires histograms of quantized detection times, depth estimates can be refined via gradient-based optimization. This allows benchmarking in simulations with floating-point-resolution detection times, providing insights into performance across various lighting conditions, depths, and dead times, as demonstrated in \Cref{sec:numexp}.
Moreover, there are other advantages, including improved accuracy for synchronous measurements and faster computation for free-running meausurements.

\subsection{Ranging as Matched Filtering} \label{sec:mf}

\input{figs/fig-baselines}

For simplicity, we write the estimators in terms of the time-of-flight $\tau$ which can be converted to the depth by $z = c \tau / 2$.

\subsubsection{Synchronous Ranging}

The ML estimator of $\tau$ for the synchronous mode is
\begin{equation}
    \tauhat^{\sc} = \argmax_\tau h(\tau) \oplus u(\tau),
\end{equation}
where $\oplus$ denotes correlation, and
\begin{align}
    h(t) &= \sum_{i = 1}^N \delta(t - X_i), \quad t \in [0, t_r), \label{eq:histDef} \\
    u(t) &= \log (S f(t) + b) - S F(t). \label{eq:syncU}
\end{align}
When the detection times are quantized, $h(t)$ is effectively the histogram (see Supplement).
This ML estimator is different from taking the maximum of a Coates-corrected histogram, which is the ML estimate of the photon arrival intensity~\cite{guptaAsynchronousSinglePhoton3D2019}, but it is not necessarily ML estimation of the depth.
As shown in \Cref{fig:base-rmse}, the ML estimator achieves more accurate ranging than Coates's correction.

\subsubsection{Free-running Ranging}

The ML estimator of $\tau$ for the free-running mode is
\begin{equation} \label{eq:freeDepth}
    \tauhat^{\fr} = \argmax_\tau h(\tau) \oplus u(\tau) + g(\tau) \oplus v(\tau),
\end{equation}
where $v(\tau) = S F(\tau)$, and
\begin{align}
    g(t) &= \sum_{i = 1}^N \delta(t - (X_i + \td) \bmod \tr), \quad t \in [0, t_r)
\end{align}
is $h(t)$ that is cyclic-shifted in time by $\td$.
\RV{\Cref{fig:base-rmse}} shows that the ML depth estimator matches the accuracy of \citet{rappDeadTimeCompensation2019}'s method in significantly less time.
Moreover, when $S$ and $B$ are not known, their method requires absolute detection times for flux estimation, whereas our joint ML estimator uses only relative detection times. 

\subsection{Simulation Results} \label{sec:numexp}

We benchmark the errors of the proposed ML estimators for the three detector modes using simulation in which ground truths are known.
We use root-mean-square error (RMSE) as the metric, where $\rmse(\what{\theta}, \theta)$ denotes the RMSE for the estimator $\what{\theta}$ of the parameter $\theta$.
Unless stated otherwise, the following settings are used for all of the simulation results: $\tr = \SI{100}{\nano\second}$ and $\nr = 100$, resulting in the acquisition time $\nr\tr = \SI{10}{\micro\second}$.
The depth is $z = z_{\max} / 2 = \SI{7.49}{\meter}$.
The pulse is Gaussian with width $w = \SI{0.1}{\nano\second}$. The dead time is $t_d = \SI{20}{\nano\second}$. All error statistics are computed from 10000 Monte Carlo trials.
The ML estimators jointly estimate the signal flux, the background flux, and the depth with no prior information about any of the three parameters.
Except for the results previously shown in \Cref{fig:baselines}, the detection times are not quantized.
For any matched filtering step, the detection times and the filters are quantized into uniform \SI{10}{\pico\second} bins, and the estimates from \Cref{alg:mlEst} are further refined by maximizing the log likelihood using the L-BFGS-B algorithm~\cite{zhu1997algorithm}.

\subsubsection{Synchronous Performance is Depth Dependent} \label{sec:numexp-depth}

\input{figs/fig-numexp-2}

Pile-up in synchronous mode reduces estimation accuracy for $S$ and $z$ at longer depths, while free-running performance remains consistent across depths.
As shown in the left column of \Cref{fig:num-exp-2}, the RMSEs of signal flux and depth estimates remain constant with depth in free-running mode but increase in synchronous mode after an initially similar RMSEs.
In simulation, we fix $S = 1$ and $\sbr = 0.1$. We vary the depth: $z \in [0.1 z_{\max}, 0.9 z_{\max}] = [\SI{1.50}{\meter}, \SI{13.50}{\meter}]$.
The RMSE of background flux estimates remains nearly constant in all SPL modes.

\subsubsection{Free-running Wins Even At Long Dead Time} 

The RMSEs of $S$, $B$, and $z$ estimates in free-running mode increase with dead time, while those in synchronous mode remain almost constant. However, the free-running RMSEs remain lower even at the longest dead time.
As shown in the right column of \Cref{fig:num-exp-2}, RMSEs for all estimates from free-running measurements are lower than those from synchronous measurements across $\td \in [\SI{0}{\nano\second}, \SI{90}{\nano\second}]$, \ie, covering up to 90\% of $\tr$.
Only the free-running depth RMSE approaches the synchronous value at the highest $\td$. 
As in \Cref{sec:numexp-depth}, the simulation uses $S = 1$ and $\sbr = 0.1$.

\subsubsection{Free-running is Optimal at Higher Flux}

\input{figs/fig-numexp-1}

Similar to the optimal flux for synchronous mode \cite{guptaPhotonFloodedSinglePhoton3D2019}, an optimal flux $\Lambda^*$ minimizing $\rmse(\what{z}, z)$ at a specific SBR also exists for free-running mode. 
\Cref{fig:optflux} shows $\Lambda^*$ across $\sbr \in [0.1, 10]$. 
The free-running $\Lambda^*$ is higher than that in synchronous mode across the SBR range, even though free-running $\Lambda^*$ decreases with SBR whereas synchronous $\Lambda^*$ remains relatively unchanged.
We observe a similar phenomenon for $S$ estimation when the metric is the normalized RMSE (NRMSE), defined as $\rmse(\what{\theta}, \theta) / \theta$ for an estimate $\what{\theta}$ of parameter $\theta$. 
However, the optimal flux minimizing $\nrmse(\what{S}, S)$ at a fixed SBR differs from that for the depth.  
Similar to synchronous SPL \cite{guptaPhotonFloodedSinglePhoton3D2019}, controlled flux attenuation can improve estimation of $S$ and $z$ for free-running mode.

RMSEs from free-running estimates remain consistently lower than those from synchronous estimates across various total flux levels and SBRs. As shown in \Cref{fig:numexp-1-s,fig:numexp-1-b,fig:numexp-1-z}, the free-running mode outperforms the synchronous mode over a wide range of conditions. We vary the signal flux $S$ from 0.01 to 10 and set SBR to 0.1, 0.5, and 1, resulting in total flux ranging from 0.02 (at $S = 0.1$ and $\sbr = 1$) to 110 (at $S = 10$ and $\sbr = 0.1$). In the low-flux regime, RMSEs for both free-running and synchronous modes converge to those of an ideal system, confirming that dead-time effects become negligible.
At $\sbr = 0.1$, synchronous SPL fails for $S > 0.2$, while free-running SPL maintains accurate estimates, achieving the lowest $\nrmse(S, \what{S})$ and $\rmse(z, \what{z})$ at $S \approx 1$. 
For background flux estimation, NRMSE increases as SBR decreases due to higher $B$, leading to more background photon detections. 

%% file: figs/fig-baselines.tex
\begin{figure}[t]
    \centering
    \begin{subfigure}{0.48\linewidth}
        \centering
        \caption{Depth Error} \label{fig:base-rmse}
        \includegraphics[width=\linewidth]{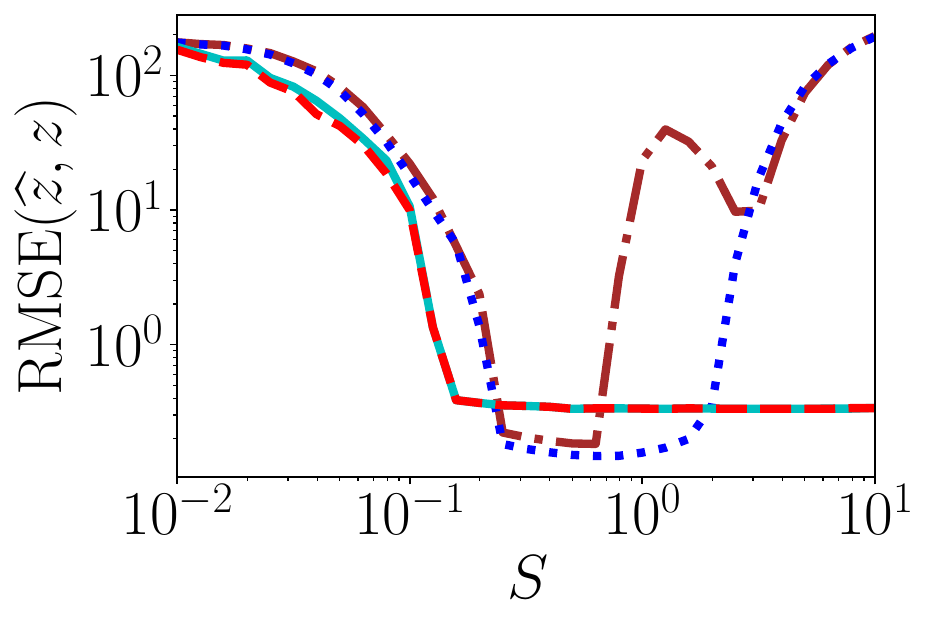}
    \end{subfigure}
    \begin{subfigure}{0.49\linewidth}
        \centering
        \caption{Running Time} \label{fig:base-time}
        \includegraphics[width=\linewidth]{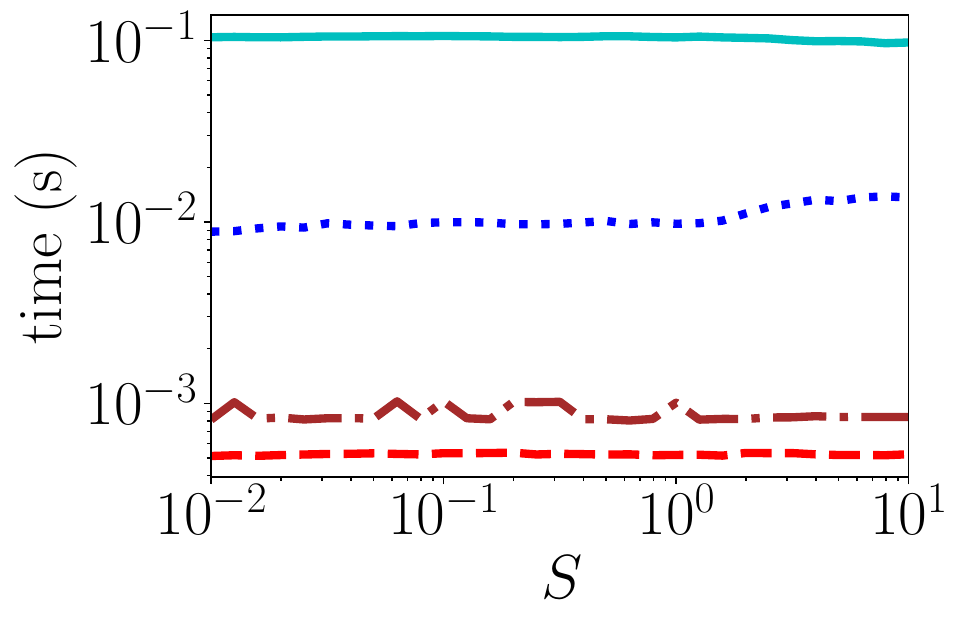}
    \end{subfigure}
    \includegraphics[width=\linewidth]{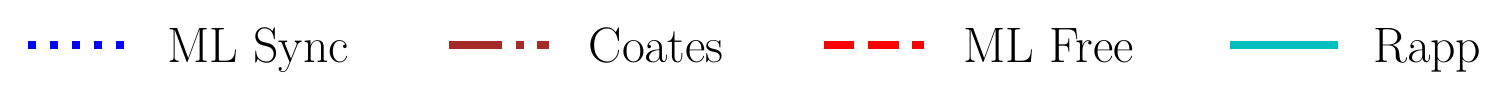}
    \caption{\RV{\textbf{(a) RMSE} and \textbf{(b) average running time} of depth estimators at varying $S$ and $\sbr = 0.5$. \Cref{sec:numexp} describes other SPL settings.
    ``ML Sync'' is the joint ML estimator~\eqref{eq:mlEst} for synchronous mode, outperforming Coates's correction when $S \in [0.6, 1.2]$.
    ``ML Free'' is~\eqref{eq:freeDepth}.
    Both ``ML Free'' and \citet{rappDeadTimeCompensation2019}'s method have access to ground truth $S$ and $B$.
    The histogram bin size is \SI{10}{\pico\second} for synchronous mode but \SI{100}{\pico\second} for free-running mode to limit ``Rapp'' running time.}
    }
    \label{fig:baselines}
    \vspace{-5mm}
\end{figure}

%% file: figs/fig-numexp-2.tex
\def\colFigWidth{3.5cm}
\def\figTextSize{\small}

\begin{figure}[tb]
\centering
\begin{tabular}{c@{\,}c@{\,}c}
    & \hspace{1.5mm} \figTextSize Vary Depth & \hspace{1.5mm} \figTextSize Vary Dead Time 
    \\
    \vspace{-1ex}
    \rotatebox[origin=l]{90}{\hspace{7mm} \figTextSize $\rmse(\what{S}, S)$}
    & 
    \begin{subfigure}[t]{\colFigWidth}
        \centering
        \includegraphics[width=\colFigWidth]{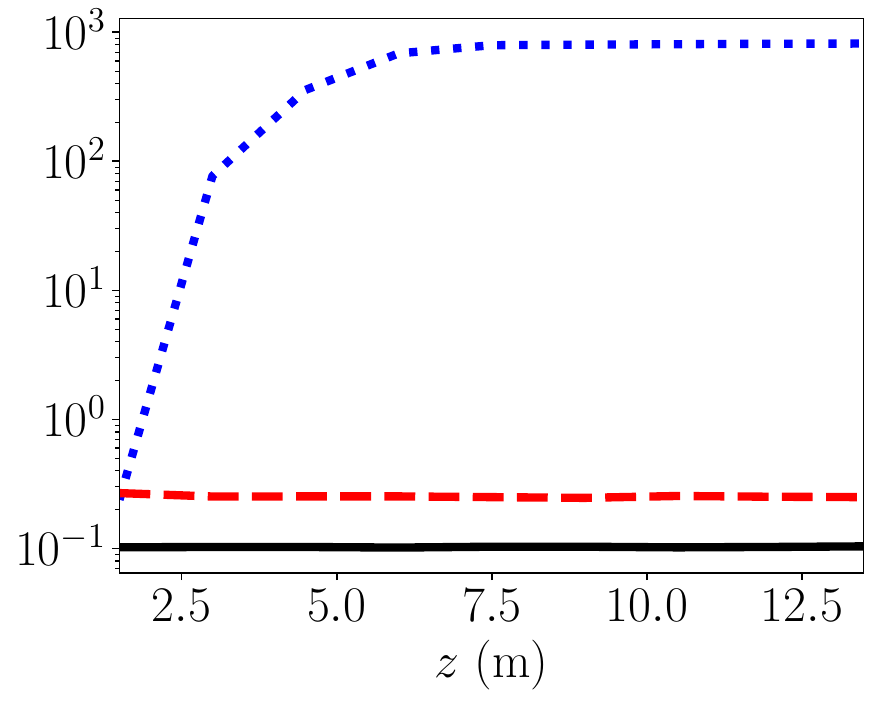}
    \end{subfigure}
    & 
    \begin{subfigure}[t]{\colFigWidth}
        \centering
        \includegraphics[width=\colFigWidth]{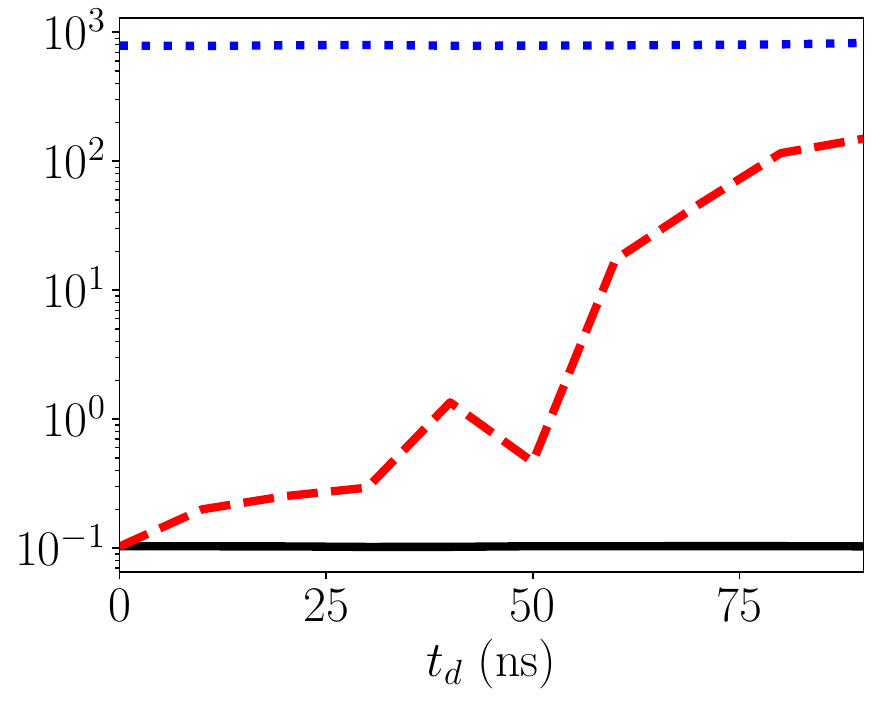}
    \end{subfigure}
    \\
    \rotatebox[origin=l]{90}{\hspace{7mm} \figTextSize $\rmse(\what{B}, B)$}
    & 
    \begin{subfigure}[t]{\colFigWidth}
        \centering
        \includegraphics[width=\colFigWidth]{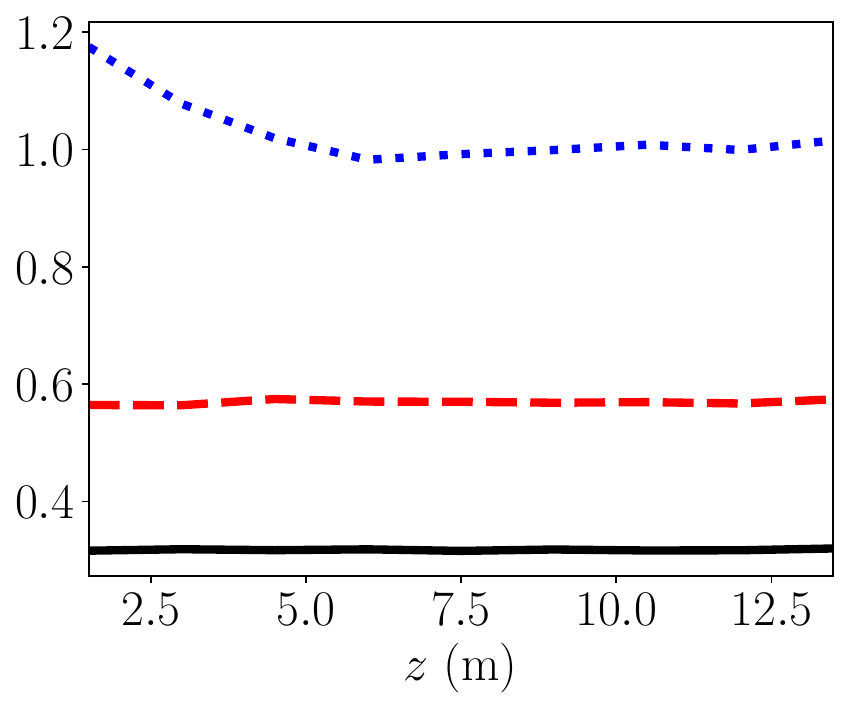}
    \end{subfigure}
    & 
    \begin{subfigure}[t]{\colFigWidth}
        \centering
        \includegraphics[width=\colFigWidth]{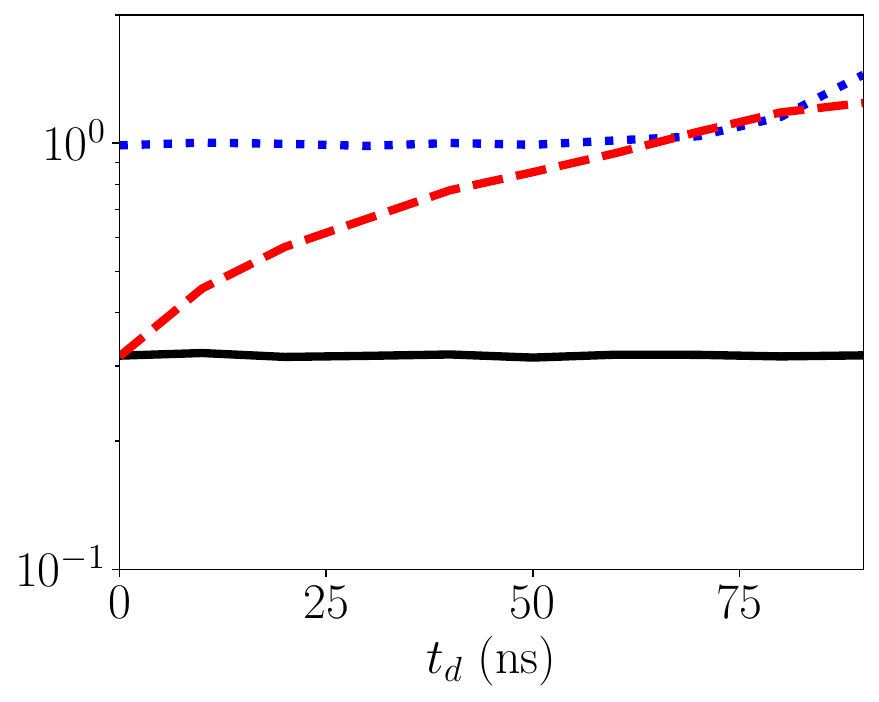}
    \end{subfigure}
    \\
    \rotatebox[origin=l]{90}{\hspace{7mm} \figTextSize $\rmse(\what{z}, z)$}
    & 
    \begin{subfigure}[t]{\colFigWidth}
        \centering
        \includegraphics[width=\colFigWidth]{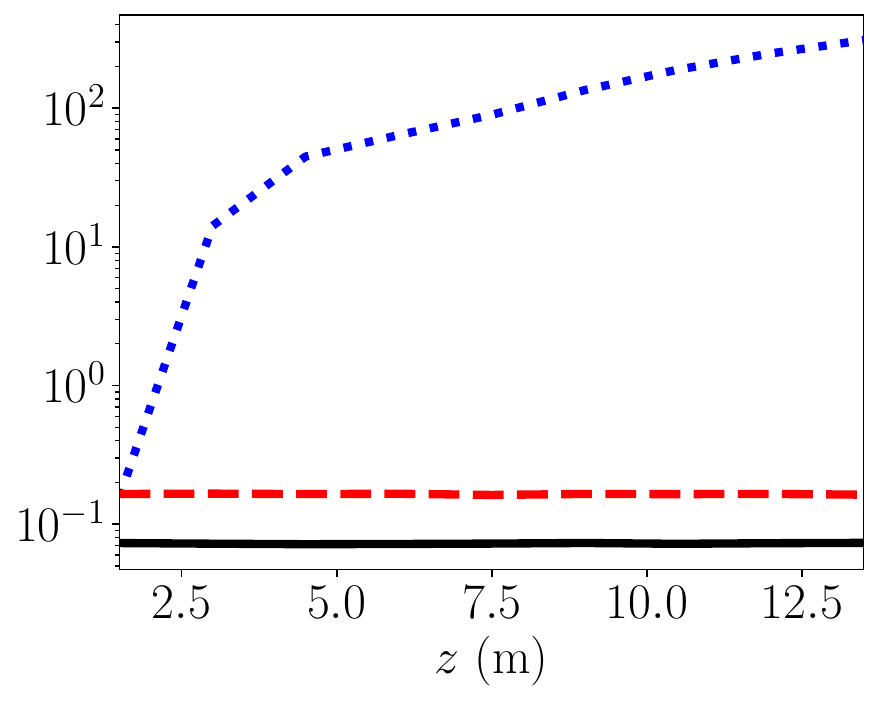}
    \end{subfigure}
    & 
    \begin{subfigure}[t]{\colFigWidth}
        \centering
        \includegraphics[width=\colFigWidth]{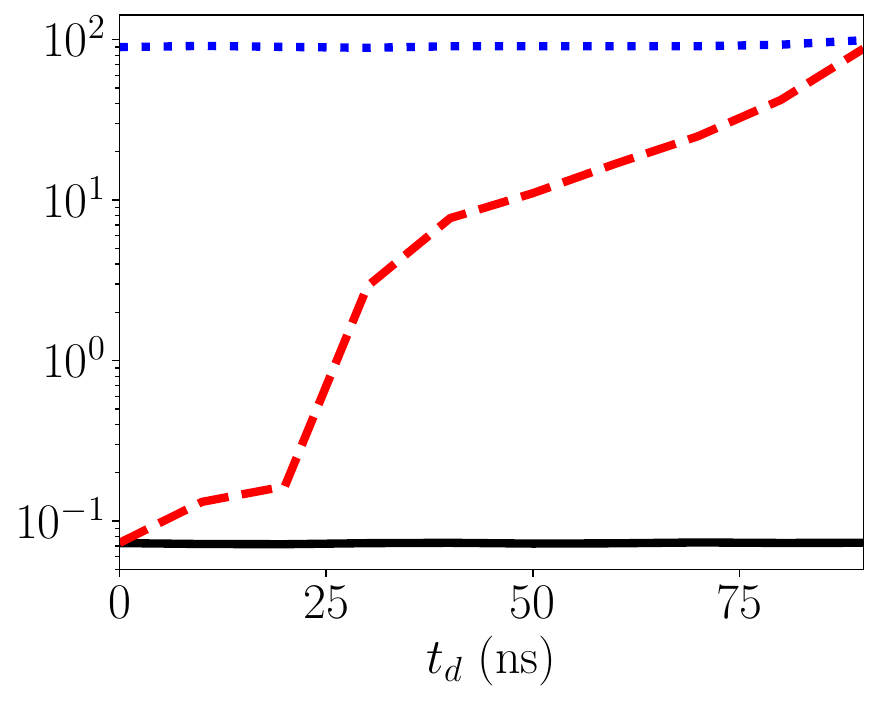}
    \end{subfigure}
    \\
    \multicolumn{3}{l}{}
    \hspace{-2mm}
    \begin{subfigure}[t]{0.93\columnwidth}
        \flushleft
        \includegraphics[width=0.93\columnwidth]{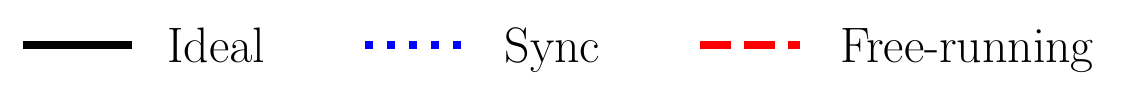}
    \end{subfigure}
\end{tabular}
\caption{RMSEs of joint ML estimates for the three detector modes at different depths \textbf{(Left)} and dead times \textbf{(Right)}.
}
\label{fig:num-exp-2}
\vspace{-5mm}
\end{figure}

%% file: figs/fig-numexp-1.tex
\begin{figure*}[htb]
    \centering
    \begin{subfigure}{0.234\linewidth}
        \centering
        \caption{Optimal flux $\Lambda^*$} \label{fig:optflux}
        \includegraphics[width=\linewidth]{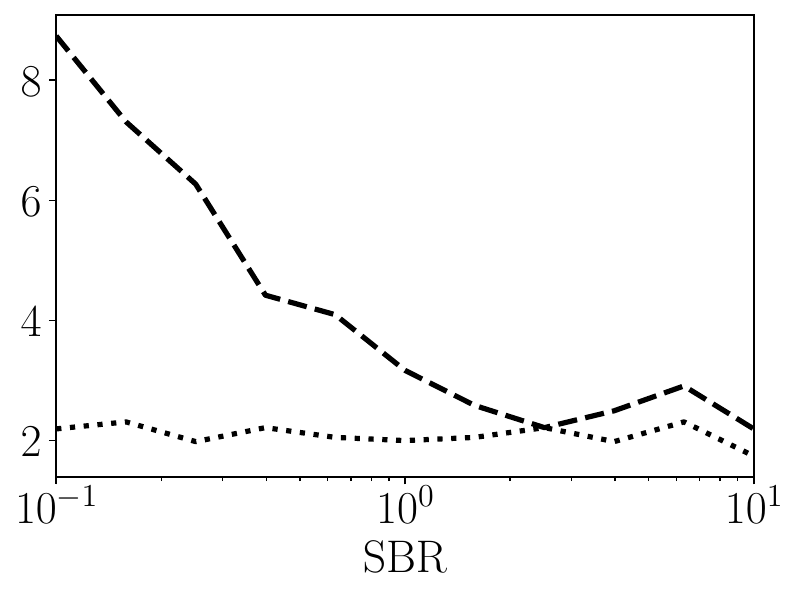}
    \end{subfigure}
    \begin{subfigure}{0.25\linewidth}
        \centering
        \caption{$\nrmse(\what{S}, S)$} \label{fig:numexp-1-s}
        \includegraphics[width=\linewidth]{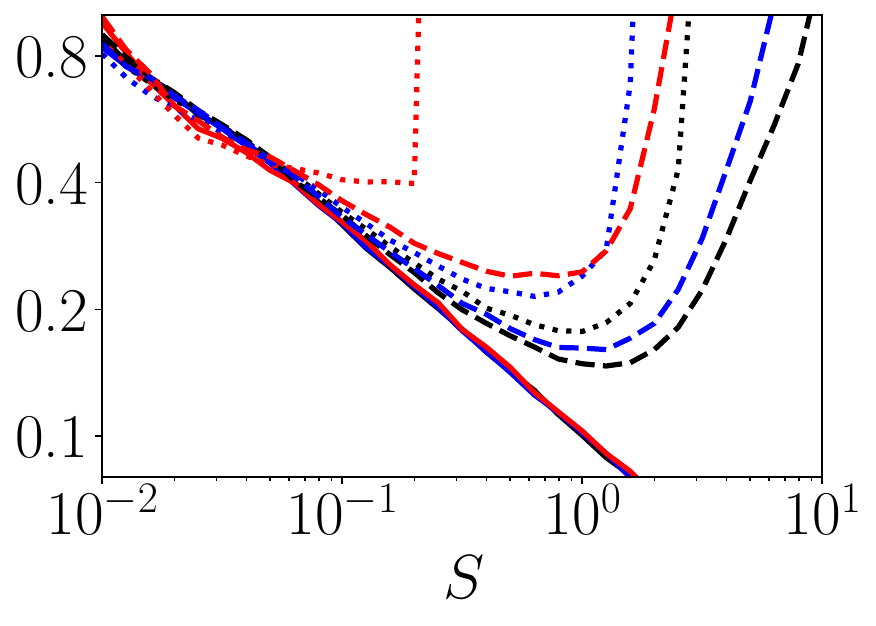}
    \end{subfigure}
    \begin{subfigure}{0.25\linewidth}
        \centering
        \caption{$\nrmse(\what{B}, B)$} \label{fig:numexp-1-b}
        \includegraphics[width=\linewidth]{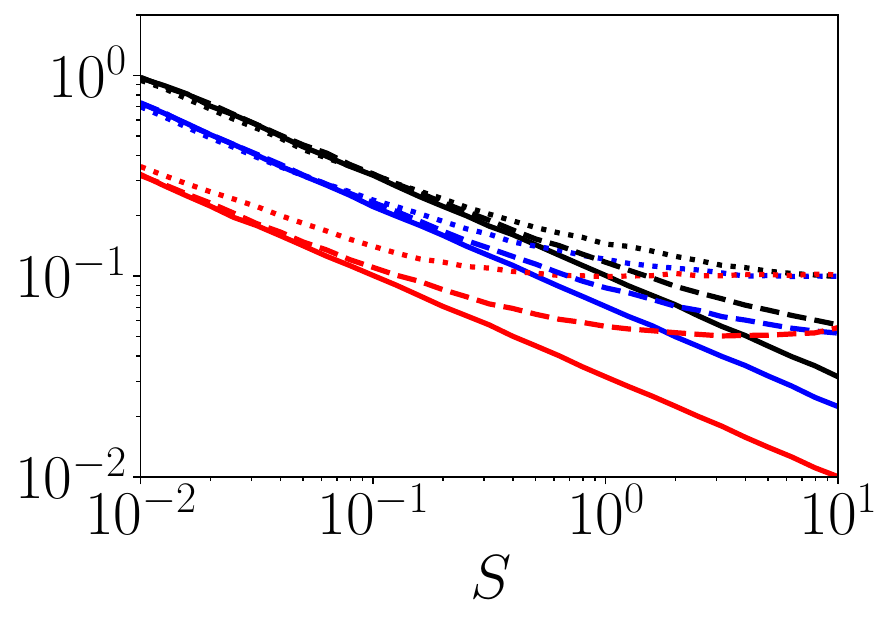}
    \end{subfigure}
    \begin{subfigure}{0.25\linewidth}
        \centering
        \caption{$\rmse(\what{z}, z)$} \label{fig:numexp-1-z}
        \includegraphics[width=\linewidth]{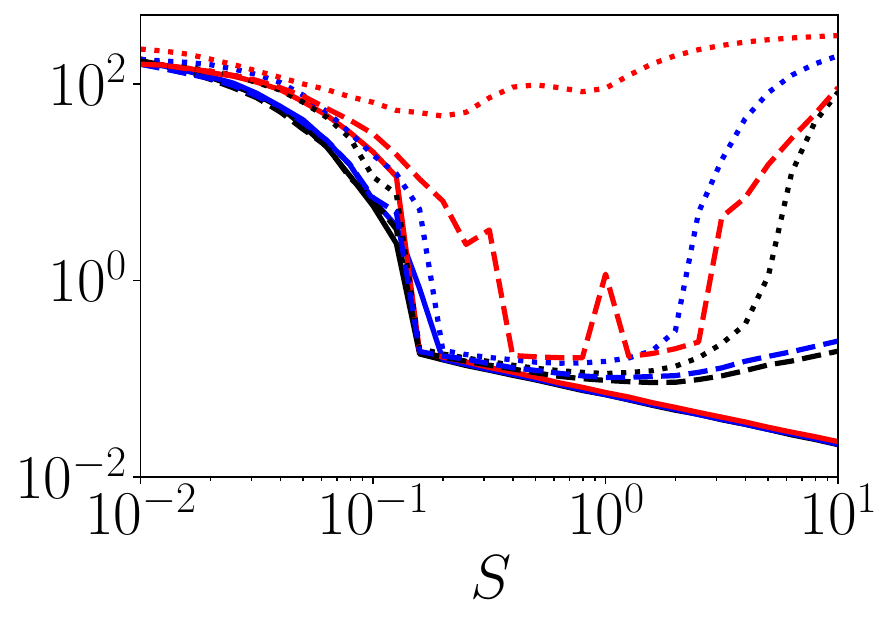}
    \end{subfigure}
    \includegraphics[width=0.9\linewidth]{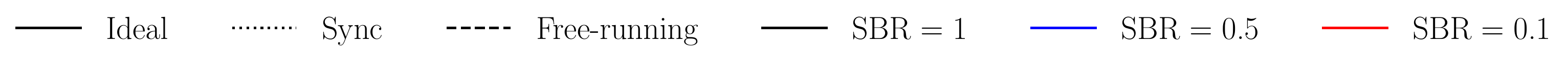}
    \caption{
    Optimal flux and errors of joint ML estimators at various flux levels and SBRs.
    \textbf{(a) Optimal flux}: the optimal flux $\Lambda^*$ minimizing $\rmse(\what{z}, z)$ for free-running and synchronous modes at each SBR ranging from 0.1 to 10.
    \textbf{(b)-(d) Errors at different flux levels}: the errors --- NRMSE for $S$ and $B$ estimates and RMSE for $z$ estimates --- of ML estimators for ideal, synchronous, and free-running measurements at $\sbr \in \{0.1, 0.5, 1\}$ and $S \in [0.01, 10]$
    }
    \label{fig:numexp-1}
    \vspace{-3mm}
\end{figure*}

%% file: 04_regularization.tex
\section{Score-based SPL Depth Regularization}
\label{sec:regularization}

The ML estimators operate on single-pixel measurements. We can acquire multi-pixel data by raster scanning the laser or using an array detector. While pixel-wise estimation enables 3D reconstruction, regularization can further improve accuracy by leveraging spatial structures. 
Since measurements from different pixels are statistically independent, the sum of log likelihoods described in \Cref{sec:mle} across all pixels naturally serve as the data-fidelity term required in many regularization approaches.
Particularly, we aim to combine SPL log likelihoods with a learned prior.
Diffusion models for inverse problems~\cite{songPseudoinverseGuidedDiffusionModels2022,chungDiffusionPosteriorSampling2023,kawarDenoisingDiffusionRestoration2022,sunProvableProbabilisticImaging2023,wuPrincipledProbabilisticImaging2024} are well-suited for SPL regularization with nonconcave log likelihood, because randomness, to some extent, allows the algorithm to avoid local optima~\cite{wuPrincipledProbabilisticImaging2024}. We introduce Score-based SPL Depth Regularization (SSDR), the first sampling-based SPL regularization method using a learned point cloud score model as the prior.
Our key insight is that 3D scores for a point cloud can be projected onto the detector's line-of-sight to effectively compute depth scores.
We focus on regularizing the depth estimates and leave the regularization of signal and background flux estimates to future work.

We assume that the SPL system raster scans the laser onto the scene with galvos and detect photons with a single-element SPAD.
Suppose the detector and the laser are approximately co-located at coordinate $\bc \in \R^3$, which can be set to $\bzero$ without loss of generality.
There are $P$ scan locations indexed by $\{1, \ldots, P\}$.
The scan angles are $\{(\theta_p, \phi_p)\}_{p = 1}^P$, where $\theta_p \in [0, \pi]$ is the polar angle and $\phi_p \in [0, \pi]$ is the azimuthal angle.
Suppose $z_p$ is the depth of point $p$. Then, the Cartesian coordinate of the point $\bx_p \in \R^3$ can be computed from the spherical coordinate $(\theta_p, \phi_p, z_p)$.
We remark that our regularization framework can be generalized to array detectors and other configurations as long as the detector's view angles are known.

\subsection{Depth Scores}

\input{figs/fig-depthsscore}

A major limitation of learning depth map distributions is their dependence on the detector’s location. Instead, we model the distribution of 3D point clouds, which is viewpoint-independent. Specifically, we use a pretrained point cloud denoising model from~\citet{luoScoreBasedPointCloud} that estimates the Stein score --- the gradient of the log probability density of a Gaussian-blurred distribution --- for 3D point clouds, which is an essential component for diffusion-based regularization.

The point cloud score model $\cS: \R^{P \times 3} \to \R^{P \times 3}$ requires the concatenation of Cartesian coordinates $\{\bx_p\}_{p = 1}^P$ corresponding to some depth estimates and returns the scores $\{\bs_p\}_{p = 1}^P$.
We define the \emph{depth score} of a point $p$ as the projection of $\bs_p$ onto the detector's line-of-sight:
\begin{equation}
    \sigma_p := \frac{\langle \bs_p, \bx_p - \bc \rangle}{\norm{\bx_p - \bc}} \label{eq:depthScore}
\end{equation}
In the Supplement, we show that $\sigma_p$ is the Stein score of the depth with an additional term.

We also find that guidance from the score model is only useful for points with large scores.
Hence, in the regularization algorithm, we use hard-thresholded depth score:
\begin{equation} \label{eq:thresDepthScore}
    \sigbar_p = \sigma_p \ones_{(-\infty, -\epsilon) \cup (\epsilon, \infty)}(\sigma_p)
\end{equation}
where the threshold $\epsilon > 0$ is a hyperparameter.

\input{figs/fig-simreg}

\subsection{Median Smoothing for Initialization}

ML depth estimates often exhibit outliers from random peaks due to background~\cite{rappFewPhotonsMany2017}. When the largest histogram peak contains signal photons, errors are millimeter-scale, dictated by pulse width. However, if the peak arises from background noise, the estimate becomes uniformly distributed within the unambiguous range, leading to meter-scale errors. This effect is observed across different detector modes.
To speed up the regularization algorithm, we initialize the depths of outliers with median smoothing.
Suppose we have pixelwise depth estimates $\{\what{z}^\pw_p\}_{p = 1}^P$ and their depth scores $\{\sigma_p\}_{p = 1}^P$.
We first identify outliers as $\cO = \{p \in \{1, \ldots, P\} | |\sigma_p| > \epsilon_{\rm{init}}\}$, where the threshold $\epsilon_{\rm{init}} > 0$ is a hyperparameter.
We set the outlier depths to the median of $K$-nearest neighbors. We define the ``distance'' between point $p$ and $q$ as $d(p, q) := \sqrt{(\theta_p - \theta_q)^2 + (\phi_p - \phi_q)^2}$.
Let $\KNN(p)$ denote the $K$-nearest neighbors of point $p$ based on $d(p, q)$.
Then, we initialize the depth estimates as
\begin{equation} \label{eq:medSmooth}
    \what{z}_p^0 = \begin{cases}
        \median \{ \what{z}_q^{\pw} | q \in \KNN(p) \} &\text{ if } p \in \cO, \\
        \what{z}_p^{\pw} &\text{ if } p \notin \cO.
    \end{cases}
\end{equation}

\subsection{Depth Regularization Algorithm}

The SSDR algorithm, described in \Cref{alg:depthReg}, is based on Plug-and-play Monte Carlo\RV{~\cite{laumont2022bayesian,sunProvableProbabilisticImaging2023}}, a framework for posterior sampling for inverse problems \RV{using} a pretrained score model.
We denote the set of relative detection times for pixel $p$ by $\cX_p$.
An ablation study is included in the Supplement.

\begin{algorithm}[ht]
\caption{Score-based SPL Depth Regularization (SSDR)} \label{alg:depthReg}
\begin{algorithmic}[1]

\STATE \textbf{Input:} 
pixelwise estimate $\{(\what{S}_p^\pw, \what{B}_p^\pw, \what{z}_p^\pw)\}_{p = 1}^P$, 
scan angles $\{(\theta_p, \phi_p)\}_{p = 1}^P$,
detection times $\{\cX_p\}_{p = 1}^P$, 
log likelihood function $\cL$, 
point cloud score model $\cS$, 
regularization strength $\alpha \in \R$, 
step size $\gamma \in \R$.

\STATE Initialize $\{\what{z}_p^0\}_{p = 1}^P$ from $\{\what{z}_p^\pw\}_{p = 1}^P$ using~\eqref{eq:medSmooth}.

\FOR{$k = 1, \ldots, k_{\max}$}

    \STATE Compute $\{\sigbar^k_p\}_{p = 1}^P$ from $\{\what{z}^k_p\}_{p = 1}^P$ using~\eqref{eq:depthScore} and~\eqref{eq:thresDepthScore}.

    \FOR{$p = 1, \ldots, P$}

        \STATE $g^k_p = \frac{\partial}{\partial z} \cL(\what{S}_p^\pw, \what{B}_p^\pw, z | \cX_p) \Big|_{z = \what{z}^k_p}$
        
        \STATE $\eps^k_p \sim \cN(0, 1)$
        
        \STATE $\what{z}^k_p = \what{z}^{k - 1}_p + \gamma (g^k_p + \alpha \sigbar^k_p) + \sqrt{2 \gamma} \eps^k_p$

    \ENDFOR
    
\ENDFOR

\STATE \textbf{Return:} Regularized depth estimate 
 $\{\what{z}^K_p\}_{p = 1}^P$.
 
\end{algorithmic}
\end{algorithm}

\vspace{-2mm}
\subsection{Simulation Results for 3D Imaging}

In this section, we demonstrate through simulations that our proposed SSDR algorithm significantly improves depth estimation accuracy. Ground truth point clouds are generated using meshes from the \emph{Greyc 3D Colored Meshes Database}~\cite{nouri2017technical}. We assume the SPL system raster scans the scene by sweeping the polar and azimuthal angles in equal discrete steps. Each 3D point corresponds to the closest intersection between the detector’s line of sight and the mesh. Signal flux is proportional to the sum of RGB color values. For all simulations, we set $B = 1$, $\tr = \SI{100}{\nano\second}$, $\nr = 100$, $\td = \SI{20}{\nano\second}$, and a \SI{0.1}{\nano\second} pulse width for all pixels.
We choose the hyperparameters by tuning them on a separate dataset (see Supplement), resulting in $\epsilon = \SI[tight-spacing=true]{4.78e-3}{\meter}$, $\epsilon_{\rm{init}} = \SI[tight-spacing=true]{1.09e-3}{\meter}$, $\gamma = \SI[tight-spacing=true]{3.74e-6}{}$, and $\alpha = \SI[tight-spacing=true]{9.70e6}{}$. We use the mean absolute error (MAE) as the metric for depth estimates, since it is less sensitive to outliers.

As shown in \Cref{fig:regsim}, 3D reconstructions from free-running measurements achieve lower depth RMSE than those from synchronous measurements, with SSDR further reducing the error. 
However, the signal flux RMSEs are similar for both synchronous and free-running modes, consistent with the results in \Cref{fig:numexp-1} at $\sbr = 0.1$ and $S = 0.1$, which approximates the average signal flux of the point clouds.
In the \emph{duck} model, most error reduction comes from median smoothing, with iterative refinement providing little additional improvement. However, in the \emph{man} model, while median smoothing yields the largest initial RMSE reduction, iterative refinement significantly improves accuracy further.

%% file: figs/fig-depthsscore.tex
\begin{figure}[ht]
    \centering
    \includegraphics[width=0.6\linewidth,trim={0 5cm 10cm 2.1cm},clip]{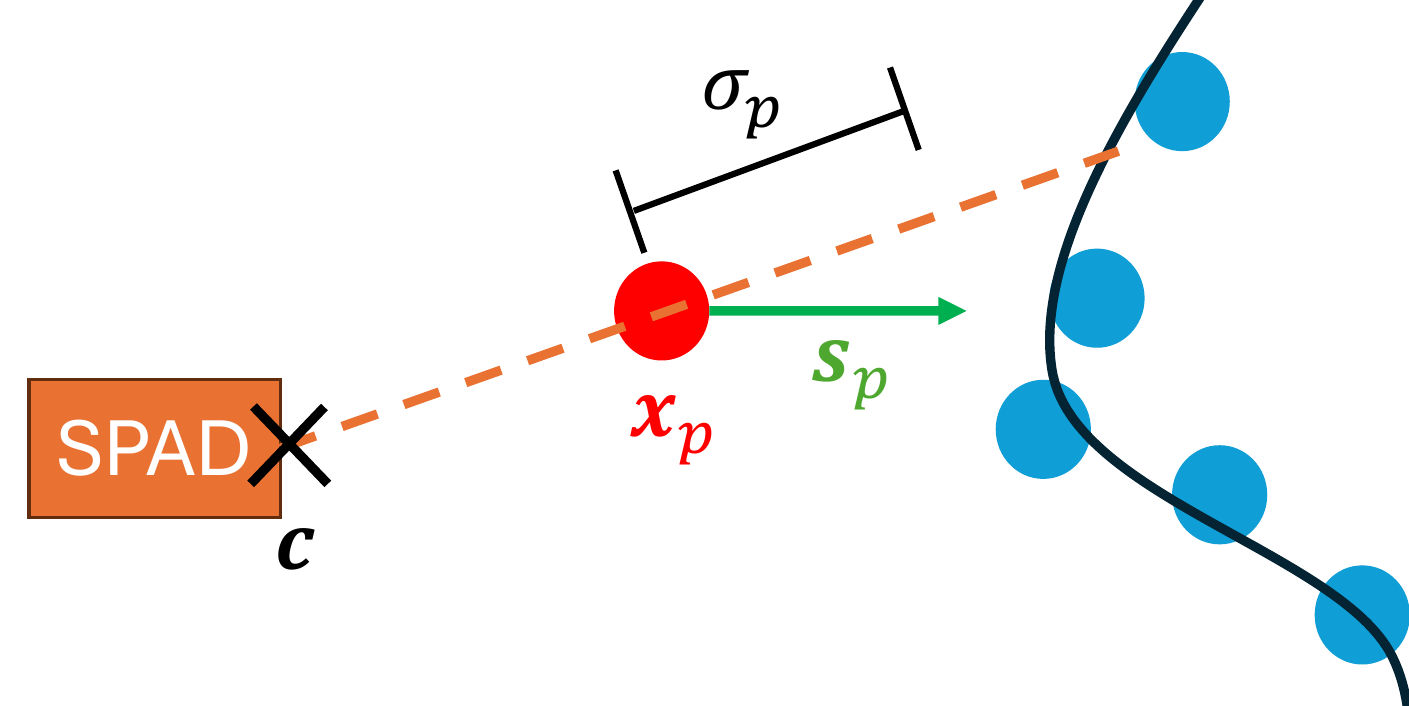}
    \caption{
    The point cloud score model assigns a score $\bs_p$ to each pixel. Projecting this score onto the line of sight $\bx_p - \bc$ yields the depth score $\sigma_p$.
    }
    \label{fig:depthScore}
\vspace{-3mm}
\end{figure}

%% file: figs/fig-simreg.tex
\begin{figure*}[ht]
    \centering
    \begin{tabular}[t]{@{}c@{\,}c@{\,}c@{\,}c@{\,}c@{\,}c@{}}
        (a) Mesh & 
        (b) Ground truth & 
        (c) Sync & 
        (d) Free & 
        (e) Free + SSDR & 
        (f) $\rmse(\what{z}, z)$
        \vspace{1em}
        \\
        \begin{subfigure}{0.1\linewidth}
            \centering
            \includegraphics[height=1.4cm,trim={4cm 3cm 2.7cm 3cm},clip]{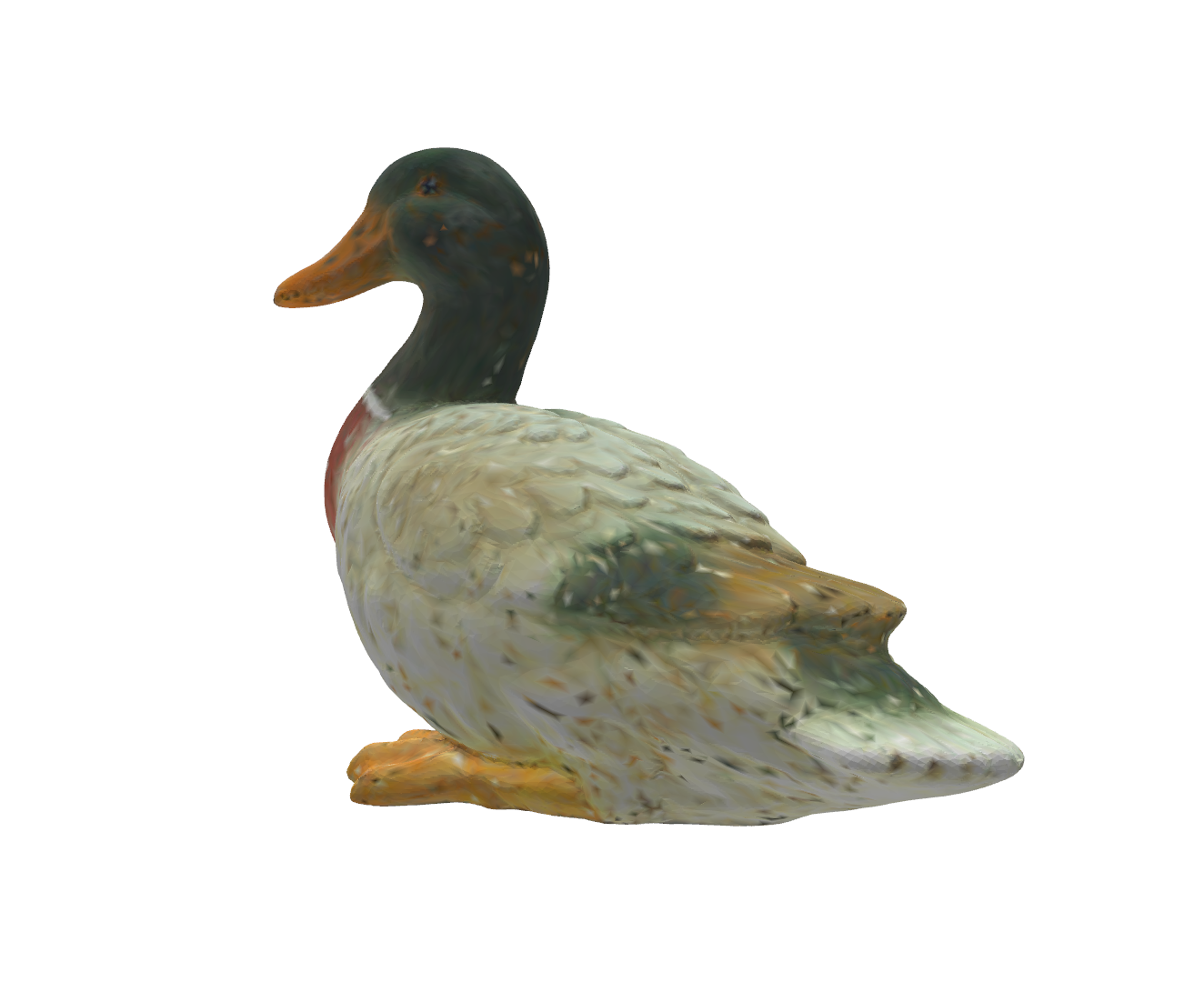}
            \vspace{0.1em}
        \end{subfigure}
         &
         \begin{subfigure}{0.17\linewidth}
            \centering
            \includegraphics[width=\linewidth,trim={3cm 1.5cm 4cm 3.5cm},clip]{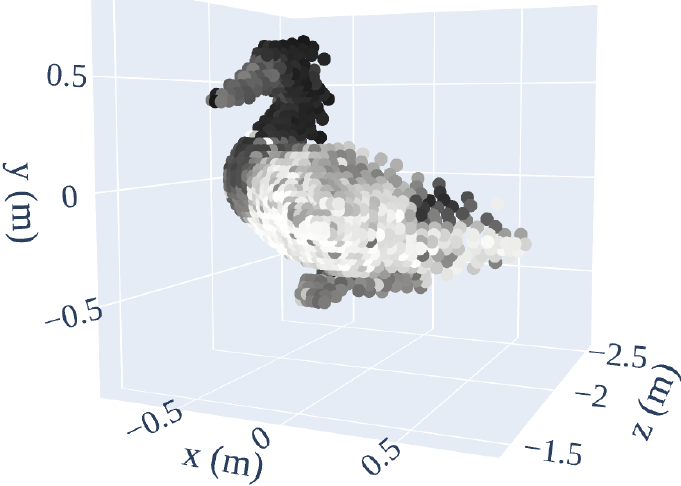}
        \end{subfigure}
        &
        \begin{subfigure}{0.17\linewidth}
            \centering
            \includegraphics[width=\linewidth,trim={1.6cm 1.5cm 5.3cm 3.5cm},clip]{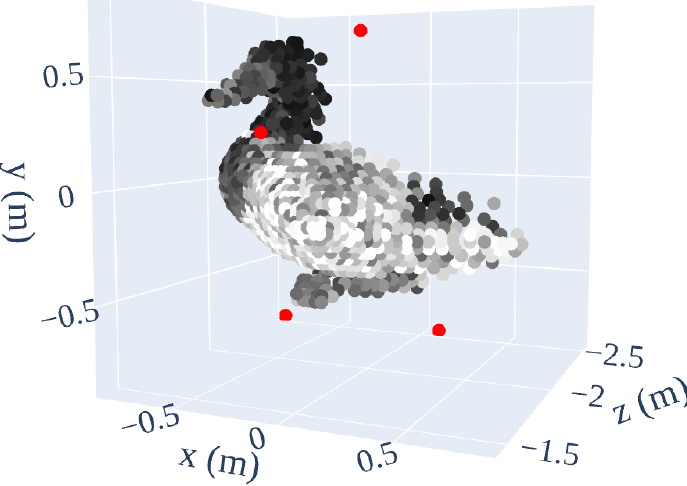}
        \end{subfigure}
        &
        \begin{subfigure}{0.17\linewidth}
            \centering
            \includegraphics[width=\linewidth,trim={1.6cm 1.5cm 5.3cm 3.5cm},clip]{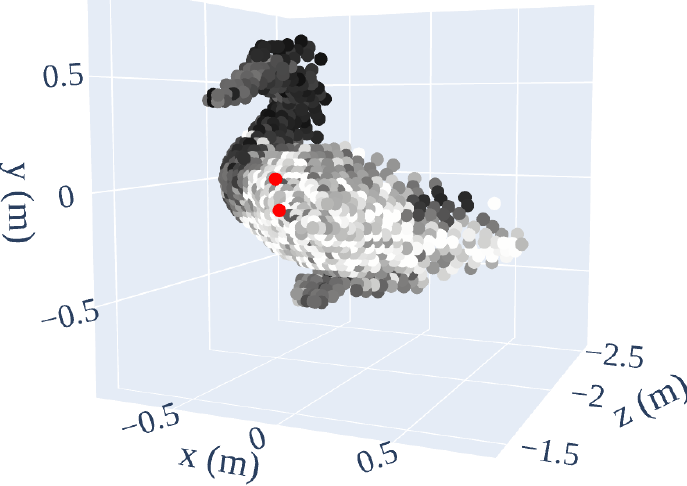}
        \end{subfigure}
        &
        \begin{subfigure}{0.17\linewidth}
            \centering
            \includegraphics[width=\linewidth,trim={1.6cm 1.5cm 5.3cm 3.5cm},clip]{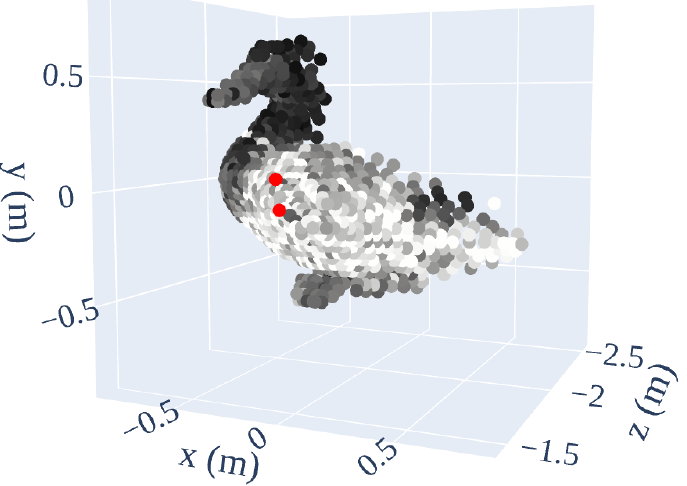}
        \end{subfigure}
        &
        \begin{subfigure}{0.17\linewidth}
            \centering
            \includegraphics[width=\linewidth, trim={0 0.45cm 0 0}, clip]{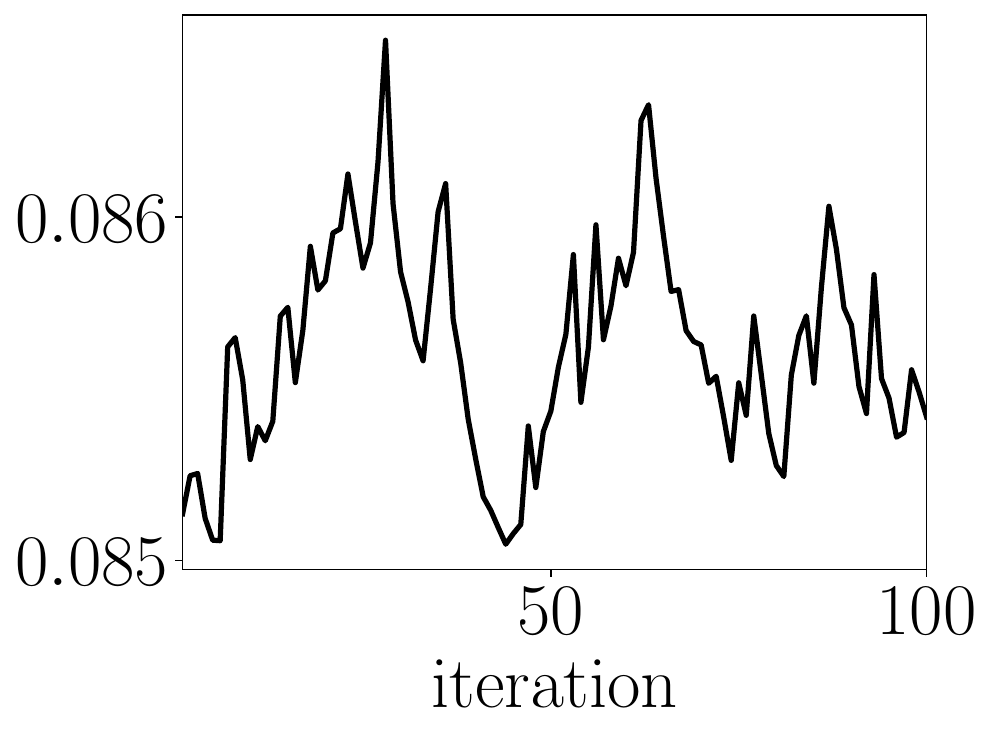}
        \end{subfigure}
        \vspace{0.3em}
        \\
        & 

        & \footnotesize{\SI[tight-spacing=true]{2.5e-2}{} $\vert$ \SI{0.084}{\meter}}
        & \footnotesize{\SI[tight-spacing=true]{2.4e-2}{} $\vert$ \SI{0.022}{\meter}}
        & \footnotesize{\SI[tight-spacing=true]{2.4e-2}{} $\vert$ \SI{0.007}{\meter}}
        
        &
        \vspace{0.5em}
        \\
        \begin{subfigure}{0.1\linewidth}
            \centering
            \includegraphics[height=1.5cm,trim={1cm 1cm 1cm 1cm},clip]{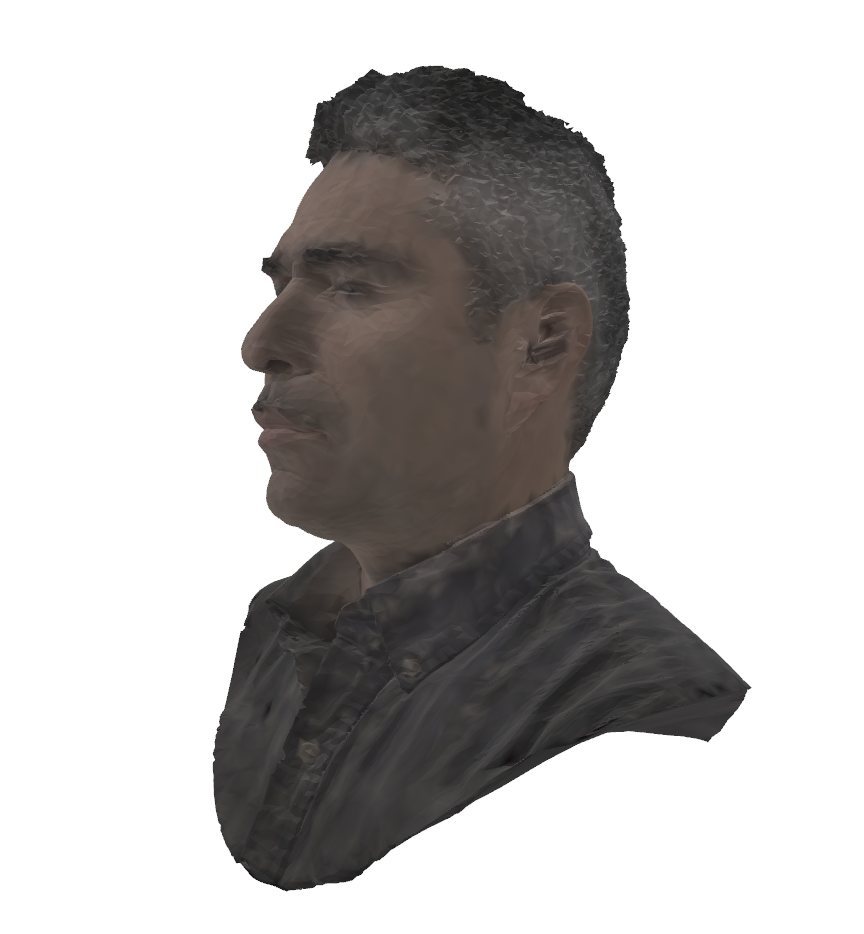}
            \vspace{1em}
        \end{subfigure}
         &
         \begin{subfigure}{0.17\linewidth}
            \centering
            \includegraphics[width=\linewidth,trim={3cm 1.5cm 4cm 3.5cm},clip]{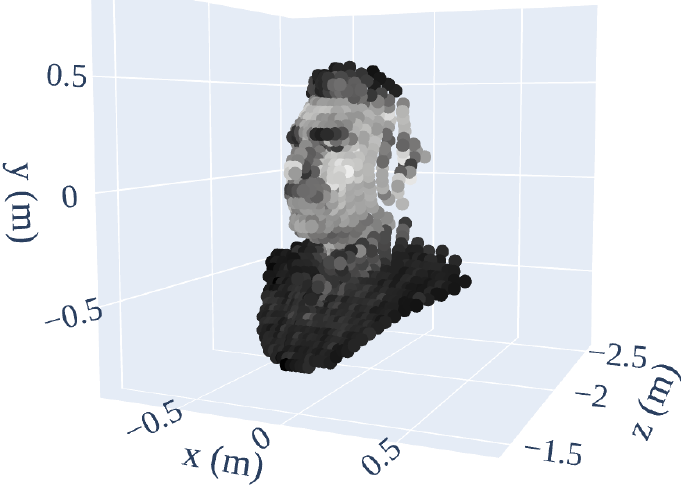}
        \end{subfigure}
        &
        \begin{subfigure}{0.17\linewidth}
            \centering
            \includegraphics[width=\linewidth,trim={1.6cm 1.5cm 5.3cm 3.5cm},clip]{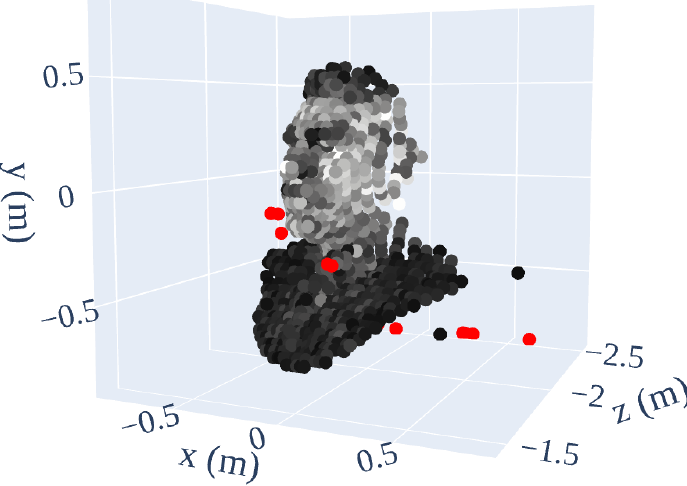}
        \end{subfigure}
        &
        \begin{subfigure}{0.17\linewidth}
            \centering
            \includegraphics[width=\linewidth,trim={1.6cm 1.5cm 5.3cm 3.5cm},clip]{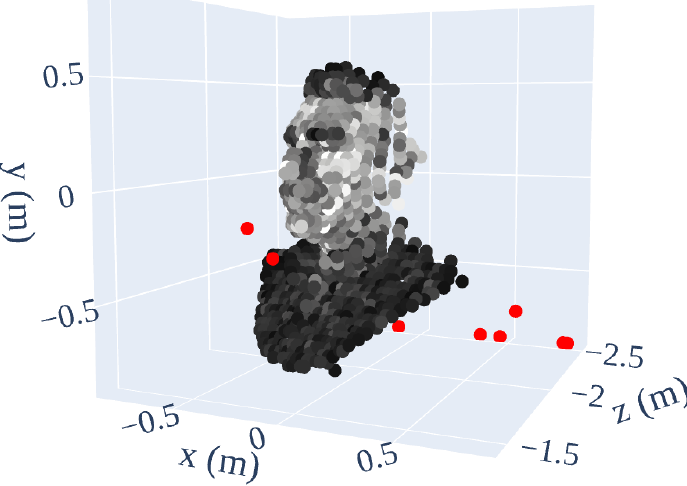}
        \end{subfigure}
        &
        \begin{subfigure}{0.17\linewidth}
            \centering
            \includegraphics[width=\linewidth,trim={1.6cm 1.5cm 5.3cm 3.5cm},clip]{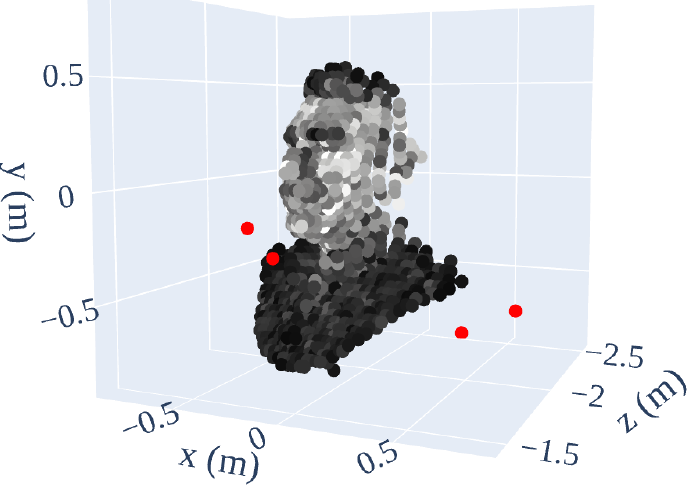}
        \end{subfigure}
        &
        \begin{subfigure}{0.17\linewidth}
            \centering
            \includegraphics[width=\linewidth, trim={0 0.45cm 0 0}, clip]{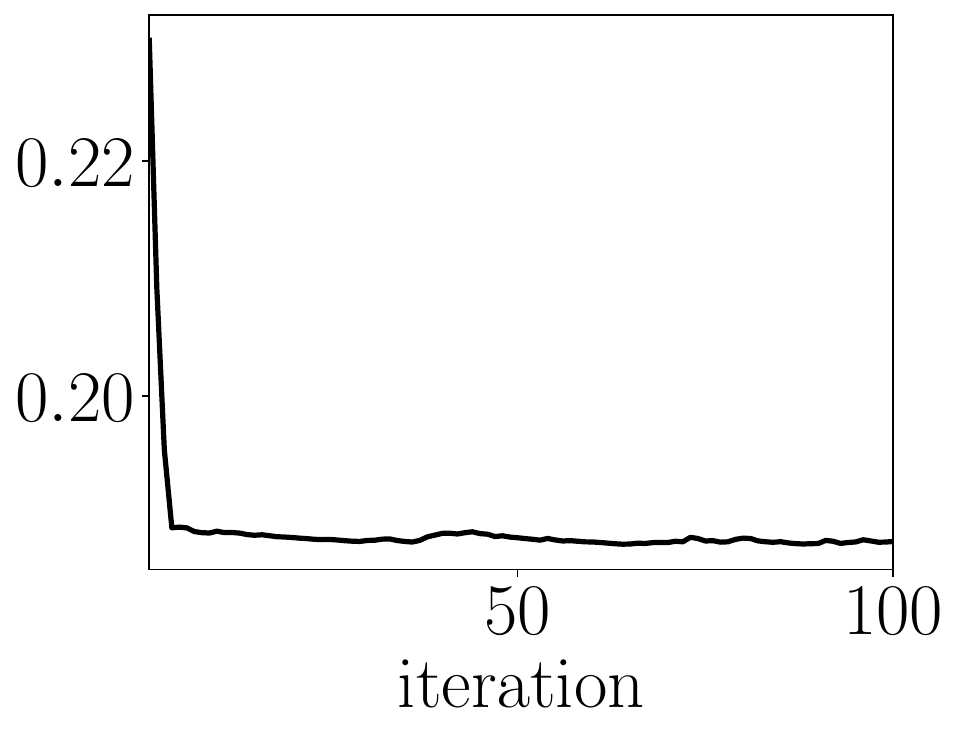}
        \end{subfigure}
        \vspace{0.2em}
        \\
        &
        \includegraphics[width=2.8cm,trim={0 0.4cm 0 0},clip]{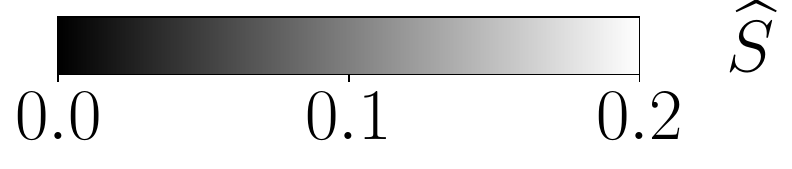}        

        & \footnotesize{\SI[tight-spacing=true]{1.8e-2}{} $\vert$ \SI{0.307}{\meter}}
        & \footnotesize{\SI[tight-spacing=true]{1.8e-2}{} $\vert$ \SI{0.294}{\meter}}
        & \footnotesize{\SI[tight-spacing=true]{1.8e-2}{} $\vert$ \SI{0.020}{\meter}}
        
        &
    \end{tabular}
    \caption{
    Reconstructions from simulated SPL measurements of 3D models. From \textbf{(a) the mesh}, we generate \textbf{(b) the ground truth point cloud} with associated signal flux.
    Below each 3D reconstruction, we report $\rmse(\what{S}, S) \ \vert \ \mae(\what{z}, z)$. 
    Pixel-wise ML estimates for simulated \textbf{(c) synchronous} and \textbf{(d) free-running} measurements are shown.
    \textbf{(e) The free-running reconstruction with SSDR} significantly reduced depth error compared to point-wise estimates.
    \textbf{(f) The error history} plot shows $\rmse(\what{z}, z)$ at each iteration of the SSDR algorithm that produces the estimate in (e).
    Points outside the plotting space are projected onto it and colored red.
    }
    \label{fig:regsim}
\end{figure*}

%% file: 05_lab.tex
\section{Experimental Results} \label{sec:lab}

\input{figs/fig-expreg}

In this section, we validate the proposed ML estimators and SSDR on experimentally collected SPL measurements in both synchronous and free-running modes.  
For each scene, we captured three measurements. First, to emulate an ideal detector, we took free-running measurements in a \textbf{low-flux} setting (dark room). The pixel-wise ML estimates of $S$ and $z$ from this serve as the reference. Then, we collected \textbf{synchronous} and \textbf{free-running} measurements under \emph{high-flux} conditions with two LED lamps illuminating the object. Additional experimental details are provided in the Supplement.

\Cref{fig:lab} presents the resulting 3D reconstructions. The pixel-wise free-running ML estimates closely match the low-flux reconstruction, aside from some out-of-bound points. In contrast, synchronous ML estimates fail, with pile-up causing most depth estimates to cluster near zero (see the $z$-axis in column (c)). This demonstrates the robustness of free-running mode in high-flux conditions. SSDR further improves free-running reconstructions by correcting out-of-bound points.
The error reduction is smaller than in simulation, likely because the point cloud score model~\cite{luoScoreBasedPointCloud} is trained on isolated 3D models. Computing scores for complex scenes with objects at different depths remains a challenge. A better score model could further improve SSDR.

\vspace{-2mm}

%% file: figs/fig-expreg.tex
\def\widthFrac{0.21}
\def\figHeight{2.5cm}

\begin{figure*}[htb]
    \centering
    \begin{tabular}{@{}c@{\,\,}c@{\,\,}c@{\,\,}c@{\,\,}c@{\,\,}c@{}}
        (a) Photo & (b) Low-flux & (c) Sync & (d) Free & (e) Free + SSDR &
        \vspace{-0.2em}
        \\
        \begin{subfigure}{0.11\linewidth}
            \centering
            \includegraphics[width=\linewidth,trim={1cm 2cm 1cm 2cm},clip]{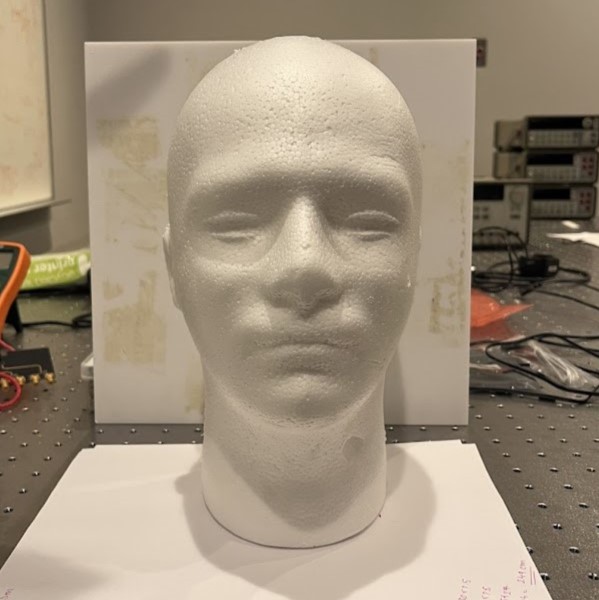}
            \vspace{0.1em}
        \end{subfigure}
         &
         \begin{subfigure}{0.18\linewidth}
            \centering
            \includegraphics[width=\linewidth,trim={3.5cm 1cm 4.4cm 3cm},clip]{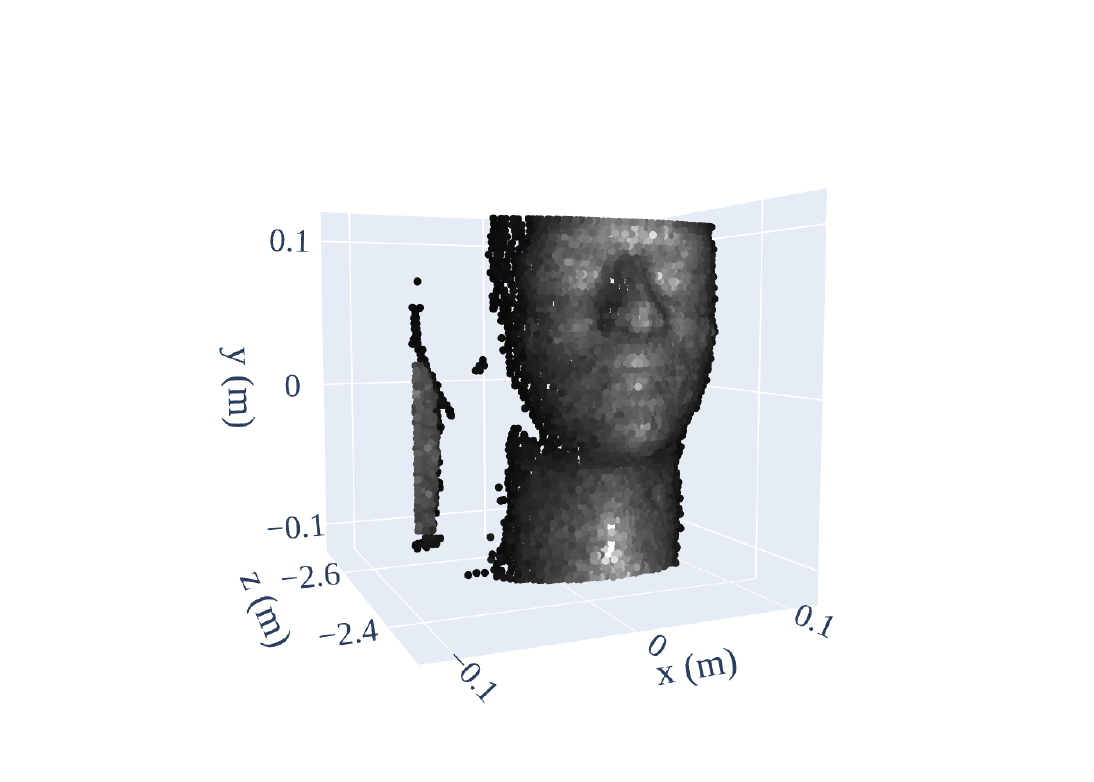}
        \end{subfigure}
        &
        \begin{subfigure}{0.18\linewidth}
            \centering
            \includegraphics[width=\linewidth,trim={2.2cm 1cm 5.5cm 2cm},clip]{figs/lab/head-sc/pointcloudOOB.pdf}
        \end{subfigure}
        &
        \begin{subfigure}{0.18\linewidth}
            \centering
            \includegraphics[width=\linewidth,trim={2.2cm 1cm 5.5cm 2cm},clip]{figs/lab/head-fr/pointcloudOOB.pdf}
        \end{subfigure}
        &
        \begin{subfigure}{0.18\linewidth}
            \centering
            \includegraphics[width=\linewidth,trim={2.2cm 1cm 5.5cm 2cm},clip]{figs/lab/head-fr/pointcloudOOB-diff.pdf}
        \end{subfigure}
        &
        \begin{subfigure}{0.07\linewidth}
            \centering
            \includegraphics[height=2.5cm]{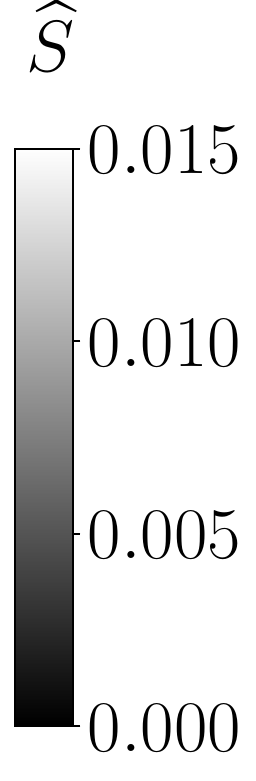}
            \vspace{1em}
        \end{subfigure}
        \\
        & & 

        \footnotesize{\SI[tight-spacing=true]{9.2e-3}{} $\vert$ \SI{2.137}{\meter}} & 
        \footnotesize{\SI[tight-spacing=true]{7.9e-4}{} $\vert$ \SI{0.136}{\meter}} & 
        \footnotesize{\SI[tight-spacing=true]{7.9e-4}{} $\vert$ \SI{0.098}{\meter}}
        
    \end{tabular}
    \caption{
    Reconstructions from experimentally collected SPL measurements.
    The pixel-wise ML estimates from \textbf{(b) low-flux}, \textbf{(c) synchronous} (average $\what{B} = 0.327)$, and \textbf{(d) free-running} (average $\what{B} = 0.852$) settings as well as \textbf{(e) SSDR-regularized free-running} reconstruction are shown.
    We report $\rmse(\what{S}, S) \ \vert \ \mae(\what{z}, z)$ for each reconstruction, where the low-flux estimate serves as the reference.
    }
    \label{fig:lab}
\vspace{-2mm}
\end{figure*}

%% file: 06_conclusion.tex
\section{Future works}
\label{sec:conclusion}

Future work may investigate the statistical properties of free-running measurements, including the distribution of detections $N$ and the Cramér-Rao lower bound. For the synchronous mode, we derived the likelihood function conditioned on the number of active repetition periods, but its distribution warrants further study. Additionally, one could compare free-running measurements to the synchronous mode with fixed gating and other gating strategies~\cite{guptaAsynchronousSinglePhoton3D2019,po2022adaptive}.  

Another direction is improving the SSDR algorithm by training the score model on a larger, more diverse dataset, modifying its architecture, or incorporating techniques from diffusion models~\cite{jalal2021robust,chung2022improving}.

%% file: appendix.tex
\maketitlesupplementary

\appendix
\input{A1_loglike}
\input{A2_estimation}
\input{A3_init}
\input{A4_depthscore}
\input{A5_simulation}
\input{A6_lab}

%% file: A1_loglike.tex
\section{Derivation of Likelihood Functions}
\label{sec:app-loglike}

\subsection{Ideal Detector}

The log likelihood for an ideal detector was previously derived in~\cite{bar-davidCommunicationPoissonRegime1969}. We include a derivation here for completeness.
Without dead time, the detector is always armed, and detection times follow an inhomogeneous Poisson process with intensity function $\lambda(t)$ described in~\eqref{eq:arrivalIntensity}. Let $\Tseq$ denote the sequence of absolute detection times. Then, following~\cite[eq. (2.33)]{snyderRandomPointProcesses2012}, the log likelihood function is
\begin{align}
    \cL^{\id} &= - \int_0^{\nr \tr} \lambda(t) \de t + \sum_{i = 1}^N \log \lambda(T_i) \\
    &\stackrel{(a)}{=} - \nr \int_0^{\tr} \lamtil(t) \de t + \sum_{i = 1}^N \log \lamtil(X_i) \\
    &= - \nr \Lambda + \sum_{i = 1}^N \log \lamtil(X_i),
\end{align}
where step (a) follows from $\lambda(t) = \lamtil(t \bmod \tr)$, where $\lamtil(t)$ is the single-period intensity~\eqref{eq:singleIntensity} in the main paper.
We remark that the log likelihood $\cL^{\id}$ is that of \emph{absolute} detection times $\Tseq$, but it can be expressed in terms of \emph{relative} detection times $\Xset$.

\subsection{Synchronous Detector}

A synchronous detector detects only the first arriving photon in a repetition period. According to~\cite[eq. (2.20)]{snyderRandomPointProcesses2012}, the probability density of the first detection time $X \in [0, \tr)$ is
\begin{align}
    p_X(x) &= \lamtil(x) \exp \left( - \int_0^x \lamtil(t) \de t \right) \\
    &= \lamtil(x) \exp (- \Phitil(x)).
\end{align}
The probability that there is a photon detection in a given repetition period is 
\begin{align}
    \P(0 \leq X < \tr)
    &= \int_0^{\tr} p_X(x) = 1 - e^{-\Lambda}.
\end{align}
Hence, the conditional probability density of $X$ given that there is a detection is
\begin{equation}
    p_{X|N}(x | 1) = \frac{1}{1 - e^{-\Lambda}} \lamtil(x) \exp (- \Phitil(x)).
\end{equation}
The synchronous detector will be inactive in some repetition periods when there are carried-over dead times from previous repetition periods.
Suppose there are $\nrp = \sum_{i = 1}^N \ones_{[\tr - \td, \tr)}(X_i)$ active repetition periods. 
Since the probability of detecting a photon in a given repetition period is $1 - e^{-\Lambda}$, the number of detections $N$ follow a binomial distribution:
\begin{equation}
    N \sim \Bin(\nrp, 1 - e^{-\Lambda}).
\end{equation}
Since the detection times are independent and identically distributed, the probability density of all detection times is
\begin{align}
    &p_{X_1, \ldots, X_N, N | \nrp}(x_1, \ldots, x_n, n | n_r') \\
    & = p_{N | \nrp}(n | n_r') p_{X_1, \ldots, X_N| N, \nrp}(x_1, \ldots, x_n | n, n_r') \\
    & = p_{N | \nrp}(n | n_r') \prod_{i = 1}^n p_{X_i| N, \nrp}(x_i | n, n_r') \\
    \begin{split}
        & = \binom{\nrp}{n} (1 - e^{-\Lambda})^n e^{- (\nrp - n) \Lambda} \\
        &\qquad \prod_{i = 1}^n \frac{1}{1 - e^{-\Lambda}} \lamtil(x_i) \exp (- \Phitil(x_i))
    \end{split} \\
    &= \binom{\nrp}{n} e^{- (\nrp - n) \Lambda} \prod_{i = 1}^n \lamtil(x_i) \exp (- \Phitil(x_i)).
\end{align}
The log likelihood conditioned on $\nrp$ is therefore
\begin{equation}
    \cL^{\sc} = - (\nrp - N) \Lambda + \sum_{i = 1}^N \log \lamtil(X_i) - \Phitil(X_i) + \text{ const}.
\end{equation}
The constant does not depend on the parameters of interest ($S$, $B$, and $z$), so it is dropped for the sake of ML estimation.

\subsection{Free-running Detector}

The free-running detector is unarmed for $\td$ whenever it detects a photon. The absolute detection times follow a self-exciting point process (SEPP) whose intensity becomes zero during dead times~\cite{snyderRandomPointProcesses2012}.
We denote the SEPP by $\{N(t), t \geq 0\}$, where $N(t)$ is the number of detections up to time $t$.
The SEPP's intensity, which itself is a stochastic process, is
\begin{equation} \label{eq:seppIntensity}
     \mu(t) = \begin{cases}
        \lambda(t), &\text{ if } t > T_{N(t)} + t_d, \\
        0, &\text{ if } T_{N(t)} < t \leq T_{N(t)} + t_d,
    \end{cases}
\end{equation}
where $T_{N(t)}$ is the most recent detection time at time $t$.
According to~\cite[Theorem 6.2.2]{snyderRandomPointProcesses2012},
the log likelihood of the \emph{absolute} detection times $\Tseq$ over the acquisition period $[0, \nr\tr)$ is
\begin{align}
    \cL^{\fr} &= - \int_0^{n_r t_r} \mu(t) \de t + \sum_{i = 1}^N \log \lambda(T_i) \\
    &= - \int_0^{n_r t_r} \mu(t) \de t + \sum_{i = 1}^N \log \lamtil(X_i). \label{eq:seppLoglike}
\end{align}
We approximate the integral:
\begin{align}
    \begin{split}
    &\int_0^{\nr \tr} \mu(t) \de t = \int_0^{T_1} \lambda(t) \de t + \sum_{i = 1}^{N - 1} \int_{T_i + \td}^{T_{i + 1}} \lambda(t) \de t \\
    &\quad + \ones_{[0, \nr\tr)}(T_N + \td) \int_{T_N + \td}^{\nr \tr} \lambda(t) \de t
    \end{split} \\
    \begin{split}
    &\quad = \Phi(T_1) + \sum_{i = 1}^{N - 1} \left( \Phi(T_{i + 1}) - \Phi(T_i + \td) \right) \\
    &\quad + \ones_{[0, \nr\tr)}(T_N + \td) \left( \Phi(\nr \tr) - \Phi(T_N + \td) \right)
    \end{split} \\
    \begin{split}
    &\quad \stackrel{(a)}{\approx} \Phi(T_1) + \sum_{i = 1}^{N - 1} \left( \Phi(T_{i + 1}) - \Phi(T_i + \td) \right) \\
    &\hspace{3em} + \left( \Phi(\nr \tr) - \Phi(T_N + \td) \right)
    \end{split} \\
    &\quad = \nr \Lambda + \sum_{i = 1}^N \Phi(T_i) - \Phi(T_i + \td) \\
    \begin{split}
    &\quad \stackrel{(b)}{=} \nr \Lambda + \sum_{i = 1}^N \Phi(T_i \bmod \tr) + \lfloor T_i / \tr \rfloor \Lambda \\
    &\hspace{3em} - \Phi(T_i \bmod \tr + \td) - \lfloor T_i / \tr \rfloor \Lambda
    \end{split} \\
    &\quad = \nr \Lambda + \sum_{i = 1}^N \Phi(X_i) - \Phi(X_i + \td).
    \label{eq:seppApprox}
\end{align}
In step (a), we replace the indicator function $\ones_{[0, \nr\tr)}(T_N + \td)$ with one. If the number of detections $N$ is large, then the approximation will have a relatively small effect on the integral.
Moreover, if the detector is active at $t = \nr\tr$, the indicator function takes the value one, and the approximation becomes an equality.
Step (b) follows from $T_i = \lfloor T_i / \tr \rfloor \tr + T_i \bmod \tr$ and $\Phi(n \tr) = n \Lambda$ for any $n \in \{0, \ldots, \nr\}$.
Substituting~\eqref{eq:seppApprox} into~\eqref{eq:seppLoglike} completes the derivation:
\begin{align}
    \cL^{\fr} &\approx - \nr \Lambda + \sum_{i = 1}^N \log \lamtil(X_i) + \Phi(X_i + \td) - \Phi(X_i). \label{eq:freeLoglike-app}
\end{align}
To write the log likelihood more explicitly in terms of the unknown parameters,
we remark that $\Phi(X_i) = \Phitil(X_i)$, where $\Phitil(t) = S F(t - 2 z / c) + bt$ is the single-period cumulative flux defined in Section 3.4.
For the other term, $\Phi(X_i + \td) = \Phitil((X_i + \td) \bmod \tr) + \ones_{[\tr, \infty)}(X_i + \td) \Lambda$, which can then be written in terms of $S$, $B$, and $z$.

%% file: A2_estimation.tex
\section{Derivation of ML Depth Estimators}
\label{sec:app-depthEst}

\subsection{Ideal Detector}

The ML depth estimator has been previously shown to be a log matched filter~\cite{bar-davidCommunicationPoissonRegime1969}.
We include a derivation of the filter here for completeness.
We rewrite the ideal log likelihood as
\begin{equation}
    \cL^{\id} = - \nr \Lambda + \sum_{i = 1}^N \log \left( S f(X_i - \tau) + b \right),
\end{equation}
where $\tau = 2 z / c$.
Maximizing $\cL^{\id}$ results in
\begin{align}
    \tauhat &= \argmax_{\tau} \sum_{i = 1}^N \log \left( S f(X_i - \tau) + b \right) \\
    &= \argmax_{\tau} \sum_{i = 1}^N \log (S f (\tau) + b) \oplus \delta(\tau - X_i) \\
    &= \argmax_{\tau} \log (S f (\tau) + b) \oplus \sum_{i = 1}^N \delta(\tau - X_i) \\
    &= \argmax_{\tau} w(\tau) \oplus h(\tau),
\end{align}
where $\oplus$ denotes correlation, the term
\begin{equation}
    w(t) := \log (S f (\tau) + b)
\end{equation}
is the matched filter, and $h(t)$ is defined in~\eqref{eq:histDef} in the main paper.

\subsection{Synchronous Detector}

We find the time-of-flight $\tau$ maximizing $\cL^{\sc}$:
\begin{align}
    \tauhat &= \argmax_\tau \sum_{i = 1}^N \log \lamtil(X_i) - \Phitil(X_i) \\
    \begin{split}
        &= \argmax_\tau \sum_{i = 1}^N \log (S f(X_i - \tau) + b) \\
        &\hspace{6em} - (S F(X_i - \tau) + bX_i)
    \end{split} \\
    &= \argmax_\tau \sum_{i = 1}^N u(\tau) \oplus \delta(\tau - X_i) \\
    &= \argmax_\tau u(\tau) \oplus h(\tau).
\end{align}
We recall that the filter $u(t)$ is
\begin{equation}
    u(t) = \log (S f (\tau) + b) - S F(t).
\end{equation}

\subsection{Free-running Detector}

We first rewrite the approximate free-running log likelihood~\eqref{eq:freeLoglike-app} as
\begin{align}
    \begin{split}
        \cL^{\fr} &\approx - \nr \Lambda + \sum_{i = 1}^N \lamtil(X_i) + \Big(\Phitil((X_i + \td) \bmod \tr) \\
        &\quad + \ones_{[\tr, \infty)}(X_i + \td) \Lambda \Big) - \Phitil(X_i).
    \end{split}
\end{align}
The terms $-\nr \Lambda$ and $\ones_{[\tr, \infty)}(X_i + \td) \Lambda$ do not depend on $\tau$, so they can be dropped for ML depth estimation.
To make the equations concise, we define $Y_i := (X_i + \td) \bmod \tr$.
We maximize this approximate log likelihood with respect to $\tau$ as follow,
\begin{align}
    \tauhat
    &= \argmax_\tau \sum_{i = 1}^N \log \lamtil(X_i) + \Phitil(Y_i) - \Phitil(X_i)  \\
    \begin{split}
    &= \argmax_\tau \sum_{i = 1}^N \log (S f(X_i - \tau) + b) \\
    &\hspace{2em} + S F(Y_i - \tau) + bY_i - S F(X_i - \tau) - b X_i
    \end{split} \\
    &\stackrel{(a)}{=} \argmax_\tau \sum_{i = 1}^N u(\tau) \oplus \delta(\tau - X_i) + v(\tau) \oplus \delta(\tau - Y_i) \\
    &= \argmax_\tau u(\tau) \oplus h(\tau) + v(\tau) \oplus g(\tau),
\end{align}
where the filter $v(t)$ is
\begin{equation}
    v(t) = S F(t).
\end{equation}

\subsection{Quantization}

In practical SPL systems, the detection times are quantized to a predefined temporal resolution, and the data are stored as a histogram.
Even in simulation, an efficient implementation of the ML depth estimators uses discrete-time correlation, which requires quantized detection times.
We will show that the histogram is equivalent to $h(t)$ defined in~\eqref{eq:histDef} of the main paper.

Suppose the period $[0, \tr)$ is partitioned into $M$ histogram bins $\{[m \Delta, (m + 1) \Delta)\}_{m = 0}^{M - 1}$, where the bin size is $\Delta = \tr / M$.
Let us consider a set of relative detection times $\Xset$.
Each detection time is assigned to a bin center:
\begin{equation}
    X_i \in \left\{(m + 1/2) \Delta \big| m \in \{0, \ldots, M - 1\}\right\}. \label{eq:binSet}
\end{equation}
We can express $\Xset$ in terms of the histogram:
\begin{equation}
    \hdel[m] = \sum_{i = 1}^N \ones_{\{(m + 1/2) \Delta\}}(X_i).
\end{equation}
Now, let us consider $h(t)$ in the main paper defined as
\begin{equation}
    h(t) = \sum_{i = 1}^N \delta(t - X_i).
\end{equation}
If the detection times are discretized, then
\begin{align}
    h(t) &= \sum_{m = 1}^{M - 1} \delta(t - (m + 1/2)\Delta) \sum_{i = 1}^N \ones_{\{(m + 1/2) \Delta\}}(X_i) \\
    &= \sum_{m = 1}^{M - 1} \delta(t - (m + 1/2)\Delta) \hdel[m].
\end{align}
At time $t = (m' + 1/2) \Delta$ for some $m' \in \{0, \ldots, M\}$, which is the $m'$-th bin center of the histogram $\hdel[m]$, we have
\begin{align}
    h(t) &= \delta(t - (m' + 1/2)\Delta) \hdel[m'].
\end{align}
This shows that there is a one-to-one correspondence between $h(t)$ and the histogram $\hdel[m]$ when $\Xset$ is quantized according to~\eqref{eq:binSet}.

Even if $\Xset$ is not quantized, we need to compute the quantized histogram $\hdel[m]$ with a bin size $\Delta$ in order to implement the depth estimators efficiently as discrete-time correlation.
For example, the ML estimator of $\tau$ for an ideal detector is implemented as
\begin{align}
    \tauhat &= (m^* + 1/2) \Delta \\
    \text{where } \ m^* &= \argmax \hdel[m] \oplus w_\Delta[m] \\
    &= \argmax \sum_{m' = -\infty}^{\infty} \hdel[m'] w_\Delta [m' - m],
\end{align}
and $w_\Delta[m] = w((m + 1/2) \Delta)$ is the discrete-time filter.
In simulation where we have access to $\Xset$ with floating-point precision, we may further improve $\tau$ estimate by maximizing the log likelihood using gradient-based methods.

%% file: A3_init.tex
\section{Initializing the ML Estimator}
\label{sec:app-init}

We initialize the alternating maximization algorithm~\eqref{alg:mlEst} for ML estimation using the following estimators.
For free-running measurements, we use the estimators for ideal measurements described below in \Cref{sec:idealInit}.

\subsection{Ideal Detector} \label{sec:idealInit}

We use the traditional log matched filtering method~\cite{snyderRandomPointProcesses2012} to obtain an initial time-of-flight estimate:
\begin{equation}
    \tauinit^{\id} = \argmax_{\tau} h(t) \oplus \log f(\tau).
\end{equation}
Inspired by the Neyman-Pearson censoring estimator~\cite{rappFewPhotonsMany2017}, we assume that the detection times in the window of size $\twin$ around $\tauinit$ are mostly due to the signal.
In this paper, we set $\twin = 4w$, where $w$ is the pulse width.
Let $\Ncl$ denote the number of detections in this window, \ie, $\Ncl = \sum_{i = 1}^N \ones_{[\tauinit - \twin / 2, \tauinit + \twin / 2]}(X_i)$.
Then, the signal flux initial estimate is
\begin{equation}
    \sinit^{\id} = \max \left( \frac{\Ncl}{\nr}, \epsilon \right),
\end{equation}
where we lower bound the signal flux estimate by a small positive number $\epsilon > 0$ to ensure that arguments of the log terms in the log likelilhoods are positive.
In our simulations and experiments, we use $\epsilon = 10^{-5}$.
We estimate the background flux as
\begin{equation}
    \binit^{\id} = \max \left( \frac{N}{\nr} - \sinit, \epsilon \right),
\end{equation}
since $N / \nr$ is an estimate of the total flux $\Lambda$, and $\Lambda = S + B$.

\subsection{Synchronous Detector}

We first use Coates's correction to estimate the photon arrival intensity~\cite{coatesCorrectionPhotonPileup1968}.
Suppose the detection times $\Xset$ are quantized into a histogram $\hdel[m]$ with bin size $\Delta$.
The Coates-corrected histogram, which is the ML estimate of the photon arrival intensity~\cite{guptaPhotonFloodedSinglePhoton3D2019}, for the $m^{\rm{th}}$ time bin is
\begin{equation}
    \what{\lambda}[m] = \log \left( \frac{N - \sum_{m' = 1}^{m - 1} \hdel[m']}{N - \sum_{m' = 1}^{m} \hdel[m']} \right).
\end{equation}
Then, we apply the log-matched filter~\cite{snyderRandomPointProcesses2012} to estimate the time-of-flight:
\begin{align}
    \tauinit^{\sc} &= (m^* + 1/2) \Delta, \\
    \text{where } \ m^* &= \argmax \what{\lambda}[m] \oplus s[m] \\
    s[m] &= \log f((m + 1/2)\Delta), 
\end{align}
where $f(t)$ is the pulse's temporal profile.
Again, the signal and background flux estimates follow from censoring of the estimated photon arrival intensity:
\begin{align}
    \sinit^{\sc} &= \max \left( \sum_{k \in \cK_{\rm{win}}} \what{\lambda}[k], \epsilon \right), \\
    \binit^{\sc} &= \max \left( \sum_{k = 1}^N \what{\lambda}[k] - \sinit^{\sc}, \epsilon \right).
\end{align}

%% file: A4_depthscore.tex
\section{Depth Score from Point Cloud Score} \label{sec:depthscore-app}

In the SSDR algorithm, the depth score $\sigma_p \in \R$ is obtained by projecting the point cloud score $\bs_p \in \R^3$ onto the detector's line of sight, which is determined by the scan angles $(\theta_p, \phi_p)$.
The trained model's point cloud score approximates the Stein score, i.e., the gradient of the log density of the distribution of 3D points \cite{luoScoreBasedPointCloud}. 
In this section, we show that $\sigma_p$ relates to the Stein score of the depth distribution corresponding to the implicit prior on 3D points in the trained model.

\subsection{From Cartesian to Spherical}

\begin{figure}[ht]
    \centering
    \includegraphics[width=0.5\linewidth,trim={1cm 0.3cm 14.5cm 1cm},clip]{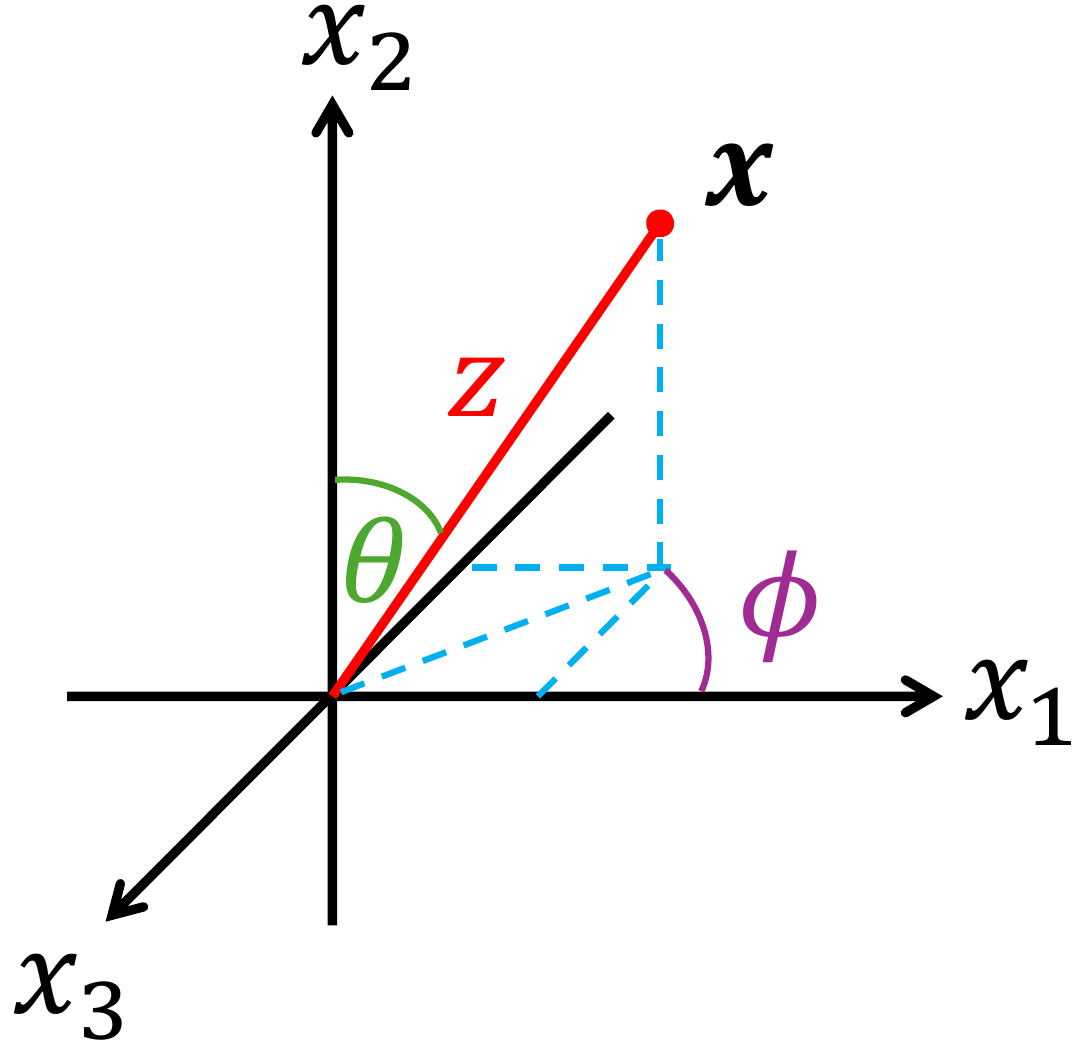}
    \caption{Relationship between Cartesian coordinate $\bx = (x_1, x_2, x_3)$ and spherical coordinate $\bxtil = (z, \theta, \phi)$.}
    \label{fig:coord-transform}
\end{figure}

We first describe the transformation between the Cartesian coordinate $\bx = (x_1, x_2, x_3)$ and the spherical coordinate $\bxtil = (z, \theta, \phi)$, where $z$ is the depth from the detector, assumed to be at the origin.
According to \Cref{fig:coord-transform}, 
\begin{align}
    x_1 &= z \sin \theta \cos \phi, \\
    x_2 &= z \cos \theta, \\
    x_3 &= -z \sin \theta \sin \phi.
\end{align}
We denote this transformation by $T$, \ie, $\bxtil = T(\bx)$, and the inverse map by $T^{-1}$.

\subsection{Spherical Coordinate Prior Distribution}

Consider a prior distribution of a 3D point in the Cartesian coordinate with density $\pi(\bx)$, representing the implicit prior in the trained point cloud score model for a single point conditioned on all others.
Since $\bxtil = T(\bx)$ and $T$ is bijective, we can compute the density of the distribution of $\bxtil$, denoted by $\pitil(\bxtil)$, using the change-of-variable formula:
\begin{equation}
    \pitil(\bxtil) = \pi(T^{-1}(\bxtil)) | \det J_{T^{-1}} (\bxtil)|,
\end{equation}
where $J_{T^{-1}}$ is the Jacobian of $T^{-1}$.
We now compute the Jacobian:
\begin{align}
    J_{T^{-1}}(\bxtil) &= \begin{bmatrix}
        \frac{\partial x_1}{\partial z} & \frac{\partial x_1}{\partial \theta} & \frac{\partial x_1}{\partial \phi} \\
        \frac{\partial x_2}{\partial z} & \frac{\partial x_2}{\partial \theta} & \frac{\partial x_2}{\partial \phi} \\
        \frac{\partial x_3}{\partial z} & \frac{\partial x_3}{\partial \theta} & \frac{\partial x_3}{\partial \phi}
    \end{bmatrix} \\
    &= \begin{bmatrix}
        \sin \theta \cos \phi & z \cos \theta \cos \phi & -z \sin \theta \sin \phi \\
        \cos \theta & -z \sin \theta & 0 \\
        - \sin \theta \sin \phi & -z \cos \theta \sin \phi & -z \sin \theta \cos \phi
    \end{bmatrix}.
\end{align}
The determinant is
\begin{equation}
    \det J_{T^{-1}} (\bxtil) = z^2 \sin \theta.
\end{equation}
Therefore, the density of the point in spherical coordinate is
\begin{align} \label{eq:sphericalDist}
    \pitil(z, \theta, \phi) &= \pi(x_1, x_2, x_3) \cdot z^2 \sin \theta.
\end{align}

\subsection{Deriving the Depth Score}

At pixel $p$, the detector's line of sight is defined by the scan angles $(\theta_p, \phi_p)$.
Given a point cloud score in Cartesian coordinate $\bs_p = [\partial \log \pi / \partial x_1, \partial \log \pi / \partial x_2, \partial \log \pi / \partial x_3]$, the Stein score of the depth $z$ can be computed from the density~\eqref{eq:sphericalDist} evaluated at these angles:
\begin{align}
    &\frac{\partial}{\partial z} \log \pitil(z, \theta_p, \phi_p) = \frac{\partial}{\partial z} \log (z^2 \sin \theta_p) + \frac{\partial \log \pi}{\partial z} \\
    &= \frac{2}{z} + \frac{\partial \log \pi}{\partial x_1} \frac{\partial x_1}{\partial z}
    + \frac{\partial \log \pi}{\partial x_2} \frac{\partial x_2}{\partial z} 
    + \frac{\partial \log \pi}{\partial x_3} \frac{\partial x_3}{\partial z} \\
    &= \frac{2}{z} + \begin{bmatrix}
        \frac{\partial \log \pi}{\partial x_1} \\
        \frac{\partial \log \pi}{\partial x_2} \\
        \frac{\partial \log \pi}{\partial x_3}
    \end{bmatrix}
    \cdot
    \begin{bmatrix}
        \sin \theta_p \cos \phi_p \\
        \cos \theta_p \\
        - \sin \theta_p \sin \phi_p
    \end{bmatrix} \\
    &= \frac{2}{z} + \bs_p \cdot \widehat{\br}_p = \frac{2}{z} + \sigma_p,
\end{align}
where $\widehat{\br}_p$ is the unit vector pointing in the direction of the scan angles $(\theta_p, \phi_p)$.
In SSDR, we use only $\sigma_p = \bs_p \cdot \widehat{\br}_p$ as the depth score, since we find that $2 / z$ often dominates the other term and does not provide good guidance.

%% file: A5_simulation.tex
\section{Additional Simulation Results}
\label{sec:app-sim}

\subsection{Running time for Free-running Estimators}

\begin{figure}[htb]
    \centering
    \begin{subfigure}{0.49\linewidth}
        \centering
        \includegraphics[width=\linewidth]{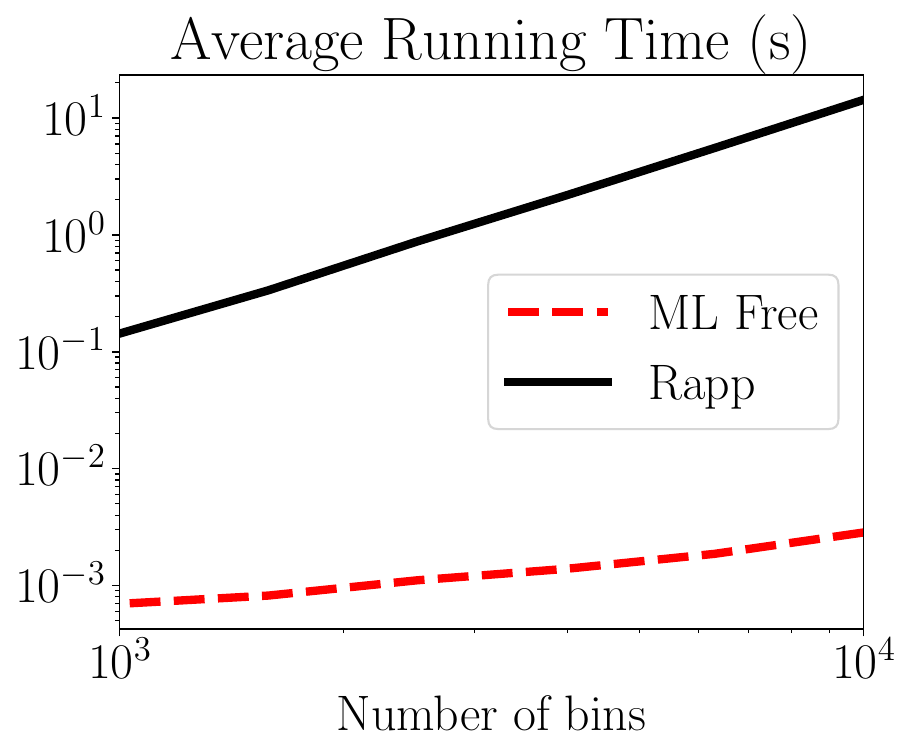}
        \caption{ML Free vs Rapp} \label{fig:rapp-time}
    \end{subfigure}
    \begin{subfigure}{0.49\linewidth}
        \centering
        \includegraphics[width=\linewidth]{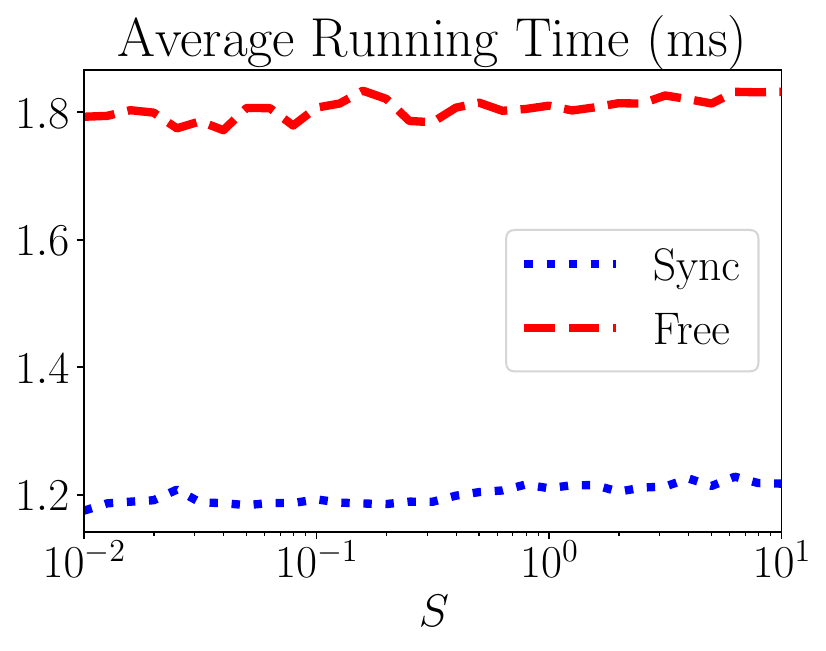}
        \caption{ML Free vs ML Sync} \label{fig:free-sync-time}
    \end{subfigure}
    \caption{\RV{(a) Running time of the free-running ML depth estimator and \citet{rappDeadTimeCompensation2019}'s method with different bin sizes averaged over 100 Monte Carlo trials.
    The setting matches that in Figure 2 of the main paper with $S = 0.1$ and $B = 1$, but the bin size and thus the number of bins vary.
    (b) Running time of free-running and synchronous ML estimators with the bin size of \SI{0.01}{\nano\second}, \ie, 10000 bins as $\tr = \SI{100}{\nano\second}$.}
    }
    \label{fig:compute-time}
\end{figure}

We provide additional results regarding the running time of depth estimator for the free-running mode.
\citet{rappDeadTimeCompensation2019}'s method requires computing the stationary distribution of the relative detection times, which involve an expensive eigenvector computation.
In contrast, the proposed ML depth estimator for the free-running mode can be implemented efficiently as matched filtering.
Both methods operate on the histogram. However, the running time of \citet{rappDeadTimeCompensation2019}'s method scales poorly with the number of time bins, as shown in \Cref{fig:rapp-time}.
At 10000 bins, the ML estimator takes \SI{2.84}{\milli\second} on average, while \citet{rappDeadTimeCompensation2019}'s method takes \SI{14.3}{\second}, or approximately 5000x that of the ML estimator.
Additionally, the ML estimator allows continuous refinement by maximizing the log likelihood with gradient-based optimization if we have access to detection times with higher numerical precision.

\RV{In \Cref{fig:free-sync-time}, we compare free-running  and synchronous ML depth estimators under the same scene parameters and number of time bins. Both estimators are matched filters with access to ground truth $S$ and $B$. Although the free-running ML estimator is slower, it is still computationally efficient, taking only \SI{1.8}{\micro\second} for 10000 time bins.}

\subsection{Hyperparameter Tuning for SSDR}

SSDR has four hyperparameters: the depth score thresholds for iterative updates ($\epsilon$) and median smoothing ($\epsilon_{\rm{init}}$), the step size ($\gamma$), and the regularization weight ($\alpha$). We tune them using the \emph{Optuna} package~\cite{optuna_2019} with the Tree-structured Parzen Estimator~\cite{bergstra2011algorithms}, minimizing the RMSE of $ z $ estimates over 200 SSDR iterations for 100 hyperparameter samples. The tuning dataset is the free-running measurement of the ``Mario'' scene in \Cref{fig:regsim-supp}, yielding the reported hyperparameters.

\subsection{Additional 3D Imaging Results}

\input{figs/fig-A-simreg}

The simulated 3D imaging results in \Cref{fig:regsim-supp} extend the main paper by including reconstructions from ideal measurements, SSDR-regularized results for ideal and synchronous measurements, and additional error metrics such as the RMSE of background flux estimates and the RMSE of depth estimates.  

For pixel-wise ML estimates, free-running measurements consistently outperform synchronous ones in estimating $S$, $B$, and $z$, while ideal measurements provide the most accurate results. However, for SSDR-regularized reconstructions, the trend is less clear. In the ``Man'' and ``Mario'' scenes, both $\mae(\what{z}, z)$ and $\rmse(\what{z}, z)$ are higher for free-running than for synchronous measurements, despite significant improvement from pixel-wise ML estimates. We attribute this to randomness in the SSDR algorithm.

\subsection{Ablation Study for SSDR}
\label{sec:app-ablation}

\input{figs/fig-A-ablation}

For the ablation study, we remove three key features from SSDR—one at a time: median smoothing for initialization, depth score hard-thresholding, and noise in the iterative update (turning it into standard gradient descent). Hyperparameters remain the same as in the main paper.
\Cref{fig:ablation} shows the RMSE of depth estimates at each SSDR iteration with and without each component. The tested scenes (“Duck,” “Man,” and “Mario”) and measurement types (“Ideal,” “Synchronous,” and “Free-running”) match those in \Cref{fig:regsim-supp}. 

Depth score thresholding has the greatest impact on RMSE reduction; without it, RMSE increases over iterations. Median smoothing is the next most influential—without it, RMSE initially decreases but then plateaus, likely due to local optima in the log-posterior density. The iterative update noise has the least effect on final RMSEs, though in some cases (e.g., “Mario” scene with ideal measurements), it helps avoid local optima, leading to improved results.
SSDR, following the Plug-and-Play Monte Carlo framework~\cite{sunProvableProbabilisticImaging2023}, also enables uncertainty quantification through multiple reconstruction samples.

\subsection{SPL Regularization Comparison}

\input{figs/fig-A-regcomparison}
\input{figs/fig-A-benchreg2}

We compare the performance of the proposed SSDR algorithm to ManiPoP, a Bayesian approach to SPL regularization~\cite{tachellaBayesian3DReconstruction2019}, for ideal measurements.
While ManiPoP can provide more accurate reconstruction and reduce outlier points, it can introduce structured artifacts and distort the point cloud's shape.

In \Cref{fig:reg-comparison}, the scene consists of the ``Duck'' model from the Greyc 3D Colored Mesh Database~\cite{nouri2017technical} with an added back pane. The detector raster scans a 99x99 grid, using SPL settings identical to Section 4.4 of the main paper, with signal flux scaled to a maximum of 0.1. SSDR slightly reduces depth errors from the pixel-wise ML reconstruction but retains some outlier points.
We remark that the score model is trained on point clouds of isolated 3D models~\cite{luoScoreBasedPointCloud}, so scenes with targets at different depths can be challenging.
ManiPoP achieves lower depth MAE but higher RMSE due to artifacts, including severe depth underestimation in a row at the bottom of the scene. The zoomed-in depth map reveals shape distortions in ManiPoP's reconstruction, whereas SSDR preserves structure more faithfully.

ManiPoP supports multi-depth imaging, detecting whether a pixel contains one or multiple targets. However, this can be a disadvantage, as shown in \Cref{fig:manipop2}. In a scene without a back pane, ManiPoP eliminates the duck's head due to its lower signal flux, also distorting the remaining structure.
Finally, we remark that the derived log likelihoods for synchronous and free-running modes can be integrated into regularization methods such as TV regularization~\cite{heideSubpicosecondPhotonefficient3D2018}, ManiPoP~\cite{tachellaBayesian3DReconstruction2019} and RT3D~\cite{tachellaRealtime3DReconstruction2019} to made them applicable to these detector modes.

%% file: figs/fig-A-simreg.tex
\begin{figure*}[htb]
    \centering
    \begin{tabular}[t]{@{}c@{\,\,}c@{\,}c@{\,}c@{\,}c@{\,}c@{\,}c@{}}
        &
        (a) Ideal & 
        (b) Sync & 
        (c) Free & 
        (d) Ideal + SSDR & 
        (e) Sync + SSDR & 
        (f) Free + SSDR
        \vspace{1em}
        \\
        \rotatebox[origin=l]{90}{\hspace{-2mm} \small Duck}
        &
        \begin{subfigure}{0.16\linewidth}
            \centering
            \includegraphics[width=\linewidth,trim={1.6cm 1.5cm 5.3cm 3.5cm},clip]{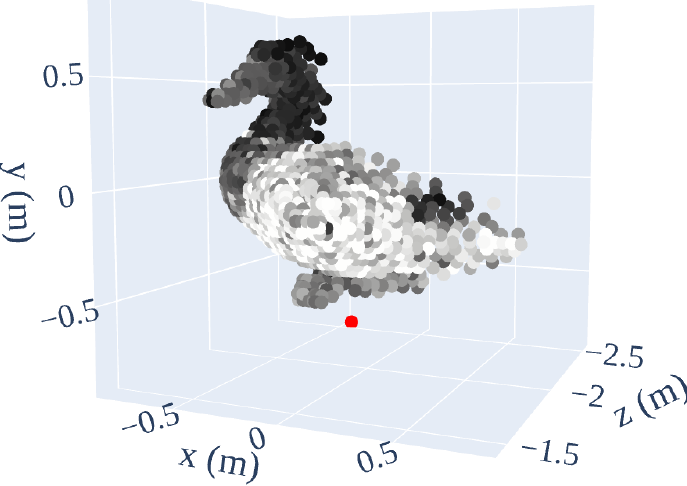}
        \end{subfigure}
        &
        \begin{subfigure}{0.16\linewidth}
            \centering
            \includegraphics[width=\linewidth,trim={1.6cm 1.5cm 5.3cm 3.5cm},clip]{figs/diff-sim/duck/oobsync-pw.pdf}
        \end{subfigure}
        &
        \begin{subfigure}{0.16\linewidth}
            \centering
            \includegraphics[width=\linewidth,trim={1.6cm 1.5cm 5.3cm 3.5cm},clip]{figs/diff-sim/duck/oobfree-pw.pdf}
        \end{subfigure}
        &
        \begin{subfigure}{0.16\linewidth}
            \centering
            \includegraphics[width=\linewidth,trim={1.6cm 1.5cm 5.3cm 3.5cm},clip]{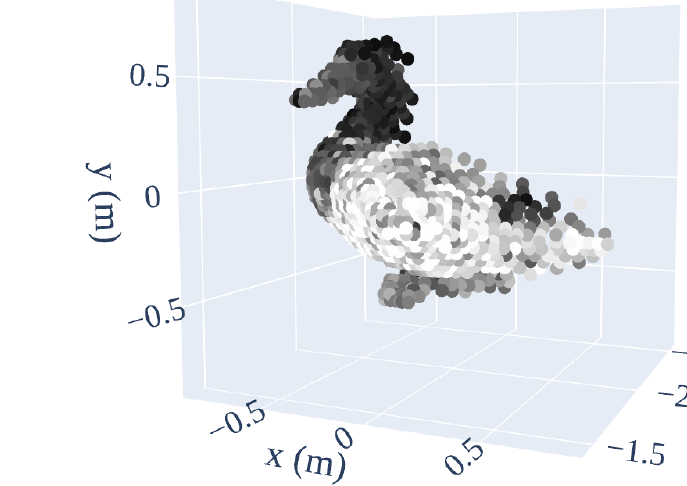}
        \end{subfigure}
        &
        \begin{subfigure}{0.16\linewidth}
            \centering
            \includegraphics[width=\linewidth,trim={1.6cm 1.5cm 5.3cm 3.5cm},clip]{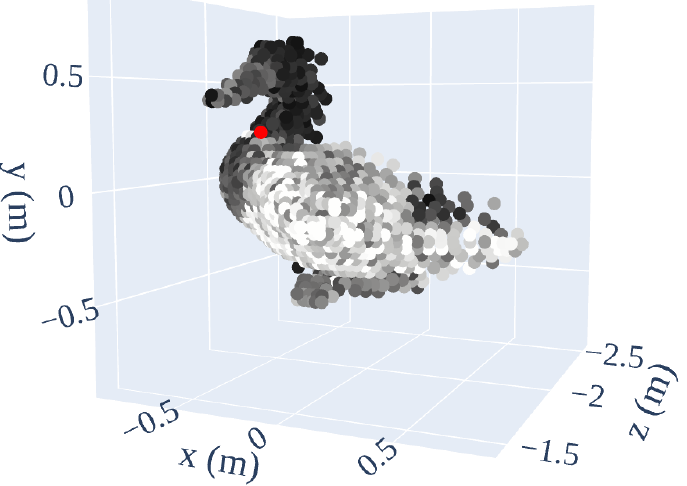}
        \end{subfigure}
        &
        \begin{subfigure}{0.16\linewidth}
            \centering
            \includegraphics[width=\linewidth,trim={1.6cm 1.5cm 5.3cm 3.5cm},clip]{figs/diff-sim/duck/oobfree-diff.pdf}
        \end{subfigure}
        \vspace{0.2em}
        \\

        &
        \begin{tabular}{@{}|@{\,}c@{\,}|@{\,}c@{\,}|@{}}
        \hline
        
        \scriptsize{$\rmse(\what{S})$} & \scriptsize{$\rmse(\what{z})$} \\
        \scriptsize{\SI[tight-spacing=true]{2.1e-02}{}} & \scriptsize{\SI{0.314}{\meter}} \\
        \hline
        
        \scriptsize{$\rmse(\what{B})$} & \scriptsize{$\mae(\what{z})$} \\
        \scriptsize{\SI[tight-spacing=true]{5.8e-02}{}} & \scriptsize{\SI{0.017}{\meter}} \\
        \hline
        \end{tabular}

        &
        \begin{tabular}{@{}|@{\,}c@{\,}|@{\,}c@{\,}|@{}}
        \hline
        
        \scriptsize{$\rmse(\what{S})$} & \scriptsize{$\rmse(\what{z})$} \\
        \scriptsize{\SI[tight-spacing=true]{2.5e-02}{}} & \scriptsize{\SI{0.884}{\meter}} \\
        \hline
        
        \scriptsize{$\rmse(\what{B})$} & \scriptsize{$\mae(\what{z})$} \\
        \scriptsize{\SI[tight-spacing=true]{9.7e-02}{}} & \scriptsize{\SI{0.084}{\meter}} \\
        \hline
        \end{tabular}

        &
        \begin{tabular}{@{}|@{\,}c@{\,}|@{\,}c@{\,}|@{}}
        \hline
        
        \scriptsize{$\rmse(\what{S})$} & \scriptsize{$\rmse(\what{z})$} \\
        \scriptsize{\SI[tight-spacing=true]{2.4e-02}{}} & \scriptsize{\SI{0.320}{\meter}} \\
        \hline
        
        \scriptsize{$\rmse(\what{B})$} & \scriptsize{$\mae(\what{z})$} \\
        \scriptsize{\SI[tight-spacing=true]{6.2e-02}{}} & \scriptsize{\SI{0.022}{\meter}} \\
        \hline
        \end{tabular}

        &
        \begin{tabular}{@{}|@{\,}c@{\,}|@{\,}c@{\,}|@{}}
        \hline
        
        \scriptsize{$\rmse(\what{S})$} & \scriptsize{$\rmse(\what{z})$} \\
        \scriptsize{\SI[tight-spacing=true]{2.1e-02}{}} & \scriptsize{\SI{0.004}{\meter}} \\
        \hline
        
        \scriptsize{$\rmse(\what{B})$} & \scriptsize{$\mae(\what{z})$} \\
        \scriptsize{\SI[tight-spacing=true]{5.8e-02}{}} & \scriptsize{\SI{0.002}{\meter}} \\
        \hline
        \end{tabular}

        &
        \begin{tabular}{@{}|@{\,}c@{\,}|@{\,}c@{\,}|@{}}
        \hline
        
        \scriptsize{$\rmse(\what{S})$} & \scriptsize{$\rmse(\what{z})$} \\
        \scriptsize{\SI[tight-spacing=true]{2.5e-02}{}} & \scriptsize{\SI{0.025}{\meter}} \\
        \hline
        
        \scriptsize{$\rmse(\what{B})$} & \scriptsize{$\mae(\what{z})$} \\
        \scriptsize{\SI[tight-spacing=true]{9.7e-02}{}} & \scriptsize{\SI{0.004}{\meter}} \\
        \hline
        \end{tabular}

        &
        \begin{tabular}{@{}|@{\,}c@{\,}|@{\,}c@{\,}|@{}}
        \hline
        
        \scriptsize{$\rmse(\what{S})$} & \scriptsize{$\rmse(\what{z})$} \\
        \scriptsize{\SI[tight-spacing=true]{2.4e-02}{}} & \scriptsize{\SI{0.085}{\meter}} \\
        \hline
        
        \scriptsize{$\rmse(\what{B})$} & \scriptsize{$\mae(\what{z})$} \\
        \scriptsize{\SI[tight-spacing=true]{6.2e-02}{}} & \scriptsize{\SI{0.007}{\meter}} \\
        \hline
        \end{tabular}

        \vspace{1em}
        
        \\
        
        \rotatebox[origin=l]{90}{\hspace{-2mm} \small Man}
        &
        \begin{subfigure}{0.16\linewidth}
            \centering
            \includegraphics[width=\linewidth,trim={1.6cm 1.5cm 5.3cm 3.5cm},clip]{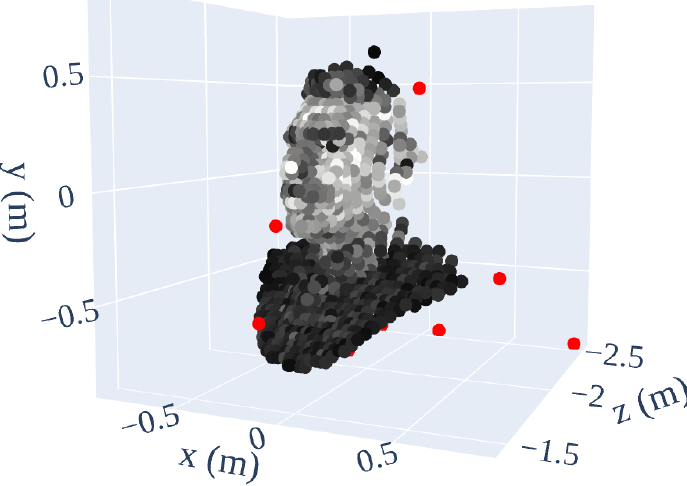}
        \end{subfigure}
        &
        \begin{subfigure}{0.16\linewidth}
            \centering
            \includegraphics[width=\linewidth,trim={1.6cm 1.5cm 5.3cm 3.5cm},clip]{figs/diff-sim/man/oobsync-pw.pdf}
        \end{subfigure}
        &
        \begin{subfigure}{0.16\linewidth}
            \centering
            \includegraphics[width=\linewidth,trim={1.6cm 1.5cm 5.3cm 3.5cm},clip]{figs/diff-sim/man/oobfree-pw.pdf}
        \end{subfigure}
        &
        \begin{subfigure}{0.16\linewidth}
            \centering
            \includegraphics[width=\linewidth,trim={1.6cm 1.5cm 5.3cm 3.5cm},clip]{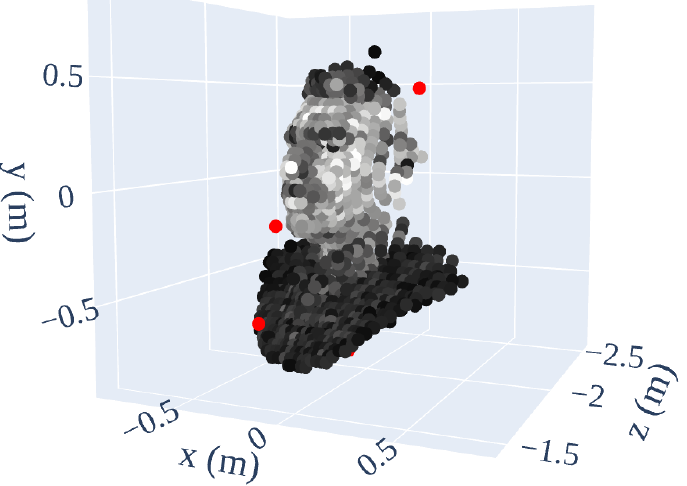}
        \end{subfigure}
        &
        \begin{subfigure}{0.16\linewidth}
            \centering
            \includegraphics[width=\linewidth,trim={1.6cm 1.5cm 5.3cm 3.5cm},clip]{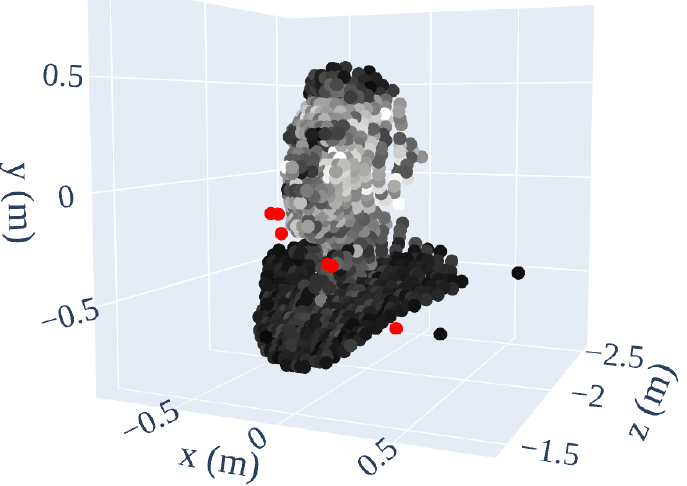}
        \end{subfigure}
        &
        \begin{subfigure}{0.16\linewidth}
            \centering
            \includegraphics[width=\linewidth,trim={1.6cm 1.5cm 5.3cm 3.5cm},clip]{figs/diff-sim/man/oobfree-diff.pdf}
        \end{subfigure}
        \vspace{0.3em}
        \\
        
        &
        \begin{tabular}{@{}|@{\,}c@{\,}|@{\,}c@{\,}|@{}}
        \hline
        
        \scriptsize{$\rmse(\what{S})$} & \scriptsize{$\rmse(\what{z})$} \\
        \scriptsize{\SI[tight-spacing=true]{1.6e-02}{}} & \scriptsize{\SI{1.117}{\meter}} \\
        \hline
        
        \scriptsize{$\rmse(\what{B})$} & \scriptsize{$\mae(\what{z})$} \\
        \scriptsize{\SI[tight-spacing=true]{5.8e-02}{}} & \scriptsize{\SI{0.152}{\meter}} \\
        \hline
        \end{tabular}

        &
        \begin{tabular}{@{}|@{\,}c@{\,}|@{\,}c@{\,}|@{}}
        \hline
        
        \scriptsize{$\rmse(\what{S})$} & \scriptsize{$\rmse(\what{z})$} \\
        \scriptsize{\SI[tight-spacing=true]{1.8e-02}{}} & \scriptsize{\SI{1.624}{\meter}} \\
        \hline
        
        \scriptsize{$\rmse(\what{B})$} & \scriptsize{$\mae(\what{z})$} \\
        \scriptsize{\SI[tight-spacing=true]{1.0e-01}{}} & \scriptsize{\SI{0.307}{\meter}} \\
        \hline
        \end{tabular}

        &
        \begin{tabular}{@{}|@{\,}c@{\,}|@{\,}c@{\,}|@{}}
        \hline
        
        \scriptsize{$\rmse(\what{S})$} & \scriptsize{$\rmse(\what{z})$} \\
        \scriptsize{\SI[tight-spacing=true]{1.8e-02}{}} & \scriptsize{\SI{1.620}{\meter}} \\
        \hline
        
        \scriptsize{$\rmse(\what{B})$} & \scriptsize{$\mae(\what{z})$} \\
        \scriptsize{\SI[tight-spacing=true]{6.6e-02}{}} & \scriptsize{\SI{0.294}{\meter}} \\
        \hline
        \end{tabular}

        &
        \begin{tabular}{@{}|@{\,}c@{\,}|@{\,}c@{\,}|@{}}
        \hline
        
        \scriptsize{$\rmse(\what{S})$} & \scriptsize{$\rmse(\what{z})$} \\
        \scriptsize{\SI[tight-spacing=true]{1.6e-02}{}} & \scriptsize{\SI{0.076}{\meter}} \\
        \hline
        
        \scriptsize{$\rmse(\what{B})$} & \scriptsize{$\mae(\what{z})$} \\
        \scriptsize{\SI[tight-spacing=true]{5.8e-02}{}} & \scriptsize{\SI{0.009}{\meter}} \\
        \hline
        \end{tabular}

        &
        \begin{tabular}{@{}|@{\,}c@{\,}|@{\,}c@{\,}|@{}}
        \hline
        
        \scriptsize{$\rmse(\what{S})$} & \scriptsize{$\rmse(\what{z})$} \\
        \scriptsize{\SI[tight-spacing=true]{1.8e-02}{}} & \scriptsize{\SI{0.115}{\meter}} \\
        \hline
        
        \scriptsize{$\rmse(\what{B})$} & \scriptsize{$\mae(\what{z})$} \\
        \scriptsize{\SI[tight-spacing=true]{1.0e-01}{}} & \scriptsize{\SI{0.016}{\meter}} \\
        \hline
        \end{tabular}

        &
        \begin{tabular}{@{}|@{\,}c@{\,}|@{\,}c@{\,}|@{}}
        \hline
        
        \scriptsize{$\rmse(\what{S})$} & \scriptsize{$\rmse(\what{z})$} \\
        \scriptsize{\SI[tight-spacing=true]{1.8e-02}{}} & \scriptsize{\SI{0.189}{\meter}} \\
        \hline
        
        \scriptsize{$\rmse(\what{B})$} & \scriptsize{$\mae(\what{z})$} \\
        \scriptsize{\SI[tight-spacing=true]{6.6e-02}{}} & \scriptsize{\SI{0.020}{\meter}} \\
        \hline
        \end{tabular}

        \vspace{1em}
        \\
        \rotatebox[origin=l]{90}{\hspace{-2mm} \small Mario}
        &
        \begin{subfigure}{0.16\linewidth}
            \centering
            \includegraphics[width=\linewidth,trim={1.6cm 1.5cm 5.3cm 3.5cm},clip]{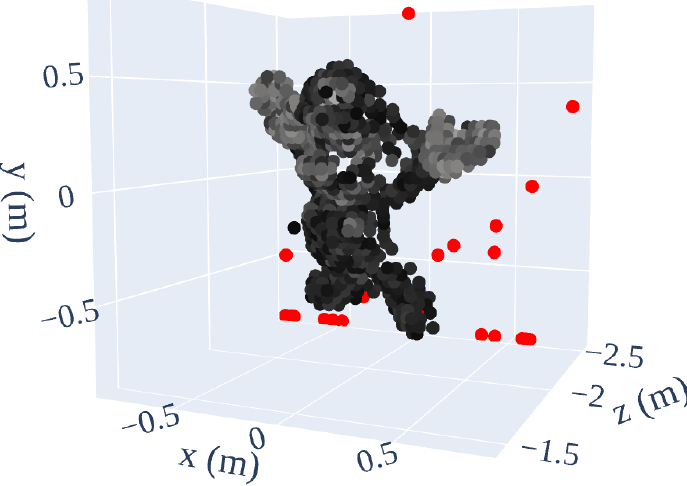}
        \end{subfigure}
        &
        \begin{subfigure}{0.16\linewidth}
            \centering
            \includegraphics[width=\linewidth,trim={1.6cm 1.5cm 5.3cm 3.5cm},clip]{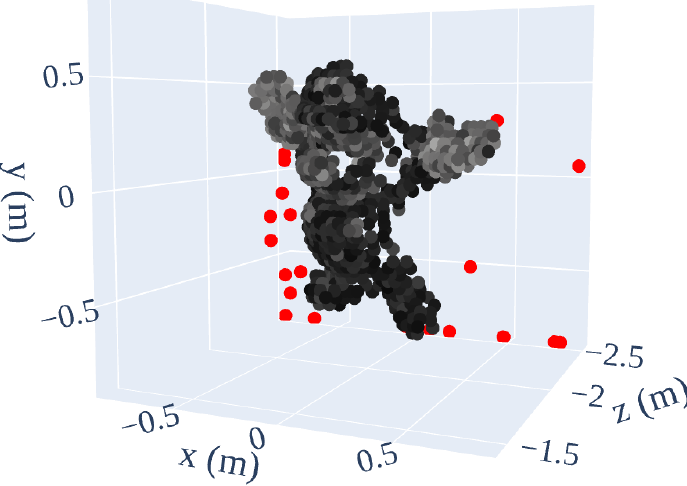}
        \end{subfigure}
        &
        \begin{subfigure}{0.16\linewidth}
            \centering
            \includegraphics[width=\linewidth,trim={1.6cm 1.5cm 5.3cm 3.5cm},clip]{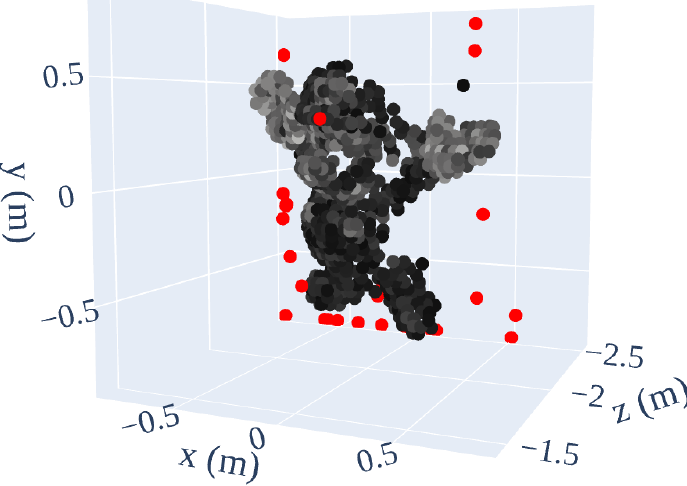}
        \end{subfigure}
        &
        \begin{subfigure}{0.16\linewidth}
            \centering
            \includegraphics[width=\linewidth,trim={1.6cm 1.5cm 5.3cm 3.5cm},clip]{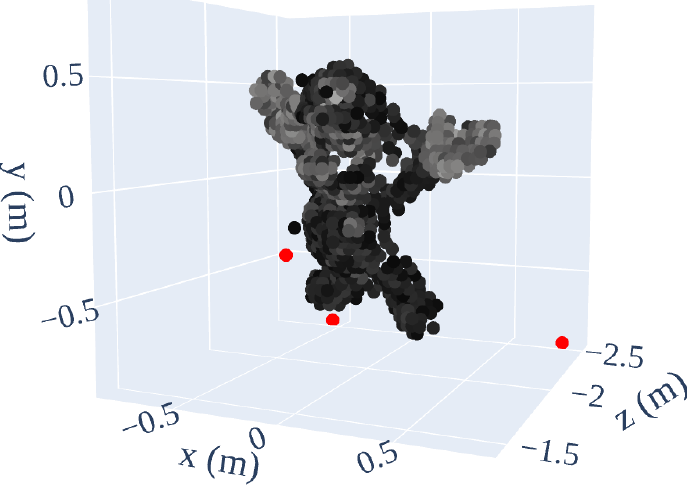}
        \end{subfigure}
        &
        \begin{subfigure}{0.16\linewidth}
            \centering
            \includegraphics[width=\linewidth,trim={1.6cm 1.5cm 5.3cm 3.5cm},clip]{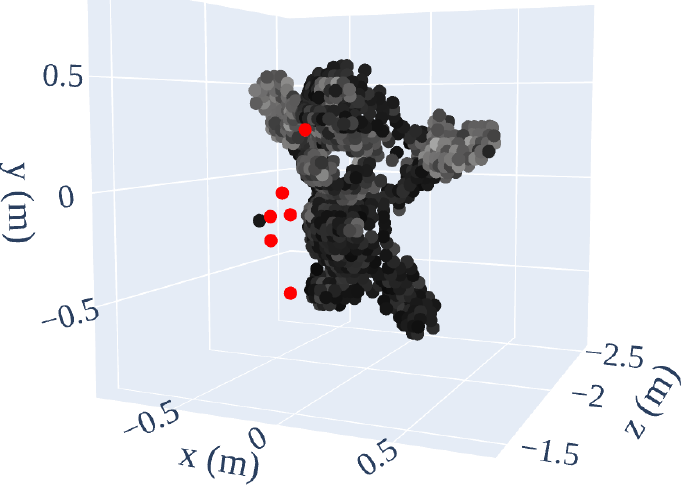}
        \end{subfigure}
        &
        \begin{subfigure}{0.16\linewidth}
            \centering
            \includegraphics[width=\linewidth,trim={1.6cm 1.5cm 5.3cm 3.5cm},clip]{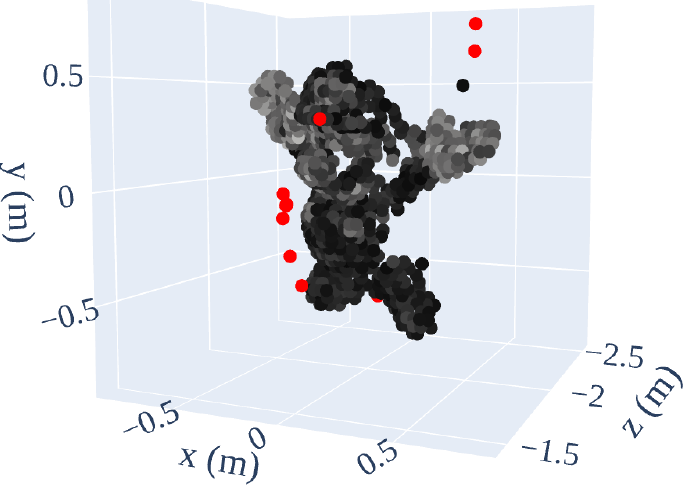}
        \end{subfigure}
        \vspace{0.3em}
        \\
        
        &
        \begin{tabular}{@{}|@{\,}c@{\,}|@{\,}c@{\,}|@{}}
        \hline
        
        \scriptsize{$\rmse(\what{S})$} & \scriptsize{$\rmse(\what{z})$} \\
        \scriptsize{\SI[tight-spacing=true]{1.4e-02}{}} & \scriptsize{\SI{1.310}{\meter}} \\
        \hline
        
        \scriptsize{$\rmse(\what{B})$} & \scriptsize{$\mae(\what{z})$} \\
        \scriptsize{\SI[tight-spacing=true]{5.9e-02}{}} & \scriptsize{\SI{0.235}{\meter}} \\
        \hline
        \end{tabular}

        &
        \begin{tabular}{@{}|@{\,}c@{\,}|@{\,}c@{\,}|@{}}
        \hline
        
        \scriptsize{$\rmse(\what{S})$} & \scriptsize{$\rmse(\what{z})$} \\
        \scriptsize{\SI[tight-spacing=true]{1.4e-02}{}} & \scriptsize{\SI{1.962}{\meter}} \\
        \hline
        
        \scriptsize{$\rmse(\what{B})$} & \scriptsize{$\mae(\what{z})$} \\
        \scriptsize{\SI[tight-spacing=true]{1.0e-01}{}} & \scriptsize{\SI{0.445}{\meter}} \\
        \hline
        \end{tabular}

        &
        \begin{tabular}{@{}|@{\,}c@{\,}|@{\,}c@{\,}|@{}}
        \hline
        
        \scriptsize{$\rmse(\what{S})$} & \scriptsize{$\rmse(\what{z})$} \\
        \scriptsize{\SI[tight-spacing=true]{1.5e-02}{}} & \scriptsize{\SI{1.860}{\meter}} \\
        \hline
        
        \scriptsize{$\rmse(\what{B})$} & \scriptsize{$\mae(\what{z})$} \\
        \scriptsize{\SI[tight-spacing=true]{6.4e-02}{}} & \scriptsize{\SI{0.397}{\meter}} \\
        \hline
        \end{tabular}

        &
        \begin{tabular}{@{}|@{\,}c@{\,}|@{\,}c@{\,}|@{}}
        \hline
        
        \scriptsize{$\rmse(\what{S})$} & \scriptsize{$\rmse(\what{z})$} \\
        \scriptsize{\SI[tight-spacing=true]{1.4e-02}{}} & \scriptsize{\SI{0.229}{\meter}} \\
        \hline
        
        \scriptsize{$\rmse(\what{B})$} & \scriptsize{$\mae(\what{z})$} \\
        \scriptsize{\SI[tight-spacing=true]{5.9e-02}{}} & \scriptsize{\SI{0.017}{\meter}} \\
        \hline
        \end{tabular}

        &
        \begin{tabular}{@{}|@{\,}c@{\,}|@{\,}c@{\,}|@{}}
        \hline
        
        \scriptsize{$\rmse(\what{S})$} & \scriptsize{$\rmse(\what{z})$} \\
        \scriptsize{\SI[tight-spacing=true]{1.4e-02}{}} & \scriptsize{\SI{0.135}{\meter}} \\
        \hline
        
        \scriptsize{$\rmse(\what{B})$} & \scriptsize{$\mae(\what{z})$} \\
        \scriptsize{\SI[tight-spacing=true]{1.0e-01}{}} & \scriptsize{\SI{0.019}{\meter}} \\
        \hline
        \end{tabular}

        &
        \begin{tabular}{@{}|@{\,}c@{\,}|@{\,}c@{\,}|@{}}
        \hline
        
        \scriptsize{$\rmse(\what{S})$} & \scriptsize{$\rmse(\what{z})$} \\
        \scriptsize{\SI[tight-spacing=true]{1.5e-02}{}} & \scriptsize{\SI{0.180}{\meter}} \\
        \hline
        
        \scriptsize{$\rmse(\what{B})$} & \scriptsize{$\mae(\what{z})$} \\
        \scriptsize{\SI[tight-spacing=true]{6.4e-02}{}} & \scriptsize{\SI{0.029}{\meter}} \\
        \hline
        \end{tabular}

        \vspace{1em}
        \\
        &
        \hspace{0.5em}
        \includegraphics[width=3.2cm,trim={0 0.4cm 0 0},clip]{figs/diff-sim/duck/cbar-horr-est.pdf}
        
    \end{tabular}
    \caption{
    Reconstructions from simulated SPL measurements using pixel-wise ML estimators for \textbf{(a) ideal}, \textbf{(b) synchronous}, and \textbf{(c) free-running} measurements. Depth-regularized SSDR reconstructions are shown in \textbf{(d)}, \textbf{(e)}, and \textbf{(f)}. Point color represents the signal flux estimate, while red points indicate those projected onto the plotting space due to axis limits. 
    Error statistics, including RMSEs of signal flux, background flux, and depth estimates, as well as the mean absolute error (MAE) of depth estimates, are provided beneath each reconstruction.
    }
    \label{fig:regsim-supp}
\end{figure*}

%% file: figs/fig-A-ablation.tex
\def\figheight{2.3cm}

\begin{figure*}[!htb]
    \centering
    \begin{tabular}[t]{@{}c@{\,\,}c@{\,}c@{\,}|@{\,}c@{\,}c@{\,}|@{\,}c@{\,}c@{}}
        &
        \multicolumn{2}{c}{Ideal}
        &
        \multicolumn{2}{c}{Synchronous}
        &
        \multicolumn{2}{c}{Free-running}
        \vspace{0.5em}
        \\
        \rotatebox[origin=l]{90}{\hspace{-2mm} \small Duck}
        &
        \begin{subfigure}{0.15\linewidth}
            \centering
            \includegraphics[height=\figheight]{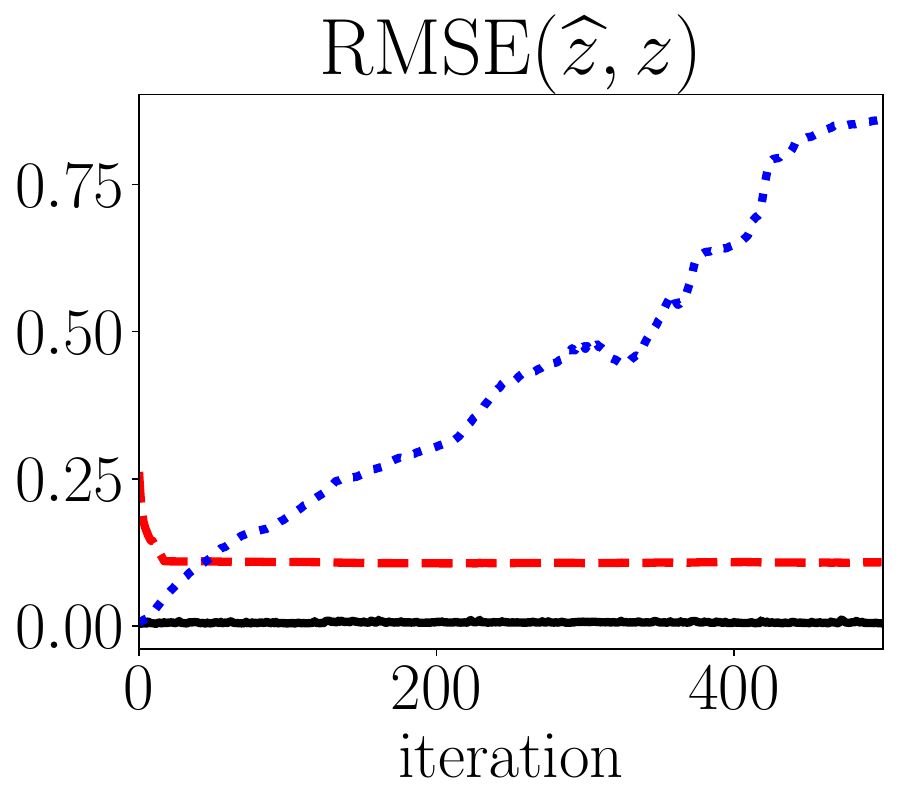}
        \end{subfigure}
        &
        \begin{subfigure}{0.16\linewidth}
            \centering
            \includegraphics[height=\figheight]{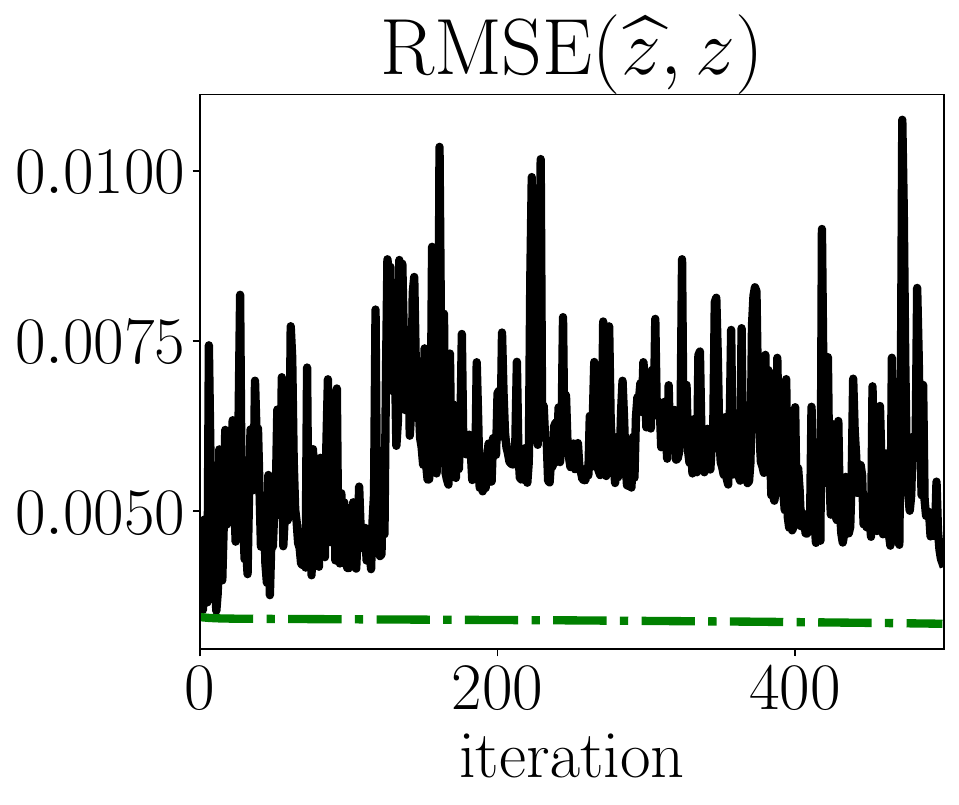}
        \end{subfigure}
        &
        \begin{subfigure}{0.15\linewidth}
            \centering
            \includegraphics[height=\figheight]{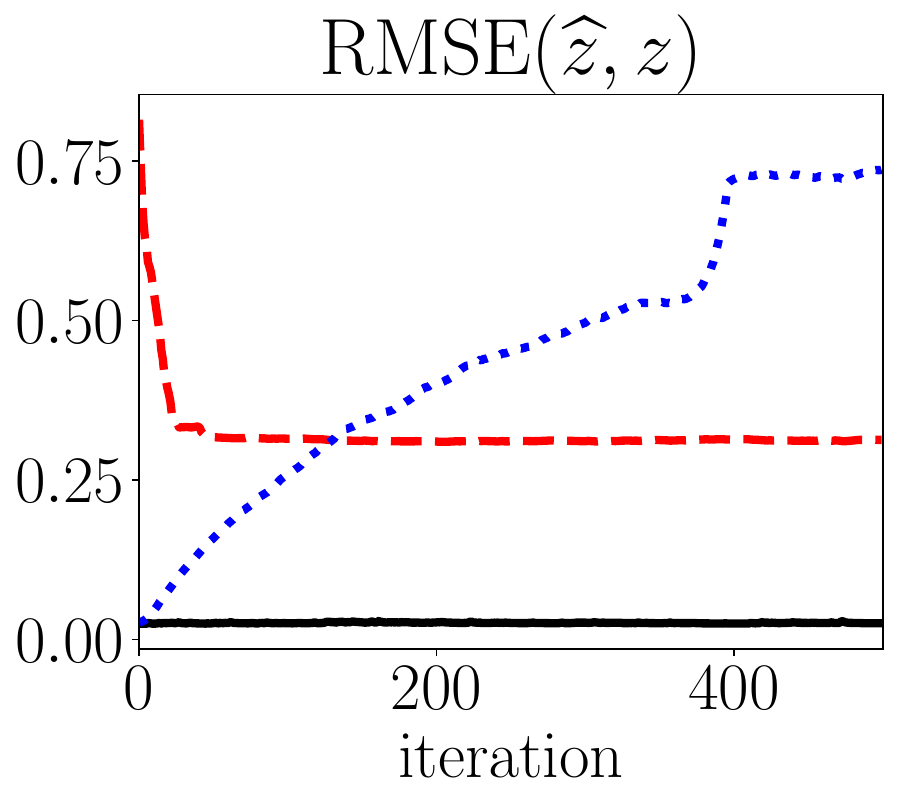}
        \end{subfigure}
        &
        \begin{subfigure}{0.16\linewidth}
            \centering
            \includegraphics[height=\figheight]{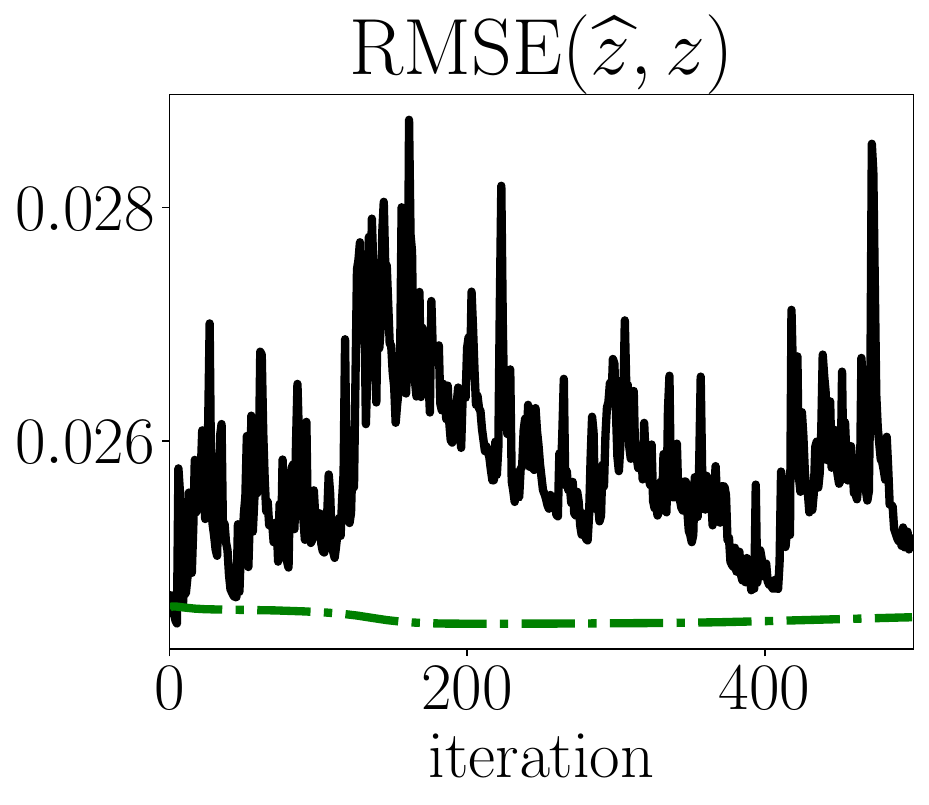}
        \end{subfigure}
        &
        \begin{subfigure}{0.15\linewidth}
            \centering
            \includegraphics[height=\figheight]{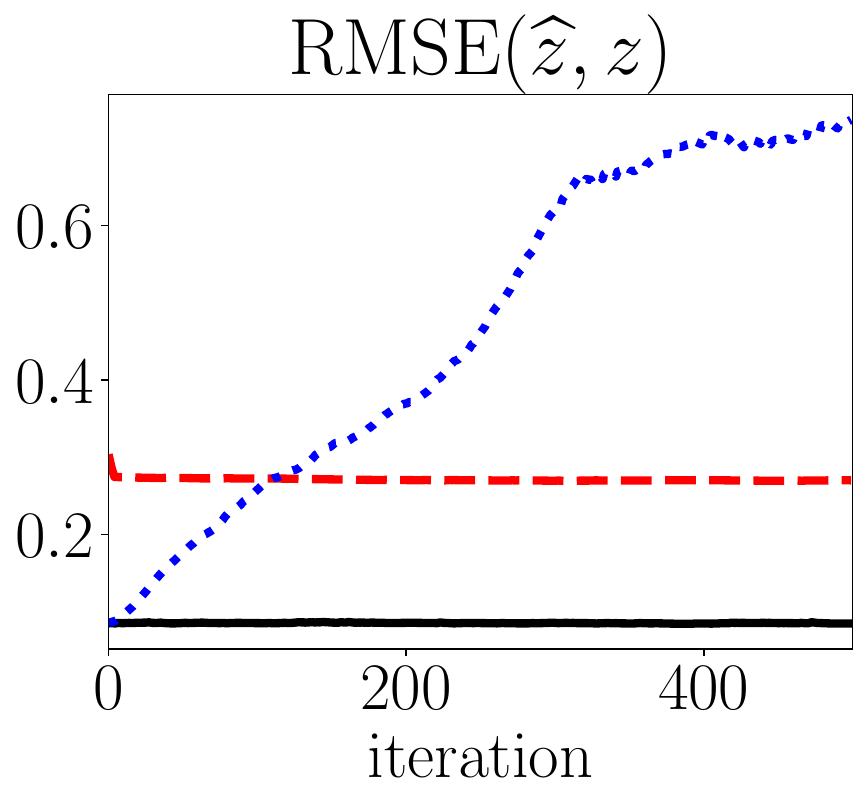}
        \end{subfigure}
        &
        \begin{subfigure}{0.16\linewidth}
            \centering
            \includegraphics[height=\figheight]{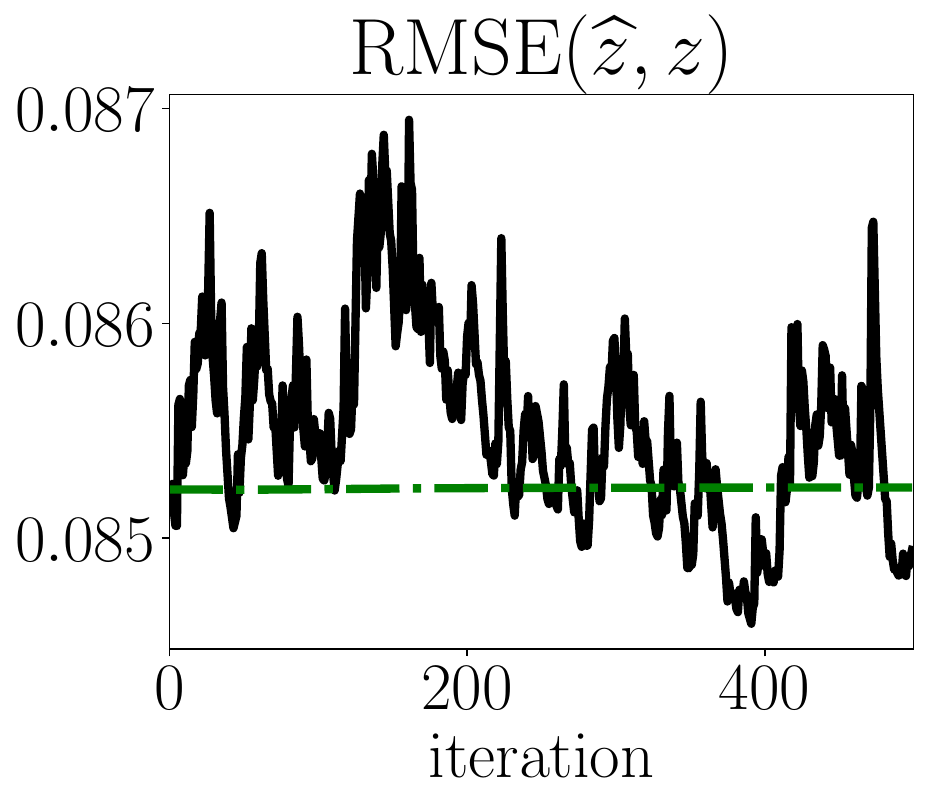}
        \end{subfigure}
        \\
        \rotatebox[origin=l]{90}{\hspace{-2mm} \small Man}
        &
        \begin{subfigure}{0.15\linewidth}
            \centering
            \includegraphics[height=\figheight]{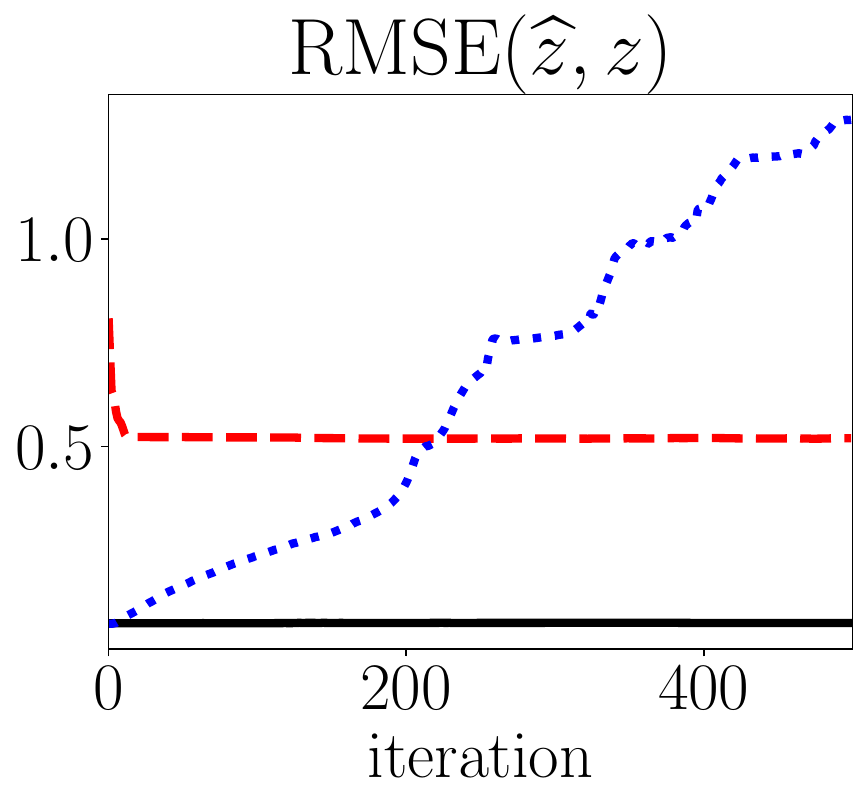}
        \end{subfigure}
        &
        \begin{subfigure}{0.16\linewidth}
            \centering
            \includegraphics[height=\figheight]{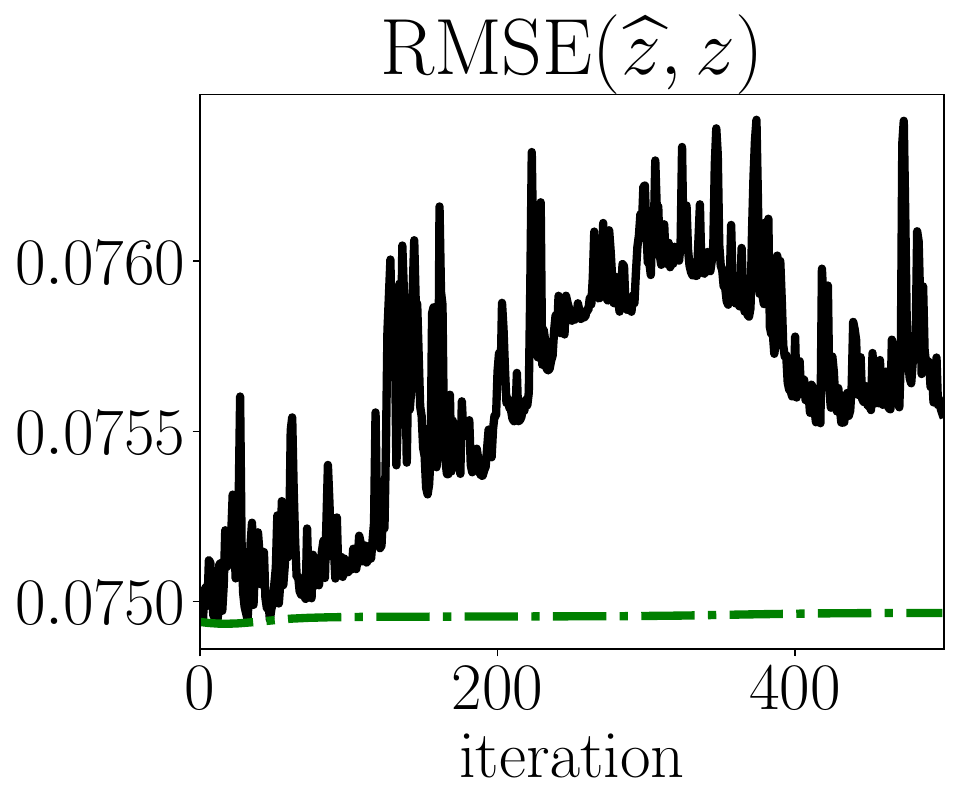}
        \end{subfigure}
        &
        \begin{subfigure}{0.15\linewidth}
            \centering
            \includegraphics[height=\figheight]{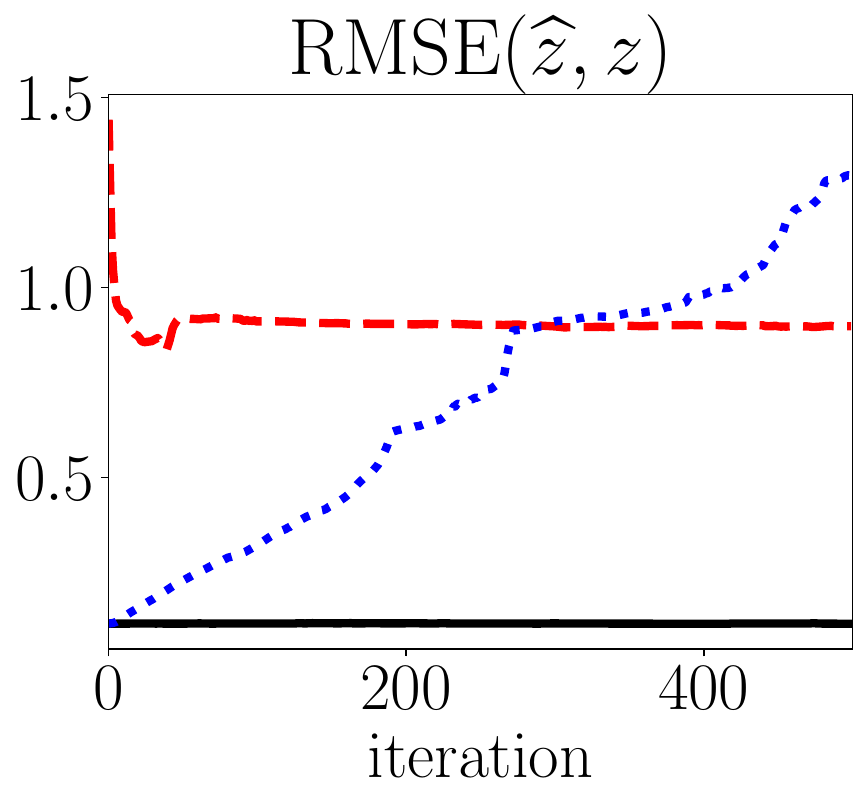}
        \end{subfigure}
        &
        \begin{subfigure}{0.16\linewidth}
            \centering
            \includegraphics[height=\figheight]{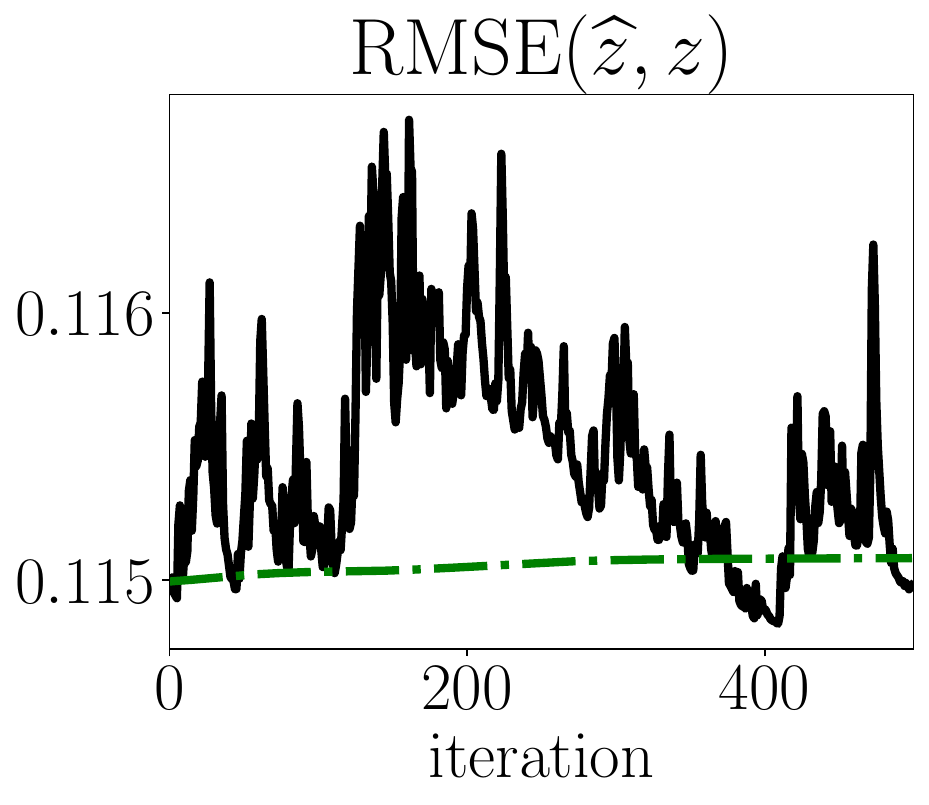}
        \end{subfigure}
        &
        \begin{subfigure}{0.15\linewidth}
            \centering
            \includegraphics[height=\figheight]{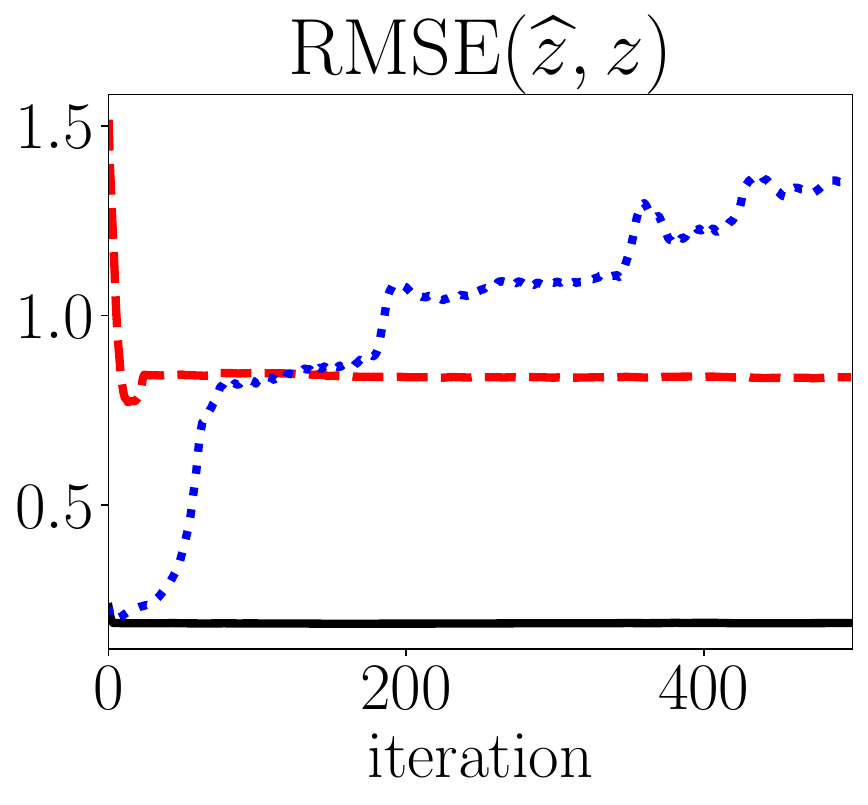}
        \end{subfigure}
        &
        \begin{subfigure}{0.16\linewidth}
            \centering
            \includegraphics[height=\figheight]{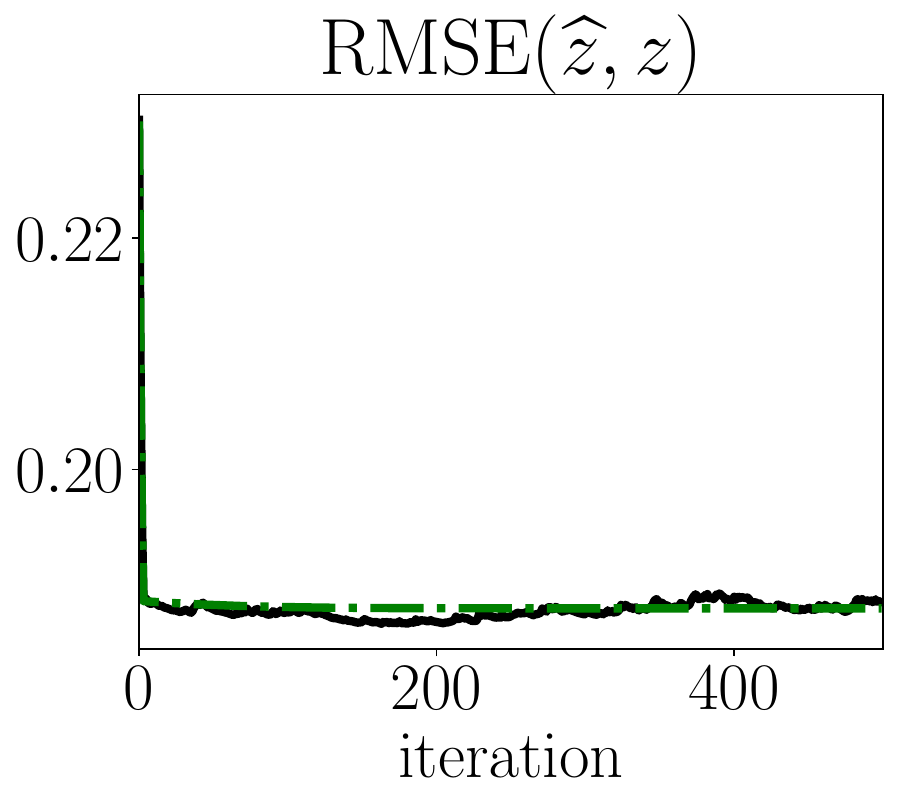}
        \end{subfigure}
        \\
        \rotatebox[origin=l]{90}{\hspace{-2mm} \small Mario}
        &
        \begin{subfigure}{0.15\linewidth}
            \centering
            \includegraphics[height=\figheight]{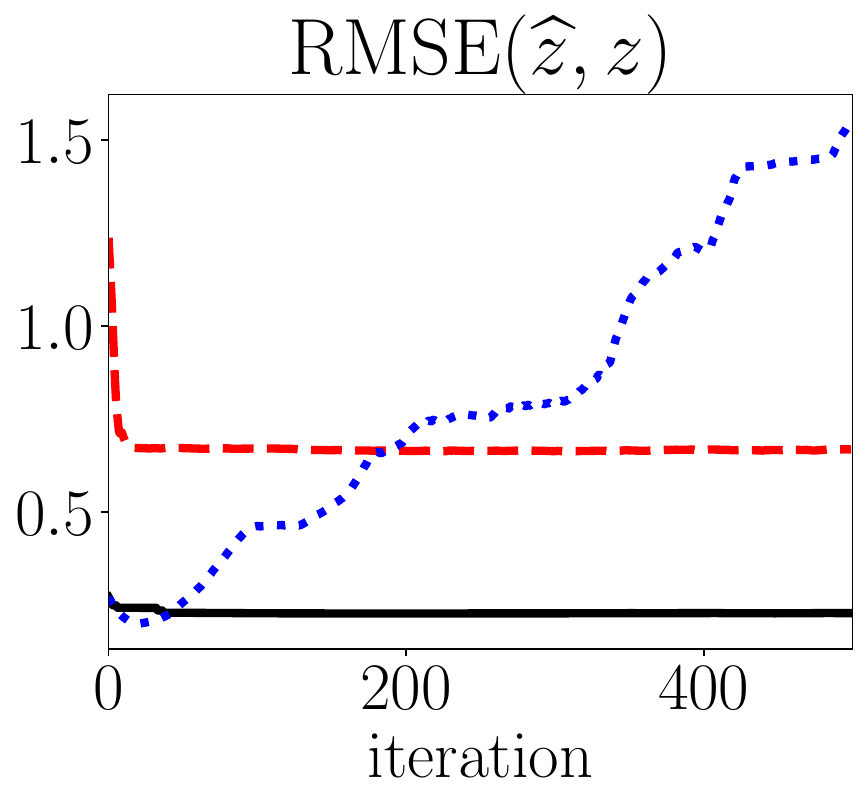}
        \end{subfigure}
        &
        \begin{subfigure}{0.16\linewidth}
            \centering
            \includegraphics[height=\figheight]{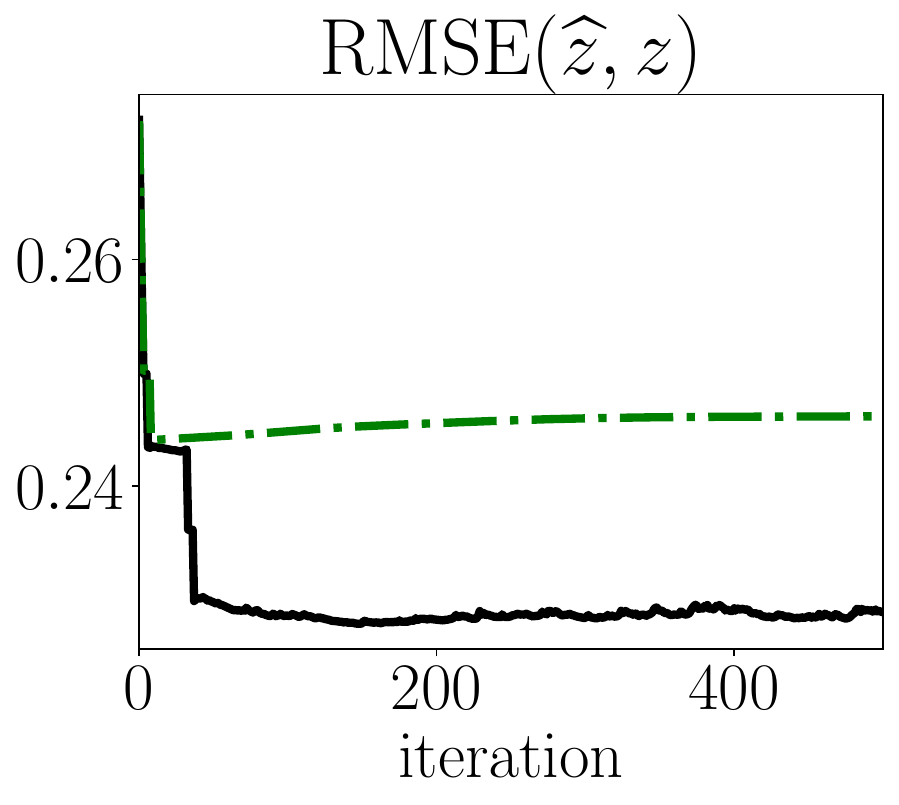}
        \end{subfigure}
        &
        \begin{subfigure}{0.15\linewidth}
            \centering
            \includegraphics[height=\figheight]{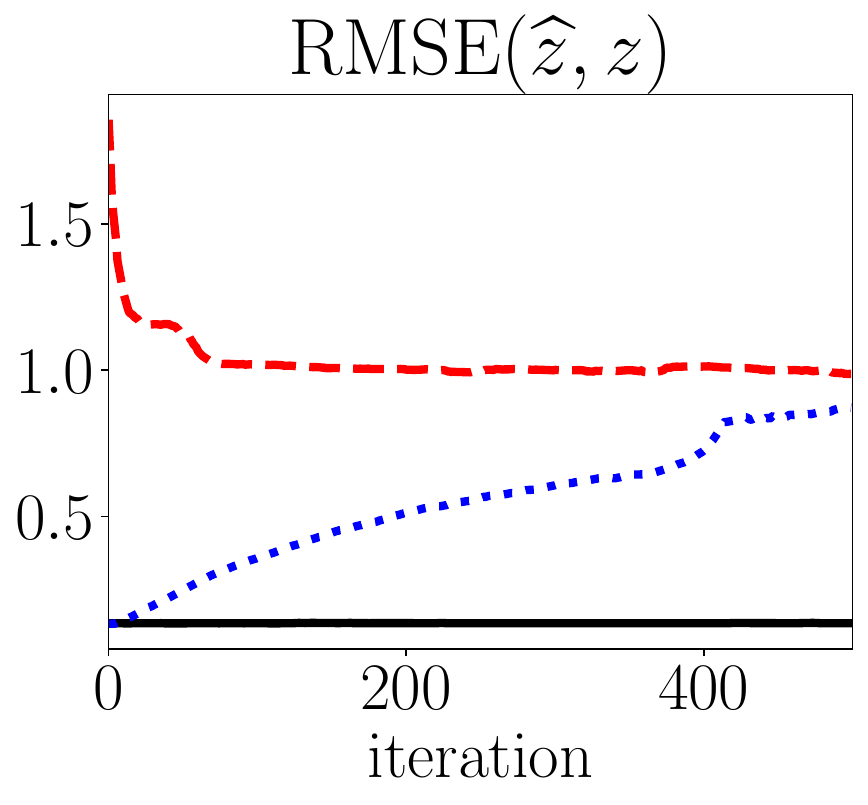}
        \end{subfigure}
        &
        \begin{subfigure}{0.16\linewidth}
            \centering
            \includegraphics[height=\figheight]{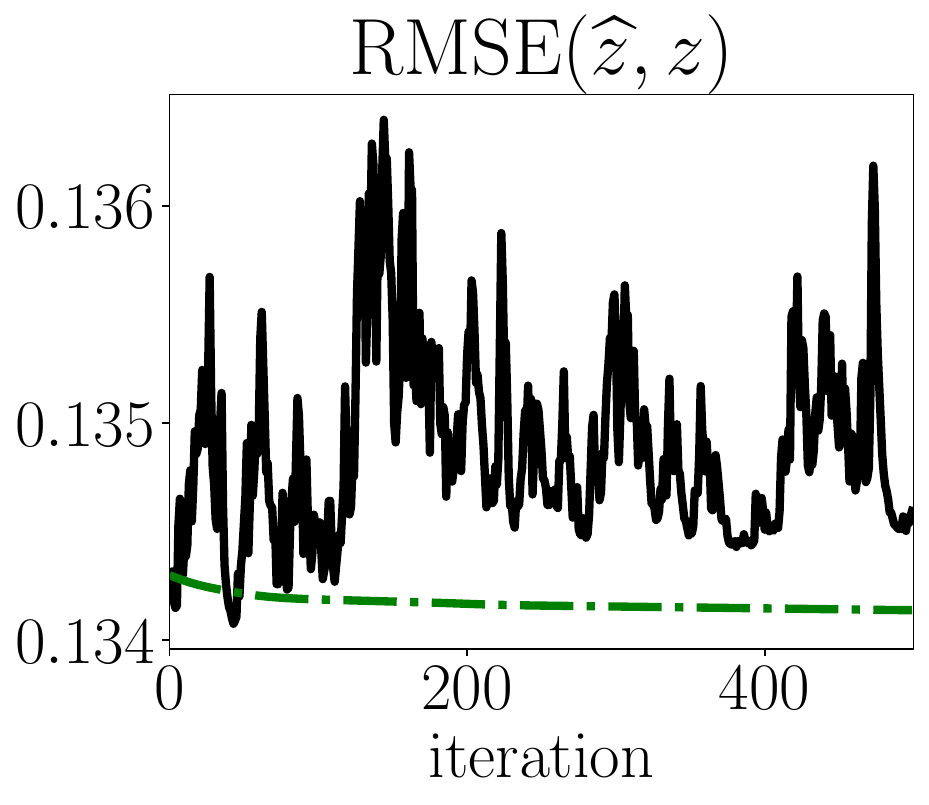}
        \end{subfigure}
        &
        \begin{subfigure}{0.15\linewidth}
            \centering
            \includegraphics[height=\figheight]{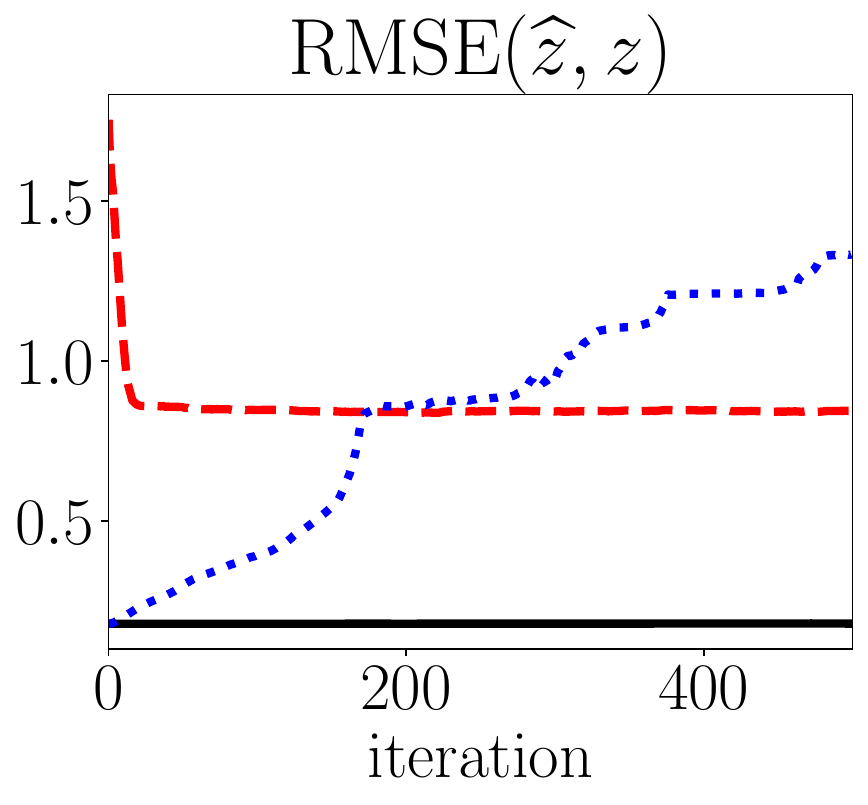}
        \end{subfigure}
        &
        \begin{subfigure}{0.16\linewidth}
            \centering
            \includegraphics[height=\figheight]{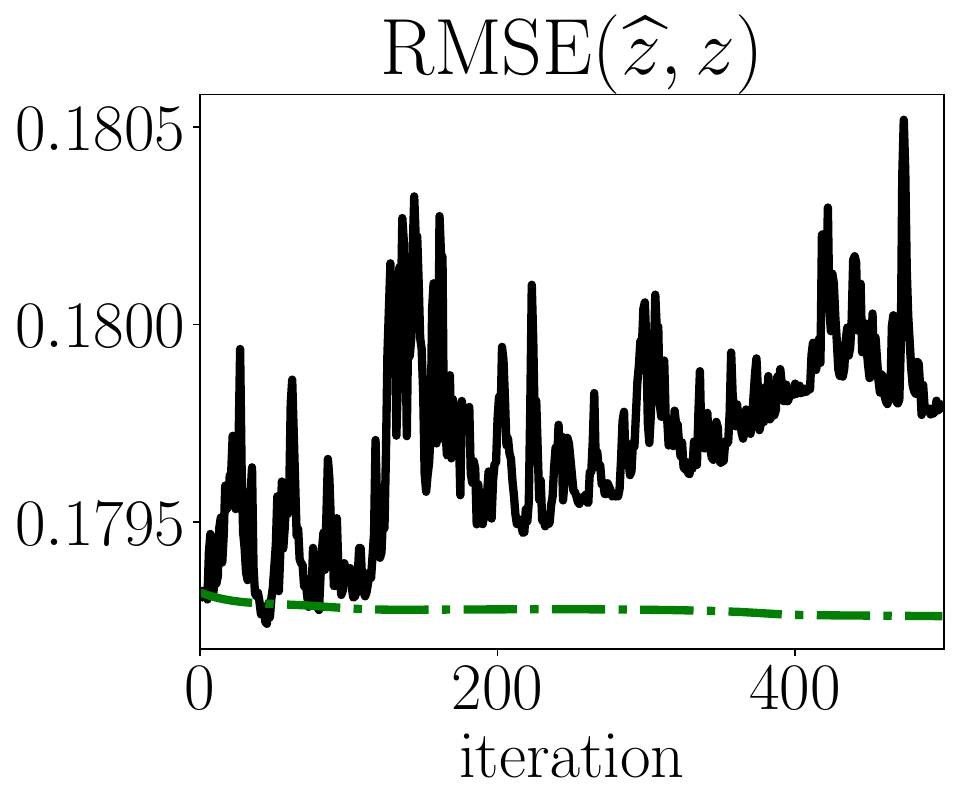}
        \end{subfigure}
    \end{tabular}
    \includegraphics[width=0.9\linewidth]{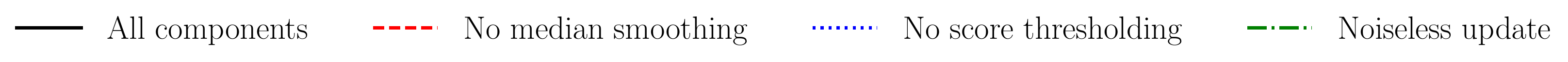}
    \caption{
    Ablation study: Depth RMSE per iteration of the SSDR algorithm under four conditions: (1) full SSDR algorithm, (2) without median smoothing, (3) without hard-thresholding the depth score, and (4) using noiseless iterative updates (i.e., standard gradient descent). Results are shown for ideal, synchronous, and free-running measurements across the three scenes from \Cref{fig:regsim-supp}.
    }
    \label{fig:ablation}
\end{figure*}

%% file: figs/fig-A-regcomparison.tex
\begin{figure*}[!htb]
    \centering
    \begin{tabular}[t]{@{}c@{\,\,}c@{}c@{}c@{}c@{}c@{}}
    & Ground Truth & Pixel-wise ML & SSDR & ManiPoP & 
    \\
    \rotatebox[origin=l]{90}{\hspace{-7mm} \small Point Cloud}
    &
    \begin{subfigure}{0.2\linewidth}
        \centering
        \includegraphics[width=\linewidth,trim={3cm 1.5cm 4cm 3.5cm},clip]{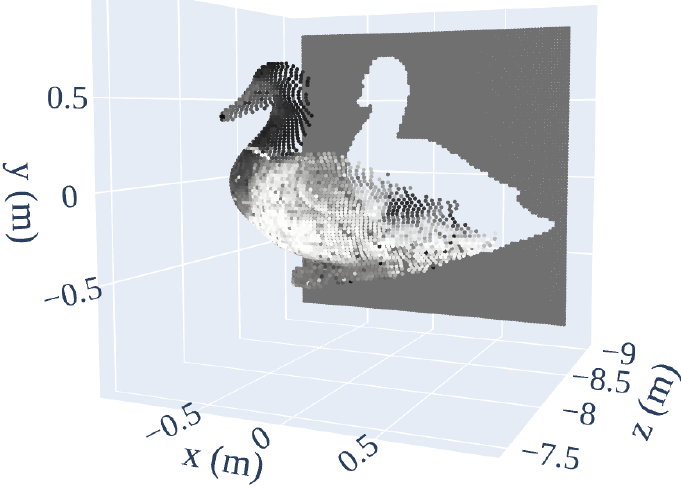}
    \end{subfigure}
    &
    \begin{subfigure}{0.2\linewidth}
        \centering
        \includegraphics[width=\linewidth,trim={3cm 1.5cm 4cm 3.5cm},clip]{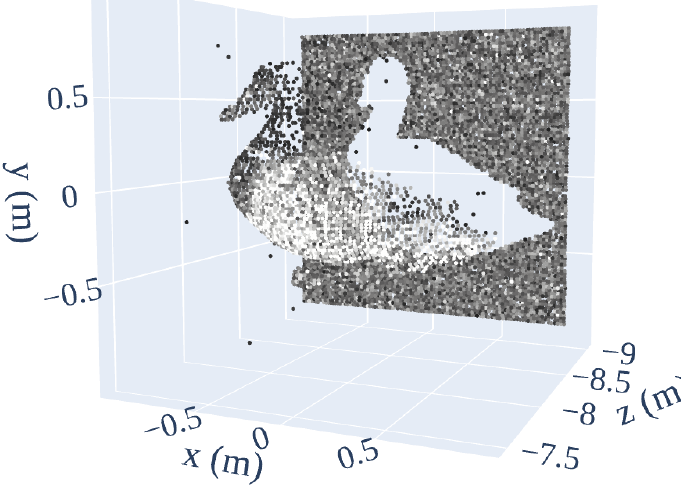}
    \end{subfigure}
    &
    \begin{subfigure}{0.2\linewidth}
        \centering
        \includegraphics[width=\linewidth,trim={3cm 1.5cm 4cm 3.5cm},clip]{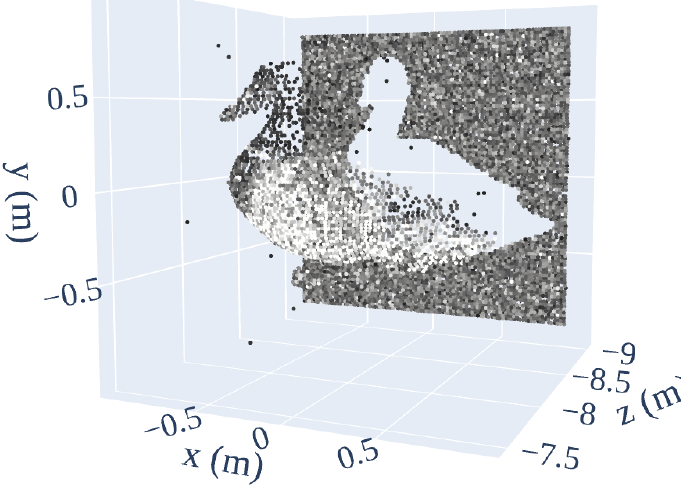}
    \end{subfigure}
    &
    \begin{subfigure}{0.2\linewidth}
        \centering
        \includegraphics[width=\linewidth,trim={3cm 1.5cm 4cm 3.5cm},clip]{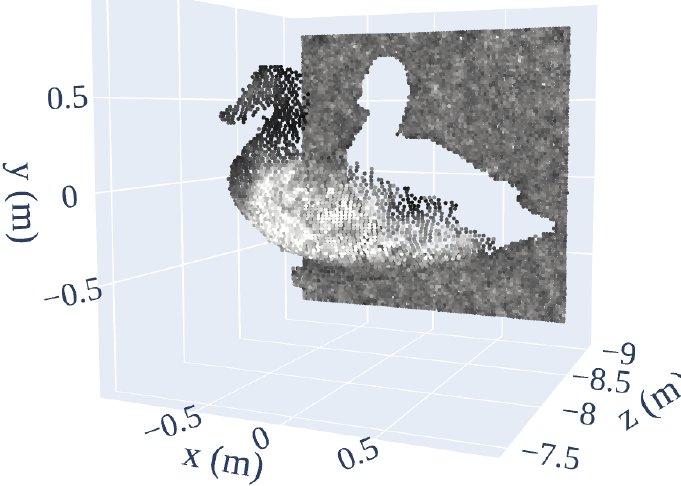}
    \end{subfigure}
    &
    \begin{subfigure}{0.1\linewidth}
        \centering
        \includegraphics[height=2.7cm]{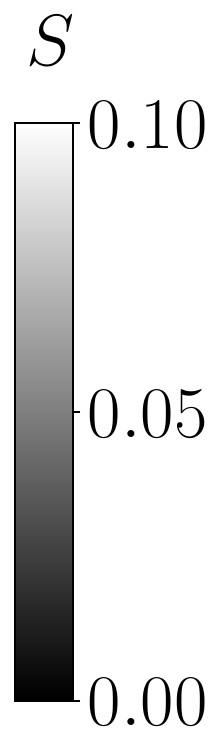}
    \end{subfigure}
    \\
    \rotatebox[origin=l]{90}{\hspace{-4mm} \small Depth} \hspace{-5mm}
    &
    \begin{subfigure}{0.2\linewidth}
        \centering
        \includegraphics[width=0.7\linewidth]{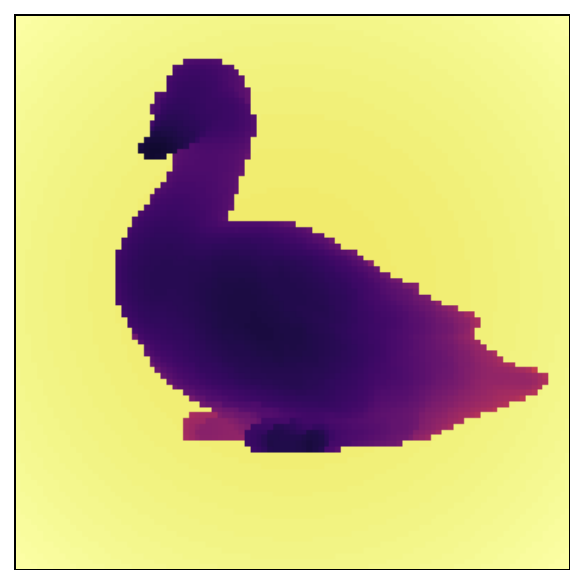}
    \end{subfigure}
    &
    \begin{subfigure}{0.2\linewidth}
        \centering
        \includegraphics[width=0.7\linewidth]{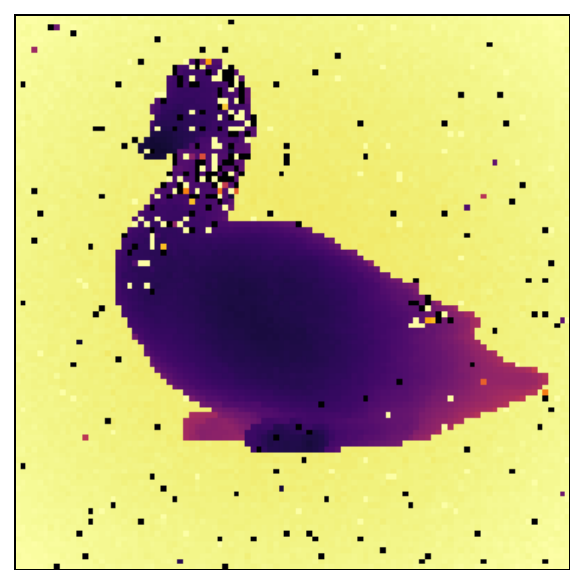}
    \end{subfigure}
    &
    \begin{subfigure}{0.2\linewidth}
        \centering
        \includegraphics[width=0.7\linewidth]{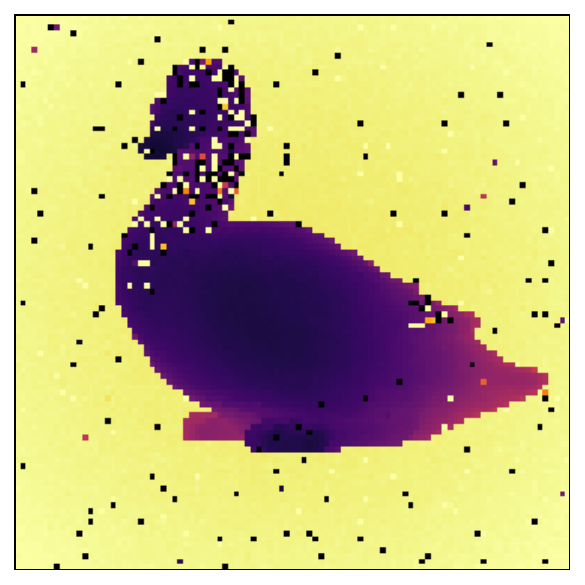}
    \end{subfigure}
    &
    \begin{subfigure}{0.2\linewidth}
        \centering
        \includegraphics[width=0.7\linewidth]{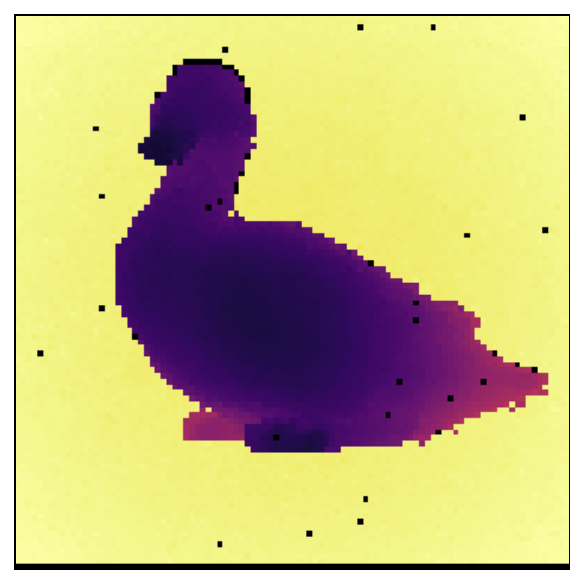}
    \end{subfigure}
    &
    \begin{subfigure}{0.1\linewidth}
        \centering
        \includegraphics[height=2.6cm]{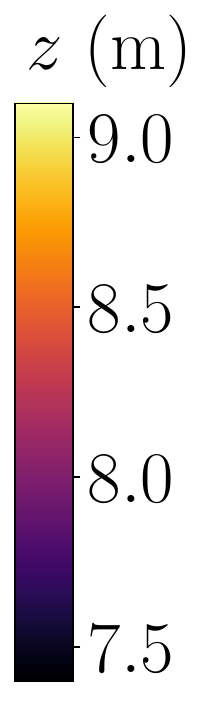}
    \end{subfigure}
    \\
    \rotatebox[origin=l]{90}{\hspace{-11mm} \small Zoom-in Depth} \hspace{-5mm}
    &
    \begin{subfigure}{0.2\linewidth}
        \centering
        \includegraphics[width=0.7\linewidth]{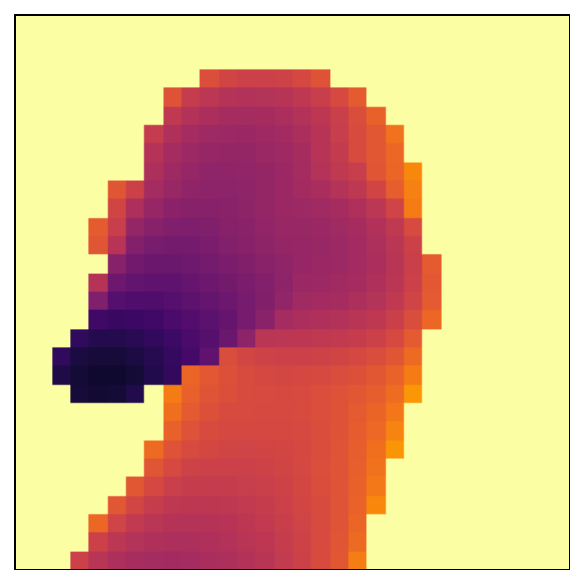}
    \end{subfigure}
    &
    \begin{subfigure}{0.2\linewidth}
        \centering
        \includegraphics[width=0.7\linewidth]{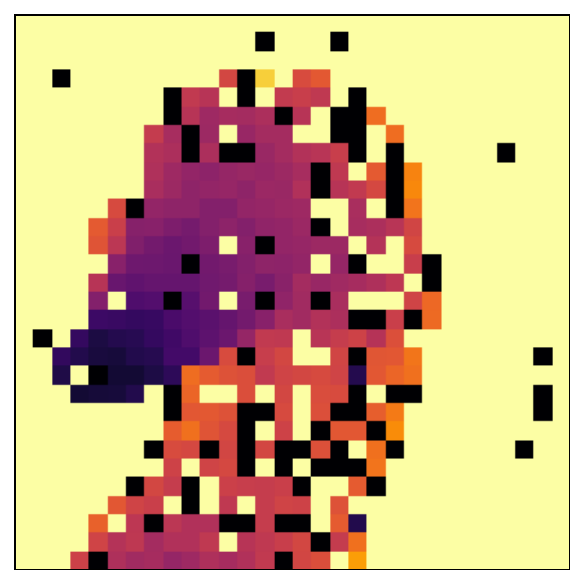}
    \end{subfigure}
    &
    \begin{subfigure}{0.2\linewidth}
        \centering
        \includegraphics[width=0.7\linewidth]{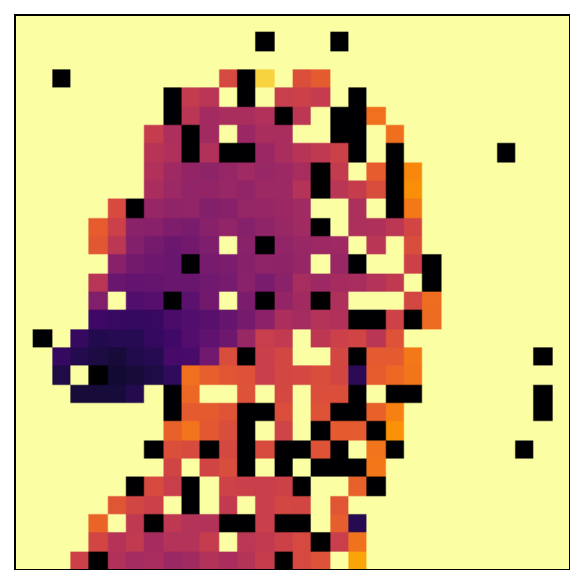}
    \end{subfigure}
    &
    \begin{subfigure}{0.2\linewidth}
        \centering
        \includegraphics[width=0.7\linewidth]{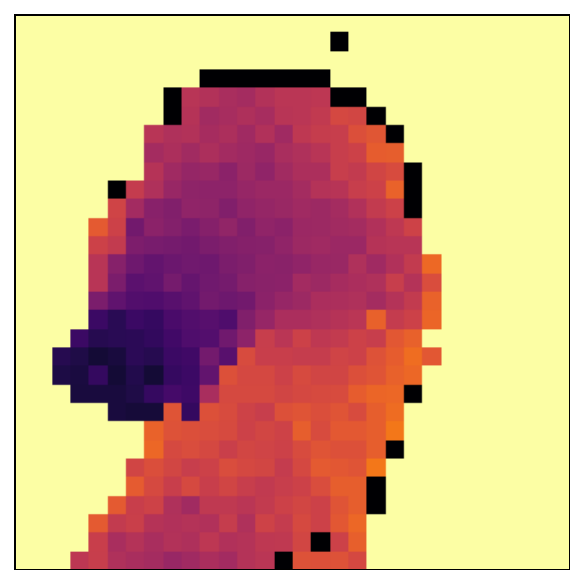}
    \end{subfigure}
    &
    \begin{subfigure}{0.1\linewidth}
        \centering
        \includegraphics[height=2.6cm]{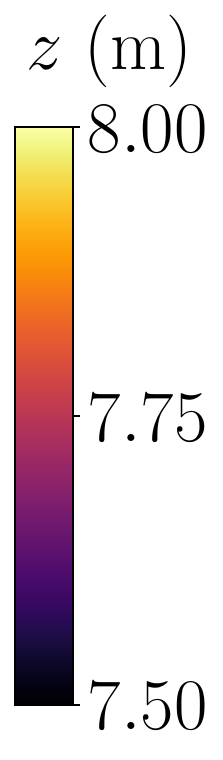}
    \end{subfigure}
    \\

    &&
    \begin{tabular}{@{}|@{\,}c@{\,}|@{\,}c@{\,}|@{}}
    \hline
    
    \scriptsize{$\rmse(\what{z})$} & \scriptsize{$\mae(\what{z})$} \\
    \scriptsize{\SI{0.876}{\meter}} & \scriptsize{\SI{0.149}{\meter}} \\
    \hline
    \end{tabular}

    &

    \begin{tabular}{@{}|@{\,}c@{\,}|@{\,}c@{\,}|@{}}
    \hline
    
    \scriptsize{$\rmse(\what{z})$} & \scriptsize{$\mae(\what{z})$} \\
    \scriptsize{\SI{0.866}{\meter}} & \scriptsize{\SI{0.146}{\meter}} \\
    \hline
    \end{tabular}

    &

    \begin{tabular}{@{}|@{\,}c@{\,}|@{\,}c@{\,}|@{}}
    \hline
    
    \scriptsize{$\rmse(\what{z})$} & \scriptsize{$\mae(\what{z})$} \\
    \scriptsize{\SI{1.086}{\meter}} & \scriptsize{\SI{0.141}{\meter}} \\
    \hline
    \end{tabular}

    \end{tabular}
    \caption{
    Comparison between SSDR and ManiPoP~\cite{tachellaBayesian3DReconstruction2019}: ManiPoP produces smoother reconstructions with lower error and regularizes signal flux estimates. However, it introduces distortions, evident in the duck's bill in the zoomed-in depth maps and depth underestimation at  boundary between the duck and the background pane.
    }
    \vspace{2em}
    \label{fig:reg-comparison}
\end{figure*}

%% file: figs/fig-A-benchreg2.tex
\begin{figure}[htb]
    \centering
    \begin{subfigure}{0.42\linewidth}
        \centering
        \caption*{\small SSDR}
        \includegraphics[width=\linewidth,trim={3cm 1.5cm 4cm 3.5cm},clip]{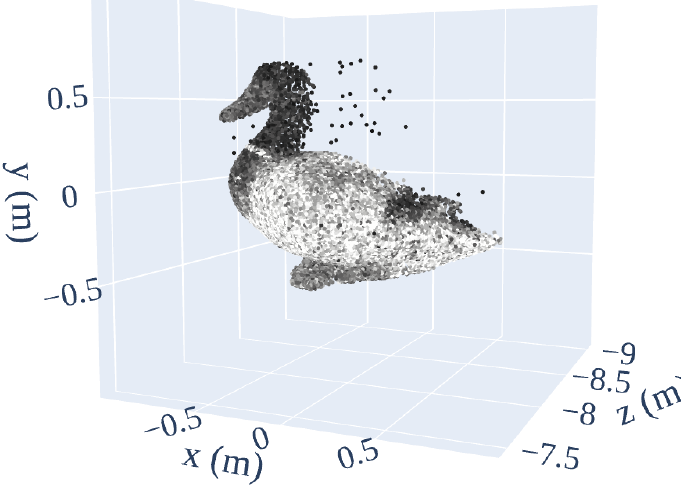}
    \end{subfigure}
    \begin{subfigure}{0.42\linewidth}
        \centering
        \caption*{\small ManiPoP}
        \includegraphics[width=\linewidth,trim={3cm 1.5cm 4cm 3.5cm},clip]{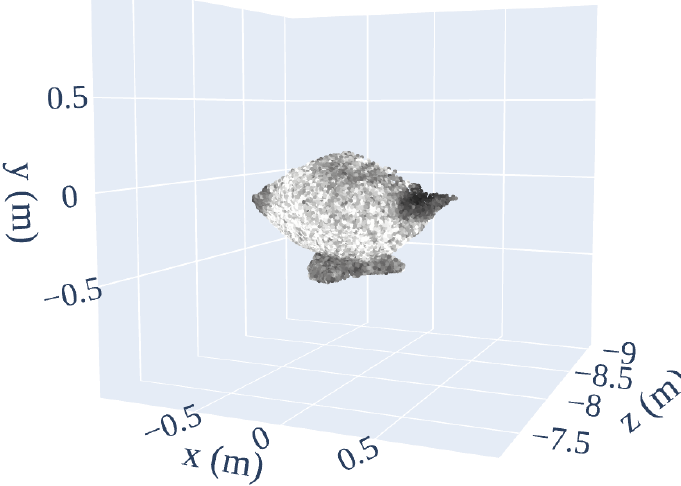}
    \end{subfigure}
    \begin{subfigure}{0.1\linewidth}
        \centering
        \includegraphics[height=2.7cm]{figs/regbench/cbar-s.pdf}
    \end{subfigure}
    \caption{Comparison between SSDR and ManiPoP~\cite{tachellaBayesian3DReconstruction2019} without a back pane:
    ManiPoP can introduce large distortions. In this case, it discards regions with low signal flux and warping the shape of the remaining structure.
    }
    \label{fig:manipop2}
\end{figure}

%% file: A6_lab.tex
\section{Additional Experimental Details}
\label{sec:app-lab}

\subsection{Experimental Setup}

\input{figs/fig-A-lab-setup}

The setup includes a pulsed laser (PicoQuant LDH-P-C-640-B) operating at \SI{10}{\mega\hertz}, resulting in \SI{100}{\nano\second} repetition period, and \SI{231}{\pico\second} pulse width. The laser is raster-scanned over a $64 \times 64$ grid using a Thorlabs GVS 202 galvo system. A FastGatedSPAD (Micro Photon Devices) detects photons at an avalanche threshold of \SI{16}{\milli\volt} and bias current of \SI{50}{\milli\ampere}. 

In synchronous mode, the SPAD gate remains open for $\ton = \SI{60}{\nano\second}$ in each repetition period, with an \SI{81}{\nano\second} hold-off time ensuring at most one detection per period. This slightly modifies the detection model, treating $\ton$ as the effective period in ML estimation while using the full \SI{100}{\nano\second} repetition period to count active repetitions $\nrp$. 
In free-running mode, the dead time is \SI{48}{\nano\second}. The SPAD output is recorded by a Swabian Instruments Time Tagger Ultra with a \SI{4}{\pico\second} bin width over a \SI{10}{\milli\second} acquisition time per pixel. Two objects at approximately \SI{2.4}{\meter} depth are measured with tunable ambient light levels using two LED lamps.
For the low-flux setting, both LED lamps and the room light are turned off.
For high-flux synchronous and free-running acquisition, both LED lamps are turned on, although the room light is still off.

\subsection{Additional Experimental Results}

\input{figs/fig-A-lab-results}

\Cref{fig:lab-supp} shows additional 3D reconstructions from experiments.
We introduce another scene and include background flux estimates, along with statistics such as the average background flux estimate and the MAE of depth estimates. Reconstructions from low-flux measurements serve as the reference for computing signal flux and depth errors.

Consistent with simulation trends, free-running measurements yield more accurate signal flux and depth estimates. In contrast, synchronous mode fails to reconstruct the scene, as pile-up causes depth underestimation, resulting in a point cloud near the detector. 
For the ``Mannequin'' scene, SSDR reduces MAE of depth estimates by 28\%.
Background estimates for high-flux measurements remain uniform across pixels, as expected, while in low-flux measurements, they correlate with signal flux estimates due to laser multiple scattering.
In the ``Dog'' scene, low reflectivity in dark regions makes estimation difficult, even in low-flux conditions. 
Both synchronous and free-running high-flux estimates remain unreliable in those regions.
For low-flux measurements, some estimated points on the Dog's dark regions appear on the cardboard, because the laser spot has a finite size and it partially illuminates the more reflective cardboard.

%% file: figs/fig-A-lab-setup.tex
\begin{figure}[htb]
    \centering
    \includegraphics[width=0.9\linewidth,trim={0 0 19.5cm 0},clip]{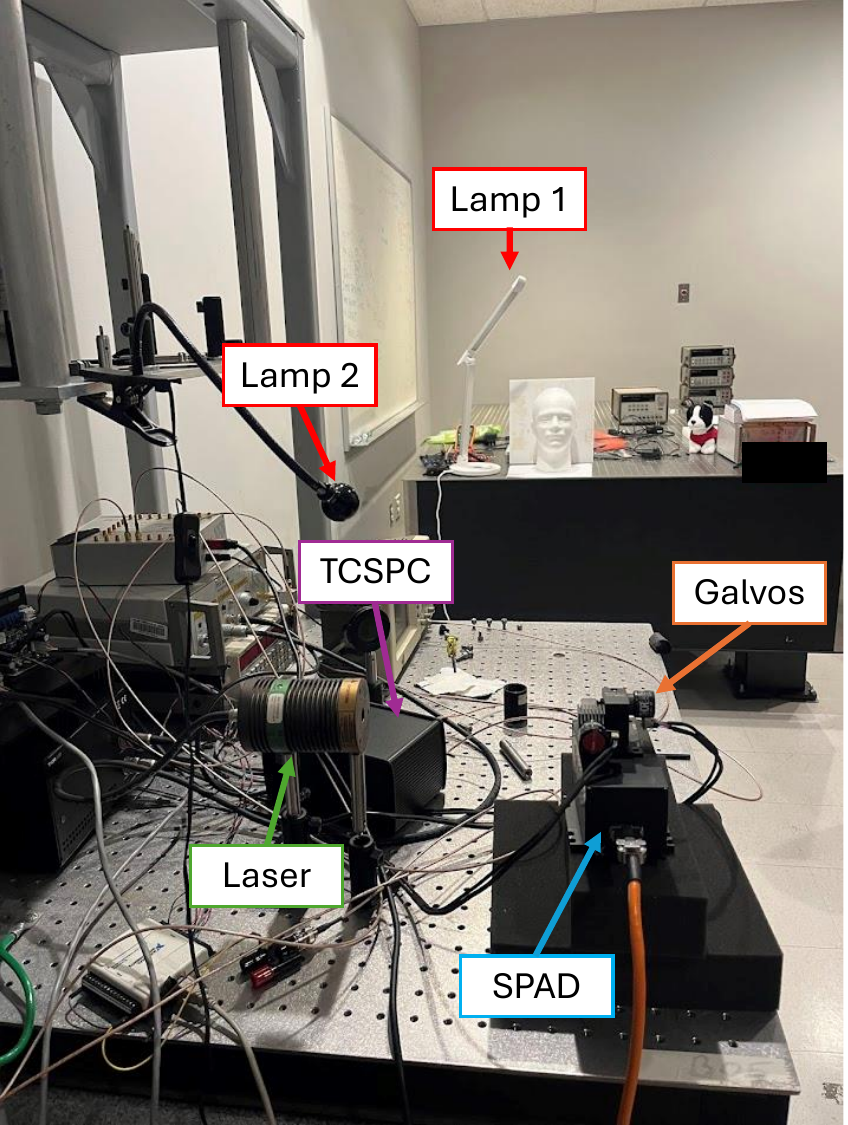}
    \caption{
    The experimental setup for acquiring SPL measurements.
    All lights are turned off for the low-flux setting.
    The two lamps are turned on while the room light is off for the synchronous and free-running high-flux settings.
    }
    \label{fig:lab-setup}
\end{figure}

%% file: figs/fig-A-lab-results.tex
\def\cbarHeight{2.3cm}

\begin{figure*}[htb]

    \centering
    \begin{tabular}[t]{@{}c@{\,}c@{\,}c@{\,}c@{\,}c@{\,}c@{\,}c@{\,}c@{\,}c@{}}
    (a) Photo & (b) Low-flux && (c) Sync && (d) Free && (e) Free + SSDR &
    
    \\
    
    \begin{subfigure}{0.12\linewidth}
        \centering
        \includegraphics[width=\linewidth]{figs/lab/head.jpg}
        \caption*{Mannequin}
    \end{subfigure}
    &
    \begin{subfigure}{0.16\linewidth}
        \centering
        \includegraphics[width=\linewidth,trim={3.5cm 1cm 4.4cm 3cm},clip]{figs/lab/head-lf/pointcloudOOB.pdf}
    \end{subfigure}
    &
    \begin{subfigure}{0.05\linewidth}
        \centering
        \includegraphics[height=\cbarHeight]{figs/lab/head-lf/cbar.pdf}
        \vspace{0.5em}
    \end{subfigure}
    &
    \begin{subfigure}{0.16\linewidth}
        \centering
        \includegraphics[width=\linewidth,trim={2.2cm 1cm 5.5cm 2cm},clip]{figs/lab/head-sc/pointcloudOOB.pdf}
    \end{subfigure}
    &
    \begin{subfigure}{0.05\linewidth}
        \centering
        \includegraphics[height=\cbarHeight]{figs/lab/head-lf/cbar.pdf}
        \vspace{0.5em}
    \end{subfigure}
    &
    \begin{subfigure}{0.16\linewidth}
        \centering
        \includegraphics[width=\linewidth,trim={2.2cm 1cm 5.5cm 2cm},clip]{figs/lab/head-fr/pointcloudOOB.pdf}
    \end{subfigure}
    &
    \begin{subfigure}{0.05\linewidth}
        \centering
        \includegraphics[height=\cbarHeight]{figs/lab/head-lf/cbar.pdf}
        \vspace{0.5em}
    \end{subfigure}
    &
    \begin{subfigure}{0.16\linewidth}
        \centering
        \includegraphics[width=\linewidth,trim={2.2cm 1cm 5.5cm 2cm},clip]{figs/lab/head-fr/pointcloudOOB-diff.pdf}
    \end{subfigure}
    &
    \begin{subfigure}{0.05\linewidth}
        \centering
        \includegraphics[height=\cbarHeight]{figs/lab/head-lf/cbar.pdf}
        \vspace{0.5em}
    \end{subfigure}
    \\
    &
    \begin{subfigure}{0.16\linewidth}
        \centering
        \includegraphics[width=\linewidth,trim={3.5cm 1cm 4.4cm 3cm},clip]{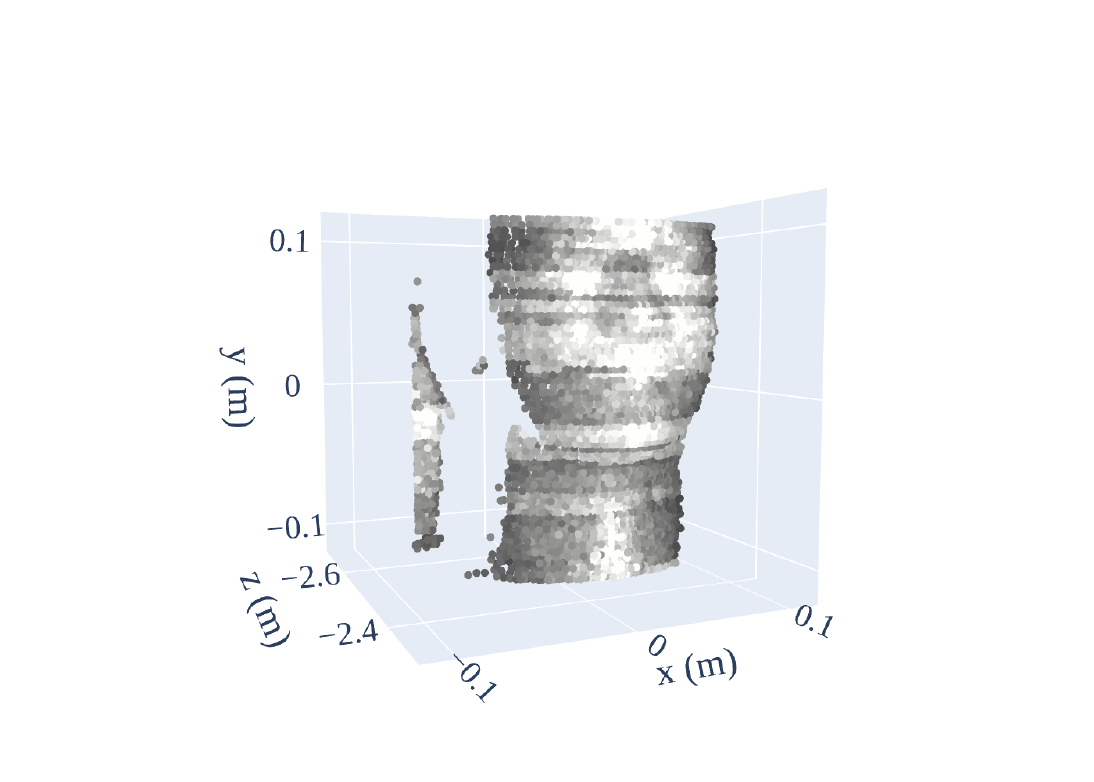}
    \end{subfigure}
    &
    \begin{subfigure}{0.05\linewidth}
        \centering
        \includegraphics[height=\cbarHeight]{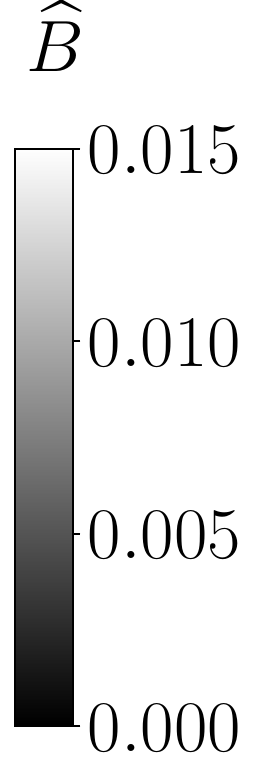}
        \vspace{0.5em}
    \end{subfigure}
    &
    \begin{subfigure}{0.16\linewidth}
        \centering
        \includegraphics[width=\linewidth,trim={2.2cm 1cm 5.5cm 2cm},clip]{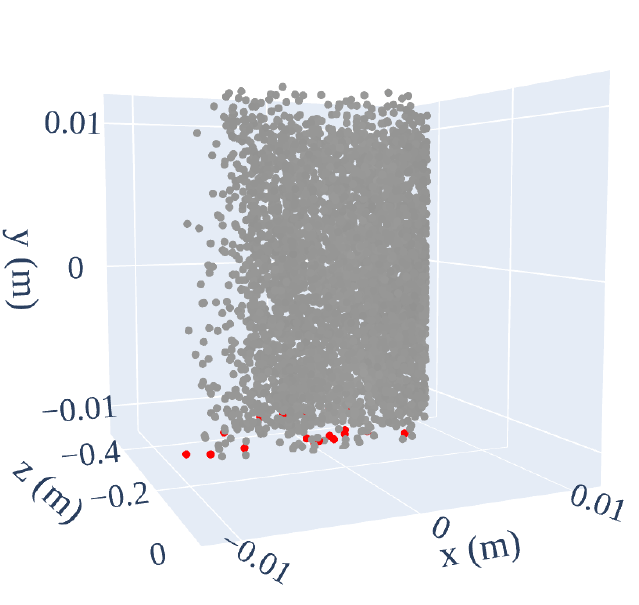}
    \end{subfigure}
    &
    \begin{subfigure}{0.05\linewidth}
        \centering
        \includegraphics[height=\cbarHeight]{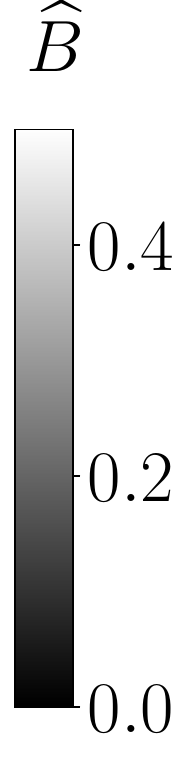}
        \vspace{0.5em}
    \end{subfigure}
    &
    \begin{subfigure}{0.16\linewidth}
        \centering
        \includegraphics[width=\linewidth,trim={2.2cm 1cm 5.5cm 2cm},clip]{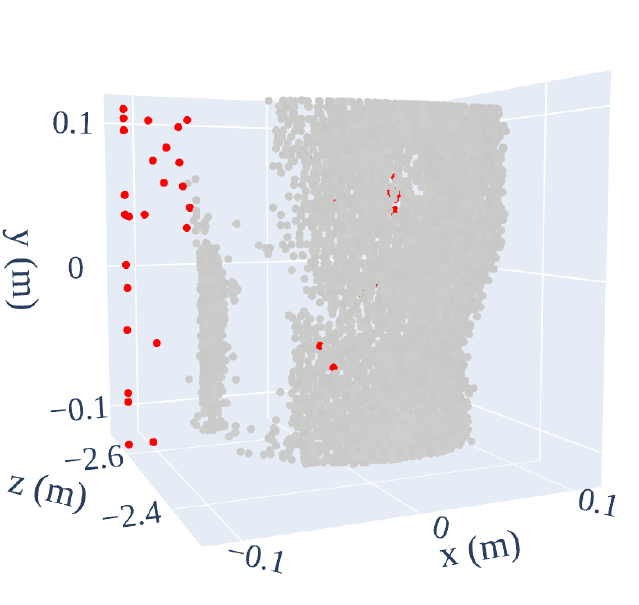}
    \end{subfigure}
    &
    \begin{subfigure}{0.05\linewidth}
        \centering
        \includegraphics[height=\cbarHeight]{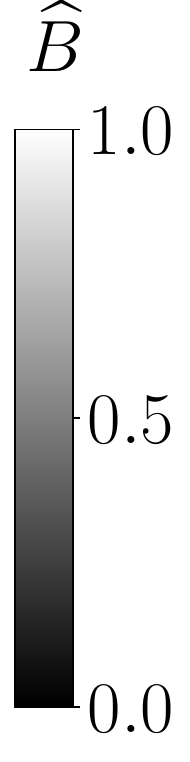}
        \vspace{0.5em}
    \end{subfigure}
    &
    \begin{subfigure}{0.16\linewidth}
        \centering
        \includegraphics[width=\linewidth,trim={2.2cm 1cm 5.5cm 2cm},clip]{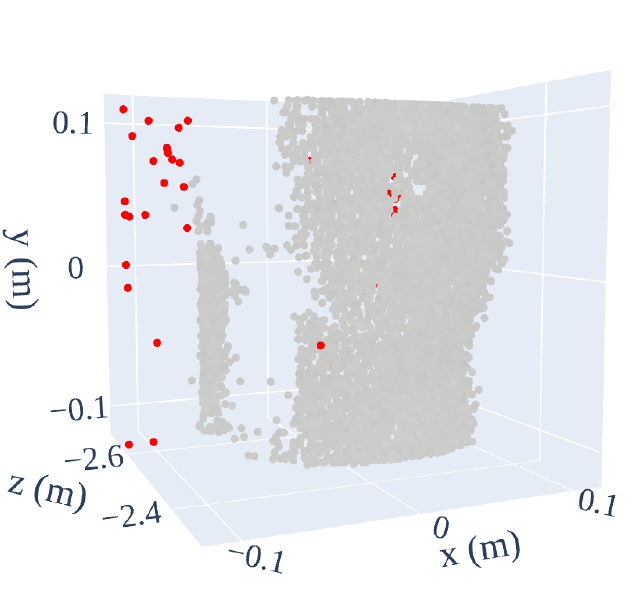}
    \end{subfigure}
    &
    \begin{subfigure}{0.05\linewidth}
        \centering
        \includegraphics[height=\cbarHeight]{figs/lab/head-fr/cbar-bg-free.pdf}
        \vspace{0.5em}
    \end{subfigure}
    \\
    &
    &&
    \begin{tabular}{@{}|@{\,}c@{\,}|@{\,}c@{\,}|@{}}
    \hline
    
    \scriptsize{$\rmse(\what{S})$} & \scriptsize{$\rmse(\what{z})$} \\
    \scriptsize{\SI[tight-spacing=true]{9.2e-03}{}} & \scriptsize{\SI{2.143}{\meter}} \\
    \hline
    
    \scriptsize{$\text{Avg. } \what{B}$} & \scriptsize{$\mae(\what{z})$} \\
    \scriptsize{\SI[tight-spacing=true]{0.326}{}} & \scriptsize{\SI{2.137}{\meter}} \\
    \hline
    \end{tabular}
    &&
    \begin{tabular}{@{}|@{\,}c@{\,}|@{\,}c@{\,}|@{}}
    \hline
    
    \scriptsize{$\rmse(\what{S})$} & \scriptsize{$\rmse(\what{z})$} \\
    \scriptsize{\SI[tight-spacing=true]{7.9e-04}{}} & \scriptsize{\SI{1.046}{\meter}} \\
    \hline
    
    \scriptsize{$\text{Avg. } \what{B}$} & \scriptsize{$\mae(\what{z})$} \\
    \scriptsize{\SI[tight-spacing=true]{0.835}{}} & \scriptsize{\SI{0.136}{\meter}} \\
    \hline
    \end{tabular}
    &&
    \begin{tabular}{@{}|@{\,}c@{\,}|@{\,}c@{\,}|@{}}
    \hline
    
    \scriptsize{$\rmse(\what{S})$} & \scriptsize{$\rmse(\what{z})$} \\
    \scriptsize{\SI[tight-spacing=true]{7.9e-04}{}} & \scriptsize{\SI{0.873}{\meter}} \\
    \hline
    
    \scriptsize{$\text{Avg. } \what{B}$} & \scriptsize{$\mae(\what{z})$} \\
    \scriptsize{\SI[tight-spacing=true]{0.835}{}} & \scriptsize{\SI{0.098}{\meter}} \\
    \hline
    \end{tabular}

    \\

    \begin{subfigure}{0.12\linewidth}
        \centering
        \includegraphics[width=\linewidth]{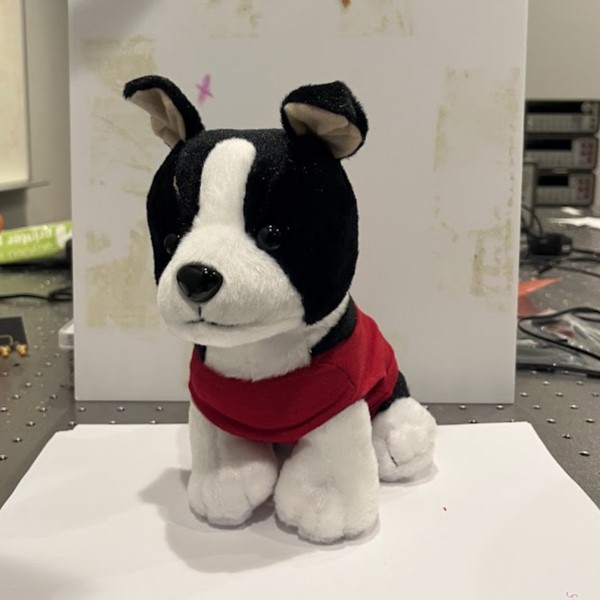}
        \caption*{Dog}
    \end{subfigure}
    &
    \begin{subfigure}{0.16\linewidth}
        \centering
        \includegraphics[width=\linewidth,trim={3.5cm 1cm 4.4cm 3cm},clip]{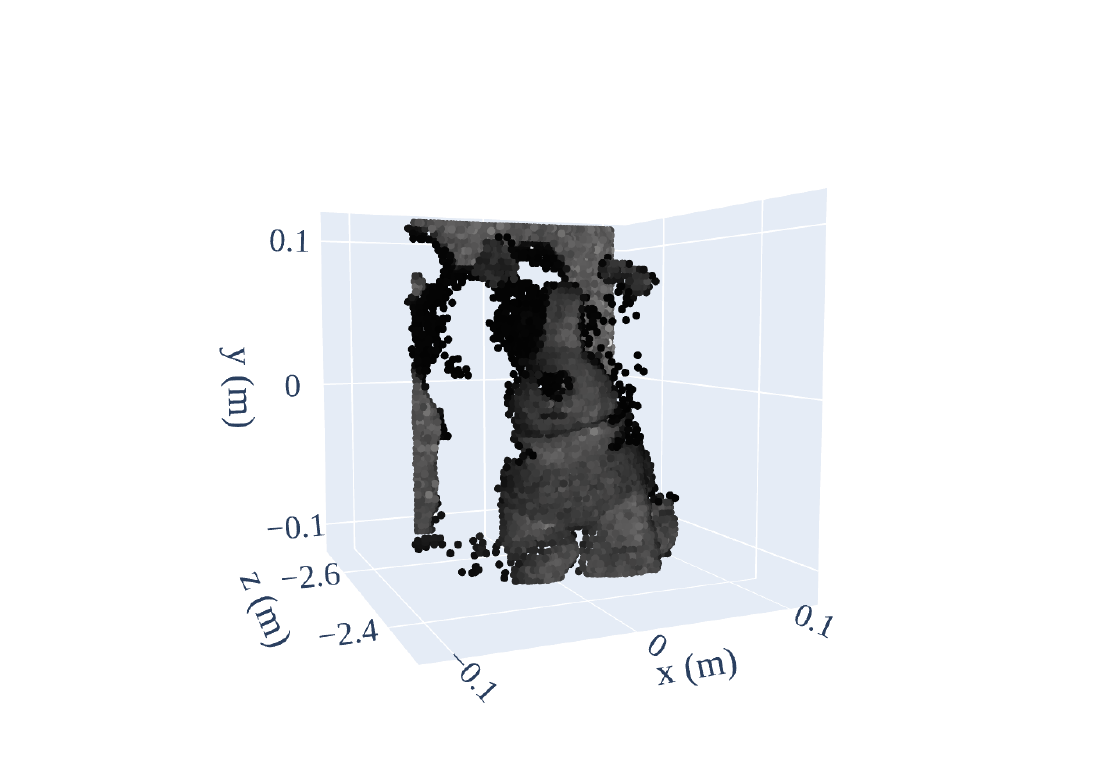}
    \end{subfigure}
    &
    \begin{subfigure}{0.05\linewidth}
        \centering
        \includegraphics[height=\cbarHeight]{figs/lab/head-lf/cbar.pdf}
        \vspace{0.5em}
    \end{subfigure}
    &
    \begin{subfigure}{0.16\linewidth}
        \centering
        \includegraphics[width=\linewidth,trim={2.2cm 1cm 5.5cm 2cm},clip]{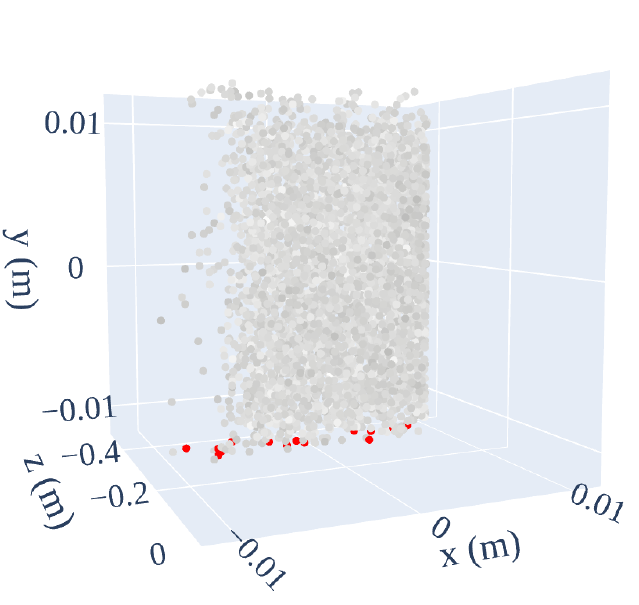}
    \end{subfigure}
    &
    \begin{subfigure}{0.05\linewidth}
        \centering
        \includegraphics[height=\cbarHeight]{figs/lab/head-lf/cbar.pdf}
        \vspace{0.5em}
    \end{subfigure}
    &
    \begin{subfigure}{0.16\linewidth}
        \centering
        \includegraphics[width=\linewidth,trim={2.2cm 1cm 5.5cm 2cm},clip]{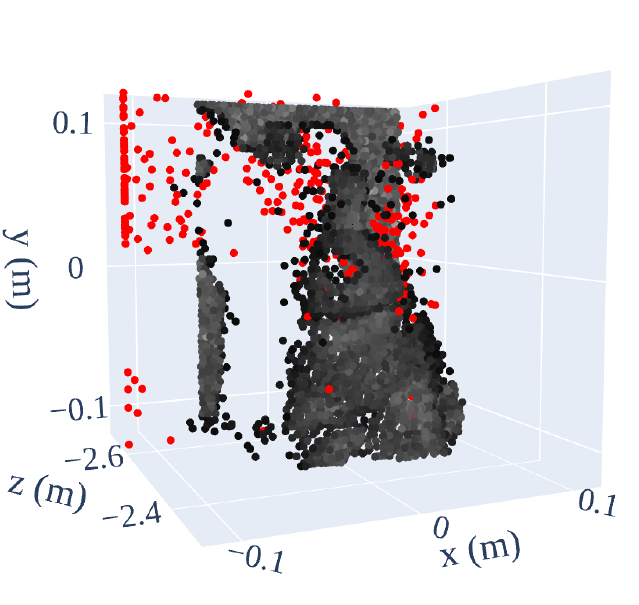}
    \end{subfigure}
    &
    \begin{subfigure}{0.05\linewidth}
        \centering
        \includegraphics[height=\cbarHeight]{figs/lab/head-lf/cbar.pdf}
        \vspace{0.5em}
    \end{subfigure}
    &
    \begin{subfigure}{0.16\linewidth}
        \centering
        \includegraphics[width=\linewidth,trim={2.2cm 1cm 5.5cm 2cm},clip]{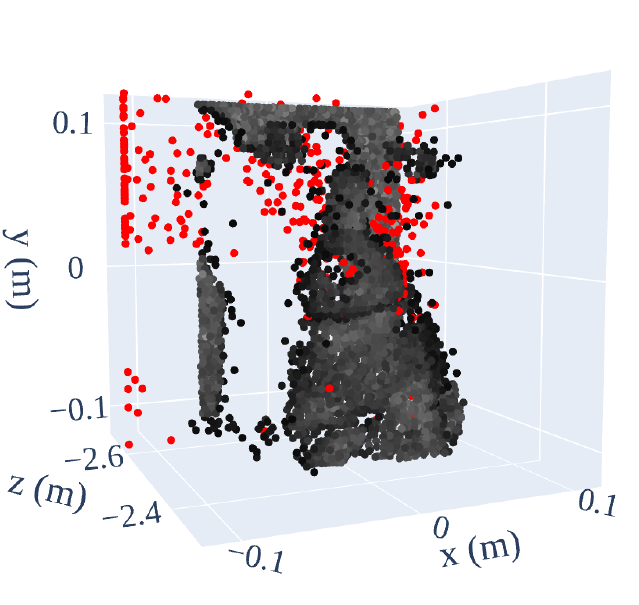}
    \end{subfigure}
    &
    \begin{subfigure}{0.05\linewidth}
        \centering
        \includegraphics[height=\cbarHeight]{figs/lab/head-lf/cbar.pdf}
        \vspace{0.5em}
    \end{subfigure}
    \vspace{-2mm}
    \\
    &
    \begin{subfigure}{0.16\linewidth}
        \centering
        \includegraphics[width=\linewidth,trim={3.5cm 1cm 4.4cm 3cm},clip]{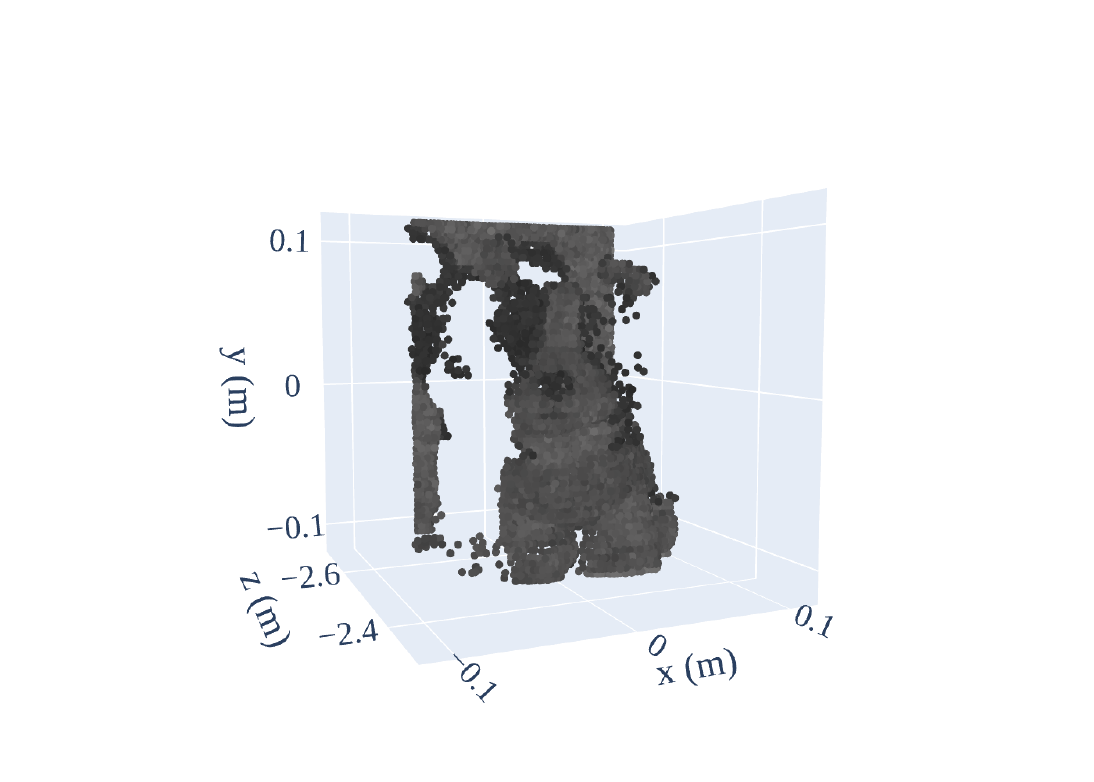}
    \end{subfigure}
    &
    \begin{subfigure}{0.05\linewidth}
        \centering
        \includegraphics[height=\cbarHeight]{figs/lab/head-lf/cbar-bg.pdf}
        \vspace{0.5em}
    \end{subfigure}
    &
    \begin{subfigure}{0.16\linewidth}
        \centering
        \includegraphics[width=\linewidth,trim={2.2cm 1cm 5.5cm 2cm},clip]{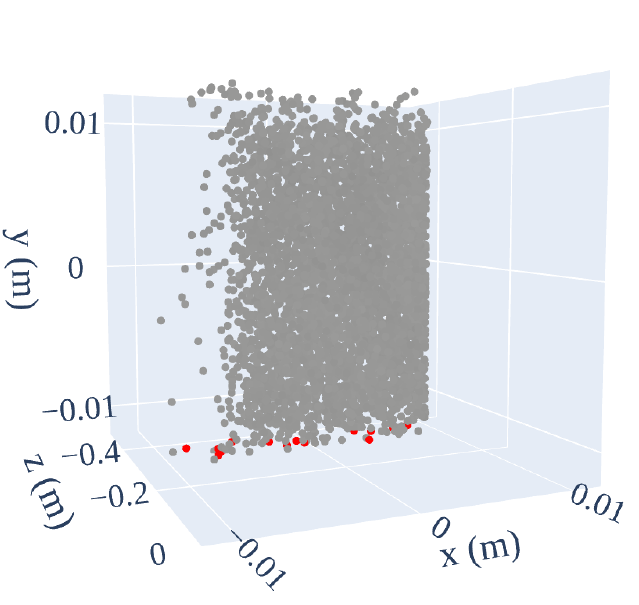}
    \end{subfigure}
    &
    \begin{subfigure}{0.05\linewidth}
        \centering
        \includegraphics[height=\cbarHeight]{figs/lab/head-sc/cbar-bg-sync.pdf}
        \vspace{0.5em}
    \end{subfigure}
    &
    \begin{subfigure}{0.16\linewidth}
        \centering
        \includegraphics[width=\linewidth,trim={2.2cm 1cm 5.5cm 2cm},clip]{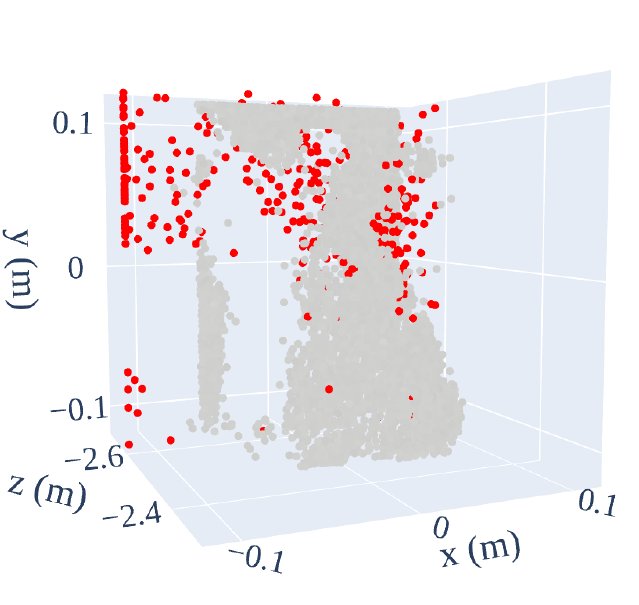}
    \end{subfigure}
    &
    \begin{subfigure}{0.05\linewidth}
        \centering
        \includegraphics[height=\cbarHeight]{figs/lab/head-fr/cbar-bg-free.pdf}
        \vspace{0.5em}
    \end{subfigure}
    &
    \begin{subfigure}{0.16\linewidth}
        \centering
        \includegraphics[width=\linewidth,trim={2.2cm 1cm 5.5cm 2cm},clip]{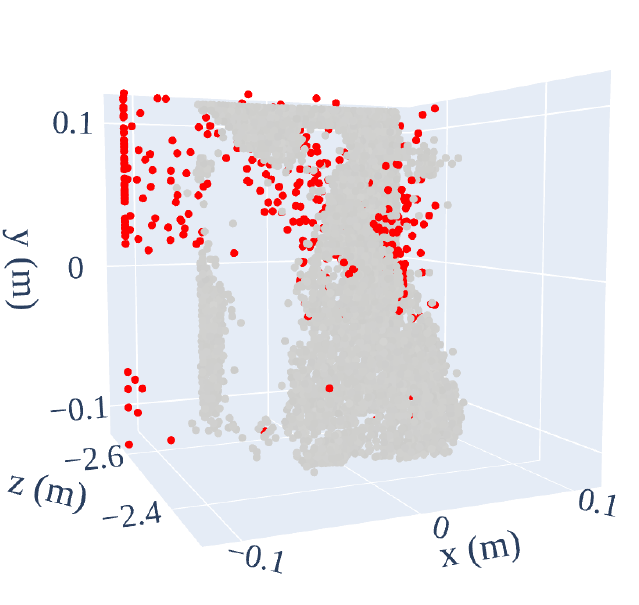}
    \end{subfigure}
    &
    \begin{subfigure}{0.05\linewidth}
        \centering
        \includegraphics[height=\cbarHeight]{figs/lab/head-fr/cbar-bg-free.pdf}
        \vspace{0.5em}
    \end{subfigure}
    \\
    &
    &&
    \begin{tabular}{@{}|@{\,}c@{\,}|@{\,}c@{\,}|@{}}
    \hline
    
    \scriptsize{$\rmse(\what{S})$} & \scriptsize{$\rmse(\what{z})$} \\
    \scriptsize{\SI[tight-spacing=true]{1.0e-02}{}} & \scriptsize{\SI{2.256}{\meter}} \\
    \hline
    
    \scriptsize{$\text{Avg. } \what{B}$} & \scriptsize{$\mae(\what{z})$} \\
    \scriptsize{\SI[tight-spacing=true]{0.327}{}} & \scriptsize{\SI{2.253}{\meter}} \\
    \hline
    \end{tabular}
    &&
    \begin{tabular}{@{}|@{\,}c@{\,}|@{\,}c@{\,}|@{}}
    \hline
    
    \scriptsize{$\rmse(\what{S})$} & \scriptsize{$\rmse(\what{z})$} \\
    \scriptsize{\SI[tight-spacing=true]{8.2e-04}{}} & \scriptsize{\SI{2.820}{\meter}} \\
    \hline
    
    \scriptsize{$\text{Avg. } \what{B}$} & \scriptsize{$\mae(\what{z})$} \\
    \scriptsize{\SI[tight-spacing=true]{0.852}{}} & \scriptsize{\SI{1.005}{\meter}} \\
    \hline
    \end{tabular}
    &&
    \begin{tabular}{@{}|@{\,}c@{\,}|@{\,}c@{\,}|@{}}
    \hline
    
    \scriptsize{$\rmse(\what{S})$} & \scriptsize{$\rmse(\what{z})$} \\
    \scriptsize{\SI[tight-spacing=true]{8.2e-04}{}} & \scriptsize{\SI{2.794}{\meter}} \\
    \hline
    
    \scriptsize{$\text{Avg. } \what{B}$} & \scriptsize{$\mae(\what{z})$} \\
    \scriptsize{\SI[tight-spacing=true]{0.852}{}} & \scriptsize{\SI{0.993}{\meter}} \\
    \hline
    \end{tabular}

    \end{tabular}

    \caption{
    Additional 3D reconstruction results from experimentally collected SPL measurements.  
    We include an additional scene and provide background flux estimates for each pixel, along with extra statistics such as the average background flux estimate and the MAE of depth estimates. Low-flux estimates serve as the ground truth for error computation.
    }
    \label{fig:lab-supp}
\end{figure*}

%% file: main.bbl
\begin{thebibliography}{58}
\providecommand{\natexlab}[1]{#1}
\providecommand{\url}[1]{\texttt{#1}}
\expandafter\ifx\csname urlstyle\endcsname\relax
  \providecommand{\doi}[1]{doi: #1}\else
  \providecommand{\doi}{doi: \begingroup \urlstyle{rm}\Url}\fi

\bibitem[Acconcia et~al.(2018)Acconcia, Cominelli, Ghioni, and
  Rech]{acconcia2018fast}
Giulia Acconcia, Alessandro Cominelli, Massimo Ghioni, and Ivan Rech.
\newblock Fast fully-integrated front-end circuit to overcome pile-up limits in
  time-correlated single photon counting with single photon avalanche diodes.
\newblock \emph{Optics Express}, 26\penalty0 (12):\penalty0 15398--15410, 2018.

\bibitem[Akiba et~al.(2019)Akiba, Sano, Yanase, Ohta, and Koyama]{optuna_2019}
Takuya Akiba, Shotaro Sano, Toshihiko Yanase, Takeru Ohta, and Masanori Koyama.
\newblock Optuna: A next-generation hyperparameter optimization framework.
\newblock In \emph{Proceedings of the 25th {ACM} {SIGKDD} International
  Conference on Knowledge Discovery and Data Mining}, 2019.

\bibitem[Altmann et~al.(2016)Altmann, Ren, McCarthy, Buller, and
  McLaughlin]{altmannRobustBayesianTarget2016a}
Yoann Altmann, Ximing Ren, Aongus McCarthy, Gerald~S. Buller, and Steve
  McLaughlin.
\newblock Robust {Bayesian} target detection algorithm for depth imaging from
  sparse single-photon data.
\newblock \emph{IEEE Transactions on Computational Imaging}, 2\penalty0
  (4):\penalty0 456--467, 2016.

\bibitem[Altmann et~al.(2020)Altmann, McLaughlin, and
  Davies]{altmannFastOnline3D2020}
Yoann Altmann, Stephen McLaughlin, and Michael~E. Davies.
\newblock Fast online {3D} reconstruction of dynamic scenes from individual
  single-photon detection events.
\newblock \emph{IEEE Transactions on Image Processing}, 29:\penalty0
  2666--2675, 2020.

\bibitem[Antolovic et~al.(2015)Antolovic, Burri, Bruschini, Hoebe, and
  Charbon]{antolovic2015nonuniformity}
Ivan~Michel Antolovic, Samuel Burri, Claudio Bruschini, Ron Hoebe, and Edoardo
  Charbon.
\newblock Nonuniformity analysis of a 65-kpixel {CMOS SPAD} imager.
\newblock \emph{IEEE Transactions on Electron Devices}, 63\penalty0
  (1):\penalty0 57--64, 2015.

\bibitem[Antolovic et~al.(2018)Antolovic, Bruschini, and
  Charbon]{antolovic2018dynamic}
Ivan~Michel Antolovic, Claudio Bruschini, and Edoardo Charbon.
\newblock Dynamic range extension for photon counting arrays.
\newblock \emph{Optics Express}, 26\penalty0 (17):\penalty0 22234--22248, 2018.

\bibitem[{Bar-David}(1969)]{bar-davidCommunicationPoissonRegime1969}
Israel {Bar-David}.
\newblock Communication under the {{Poisson}} regime.
\newblock \emph{IEEE Transactions on Information Theory}, 15\penalty0
  (1):\penalty0 31--37, 1969.

\bibitem[Beer et~al.(2018)Beer, Haase, Ruskowski, and
  Kokozinski]{beer2018background}
Maik Beer, Jan~F Haase, Jennifer Ruskowski, and Rainer Kokozinski.
\newblock Background light rejection in {SPAD}-based {LiDAR} sensors by
  adaptive photon coincidence detection.
\newblock \emph{Sensors}, 18\penalty0 (12):\penalty0 4338, 2018.

\bibitem[Bergstra et~al.(2011)Bergstra, Bardenet, Bengio, and
  K{\'e}gl]{bergstra2011algorithms}
James Bergstra, R{\'e}mi Bardenet, Yoshua Bengio, and Bal{\'a}zs K{\'e}gl.
\newblock Algorithms for hyper-parameter optimization.
\newblock \emph{Adv. Neural Inform. Process. Syst.}, 24, 2011.

\bibitem[Chen et~al.(2019)Chen, Halimi, Ren, McCarthy, Su, McLaughlin, and
  Buller]{chen2019learning}
Songmao Chen, Abderrahim Halimi, Ximing Ren, Aongus McCarthy, Xiuqin Su,
  Stephen McLaughlin, and Gerald~S Buller.
\newblock Learning non-local spatial correlations to restore sparse {3D}
  single-photon data.
\newblock \emph{IEEE Transactions on Image Processing}, 29:\penalty0
  3119--3131, 2019.

\bibitem[Chung et~al.(2022)Chung, Sim, Ryu, and Ye]{chung2022improving}
Hyungjin Chung, Byeongsu Sim, Dohoon Ryu, and Jong~Chul Ye.
\newblock Improving diffusion models for inverse problems using manifold
  constraints.
\newblock \emph{Adv. Neural Inform. Process. Syst.}, 35:\penalty0 25683--25696,
  2022.

\bibitem[Chung et~al.(2023)Chung, Kim, Mccann, Klasky, and
  Ye]{chungDiffusionPosteriorSampling2023}
Hyungjin Chung, Jeongsol Kim, Michael~Thompson Mccann, Marc~Louis Klasky, and
  Jong~Chul Ye.
\newblock Diffusion posterior sampling for general noisy inverse problems.
\newblock In \emph{Int. Conf. Learn. Represent.}, 2023.

\bibitem[Coates(1968)]{coatesCorrectionPhotonPileup1968}
P.~B. Coates.
\newblock The correction for photon `pile-up' in the measurement of radiative
  lifetimes.
\newblock \emph{Journal of Physics E: Scientific Instruments}, 1\penalty0
  (8):\penalty0 878--879, 1968.

\bibitem[Cova et~al.(2013)Cova, Ghioni, Itzler, Bienfang, and
  Restelli]{cova2013semiconductor}
Sergio Cova, Massimo Ghioni, Mark~A Itzler, Joshua~C Bienfang, and Alessandro
  Restelli.
\newblock Semiconductor-{Based} {Detectors}.
\newblock In \emph{Single-{Photon} {Generation} and {Detection}}, pages
  83--146. Academic Press, 2013.

\bibitem[Gupta et~al.(2019{\natexlab{a}})Gupta, Ingle, and
  Gupta]{guptaAsynchronousSinglePhoton3D2019}
Anant Gupta, Atul Ingle, and Mohit Gupta.
\newblock Asynchronous single-photon {3D} imaging.
\newblock In \emph{Int. Conf. Comput. Vis.}, pages 7909--7918,
  2019{\natexlab{a}}.

\bibitem[Gupta et~al.(2019{\natexlab{b}})Gupta, Ingle, Velten, and
  Gupta]{guptaPhotonFloodedSinglePhoton3D2019}
Anant Gupta, Atul Ingle, Andreas Velten, and Mohit Gupta.
\newblock Photon-flooded single-photon {3D} cameras.
\newblock In \emph{IEEE Conf. Comput. Vis. Pattern Recog.}, pages 6770--6779,
  2019{\natexlab{b}}.

\bibitem[Halimi et~al.(2017)Halimi, Tobin, McCarthy, McLaughlin, and
  Buller]{halimiRestorationMultilayeredSinglephoton2017}
Abderrahim Halimi, Rachael Tobin, Aongus McCarthy, Stephen McLaughlin, and
  Gerald~S. Buller.
\newblock Restoration of multilayered single-photon {3D} lidar images.
\newblock In \emph{{{European Signal Processing Conference}} ({{EUSIPCO}})},
  pages 708--712, 2017.

\bibitem[Halimi et~al.(2021)Halimi, Maccarone, Lamb, Buller, and
  McLaughlin]{halimi2021robust}
Abderrahim Halimi, Aurora Maccarone, Robert~A Lamb, Gerald~S Buller, and
  Stephen McLaughlin.
\newblock Robust and guided bayesian reconstruction of single-photon {3D} lidar
  data: Application to multispectral and underwater imaging.
\newblock \emph{IEEE Transactions on Computational Imaging}, 7:\penalty0
  961--974, 2021.

\bibitem[Heide et~al.(2018)Heide, Diamond, Lindell, and
  Wetzstein]{heideSubpicosecondPhotonefficient3D2018}
Felix Heide, Steven Diamond, David~B. Lindell, and Gordon Wetzstein.
\newblock Sub-picosecond photon-efficient {{3D}} imaging using single-photon
  sensors.
\newblock \emph{Scientific Reports}, 8\penalty0 (1):\penalty0 17726, 2018.

\bibitem[Heide et~al.(2019)Heide, O’Toole, Zang, Lindell, Diamond, and
  Wetzstein]{heide2019non}
Felix Heide, Matthew O’Toole, Kai Zang, David~B Lindell, Steven Diamond, and
  Gordon Wetzstein.
\newblock Non-line-of-sight imaging with partial occluders and surface normals.
\newblock \emph{ACM Trans. Graph.}, 38\penalty0 (3):\penalty0 1--10, 2019.

\bibitem[Huang et~al.(2023)Huang, Chen, Wei, and Chen]{huang2023non}
Duolan Huang, Quan Chen, Zhun Wei, and Rui Chen.
\newblock Non-line-of-sight reconstruction via structure sparsity
  regularization.
\newblock \emph{Optics Letters}, 48\penalty0 (18):\penalty0 4881--4884, 2023.

\bibitem[Ingle et~al.(2019)Ingle, Velten, and Gupta]{ingleHighFluxPassive2019}
Atul Ingle, Andreas Velten, and Mohit Gupta.
\newblock High flux passive imaging with single-photon sensors.
\newblock In \emph{IEEE Conf. Comput. Vis. Pattern Recog.}, pages 6753--6762.
  IEEE Computer Society, 2019.

\bibitem[Ingle et~al.(2021)Ingle, Seets, Buttafava, Gupta, Tosi, Gupta, and
  Velten]{a.inglePassiveInterPhotonImaging2021}
Atul Ingle, Trevor Seets, Mauro Buttafava, Shantanu Gupta, Alberto Tosi, Mohit
  Gupta, and Andreas Velten.
\newblock Passive inter-photon imaging.
\newblock In \emph{IEEE Conf. Comput. Vis. Pattern Recog.}, pages 8585--8595,
  2021.

\bibitem[Jalal et~al.(2021)Jalal, Arvinte, Daras, Price, Dimakis, and
  Tamir]{jalal2021robust}
Ajil Jalal, Marius Arvinte, Giannis Daras, Eric Price, Alexandros~G Dimakis,
  and Jon Tamir.
\newblock Robust compressed sensing mri with deep generative priors.
\newblock \emph{Adv. Neural Inform. Process. Syst.}, 34:\penalty0 14938--14954,
  2021.

\bibitem[Kawar et~al.(2022)Kawar, Elad, Ermon, and
  Song]{kawarDenoisingDiffusionRestoration2022}
Bahjat Kawar, Michael Elad, Stefano Ermon, and Jiaming Song.
\newblock Denoising diffusion restoration models.
\newblock In \emph{Adv. Neural Inform. Process. Syst.}, pages 23593--23606.
  Curran Associates, Inc., 2022.

\bibitem[Kitichotkul et~al.(2024)Kitichotkul, Rapp, and
  Goyal]{kitichotkulRoleDetectionTimes2023}
Ruangrawee Kitichotkul, Joshua Rapp, and Vivek~K Goyal.
\newblock The role of detection times in reflectivity estimation with
  single-photon lidar.
\newblock \emph{IEEE Journal of Selected Topics in Quantum Electronics},
  30\penalty0 (1):\penalty0 1--14, 2024.

\bibitem[Laumont et~al.(2022)Laumont, Bortoli, Almansa, Delon, Durmus, and
  Pereyra]{laumont2022bayesian}
R{\'e}mi Laumont, Valentin~De Bortoli, Andr{\'e}s Almansa, Julie Delon, Alain
  Durmus, and Marcelo Pereyra.
\newblock Bayesian imaging using plug \& play priors: when langevin meets
  tweedie.
\newblock \emph{SIAM Journal on Imaging Sciences}, 15\penalty0 (2):\penalty0
  701--737, 2022.

\bibitem[Legros et~al.(2021)Legros, Tachella, Tobin, Mccarthy, Meignen, Buller,
  Altmann, Mclaughlin, and Davies]{legrosRobust3DReconstruction2021}
Quentin Legros, Julian Tachella, Rachael Tobin, Aongus Mccarthy, Sylvain
  Meignen, Gerald~S. Buller, Yoann Altmann, Stephen Mclaughlin, and Michael~E.
  Davies.
\newblock Robust {3D} reconstruction of dynamic scenes from single-photon lidar
  using beta-divergences.
\newblock \emph{IEEE Transactions on Image Processing}, 30:\penalty0
  1716--1727, 2021.

\bibitem[Liu et~al.(2021)Liu, Wang, Li, Shi, Fu, and Qiu]{liu2021non}
Xintong Liu, Jianyu Wang, Zhupeng Li, Zuoqiang Shi, Xing Fu, and Lingyun Qiu.
\newblock Non-line-of-sight reconstruction with signal--object collaborative
  regularization.
\newblock \emph{Light: Science \& Applications}, 10\penalty0 (1):\penalty0 198,
  2021.

\bibitem[Luo and Hu(2021)]{luoScoreBasedPointCloud}
Shitong Luo and Wei Hu.
\newblock Score-based point cloud denoising.
\newblock In \emph{Int. Conf. Comput. Vis.}, pages 4563--4572, 2021.

\bibitem[Maccarone et~al.(2023)Maccarone, Drummond, McCarthy, Steinlehner,
  Tachella, Garcia, Pawlikowska, Lamb, Henderson, McLaughlin, Altmann, and
  Buller]{auroramaccaroneSubmergedSinglephotonLiDAR2023}
Aurora Maccarone, Kristofer Drummond, Aongus McCarthy, Ulrich~K. Steinlehner,
  Julian Tachella, Diego~Aguirre Garcia, Agata Pawlikowska, Robert~A. Lamb,
  Robert~K. Henderson, Stephen McLaughlin, Yoann Altmann, and Gerald~S. Buller.
\newblock Submerged single-photon {LiDAR} imaging sensor used for real-time
  {3D} scene reconstruction in scattering underwater environments.
\newblock \emph{Optics Express}, 31\penalty0 (10):\penalty0 16690--16708, 2023.

\bibitem[McCarthy et~al.(2009)McCarthy, Collins, Krichel, Fern{\'a}ndez,
  Wallace, and Buller]{mccarthy2009long}
Aongus McCarthy, Robert~J Collins, Nils~J Krichel, Ver{\'o}nica Fern{\'a}ndez,
  Andrew~M Wallace, and Gerald~S Buller.
\newblock Long-range time-of-flight scanning sensor based on high-speed
  time-correlated single-photon counting.
\newblock \emph{Applied Optics}, 48\penalty0 (32):\penalty0 6241--6251, 2009.

\bibitem[McCarthy et~al.(2013)McCarthy, Krichel, Gemmell, Ren, Tanner,
  Dorenbos, Zwiller, Hadfield, and Buller]{mccarthy2013kilometer}
Aongus McCarthy, Nils~J Krichel, Nathan~R Gemmell, Ximing Ren, Michael~G
  Tanner, Sander~N Dorenbos, Val Zwiller, Robert~H Hadfield, and Gerald~S
  Buller.
\newblock Kilometer-range, high resolution depth imaging via 1560 nm wavelength
  single-photon detection.
\newblock \emph{Optics Express}, 21\penalty0 (7):\penalty0 8904--8915, 2013.

\bibitem[Nouri et~al.(2017)Nouri, Charrier, and
  L{\'e}zoray]{nouri2017technical}
Anass Nouri, Christophe Charrier, and Olivier L{\'e}zoray.
\newblock Technical report: Greyc {3D} colored mesh database.
\newblock Technical report, Normandie Universit{\'e}, Unicaen, EnsiCaen, CNRS,
  GREYC UMR 6072, 2017.

\bibitem[O'Connor and Phillips(1984)]{oconnorTimecorrelatedSinglePhoton1984}
Desmond O'Connor and David Phillips.
\newblock \emph{Time-Correlated {{Single Photon Counting}}}.
\newblock Elsevier Science, 1984.

\bibitem[O'Toole et~al.(2017)O'Toole, Heide, Lindell, Zang, Diamond, and
  Wetzstein]{otooleReconstructingTransientImages2017}
Matthew O'Toole, Felix Heide, David~B. Lindell, Kai Zang, Steven Diamond, and
  Gordon Wetzstein.
\newblock Reconstructing transient images from single-photon sensors.
\newblock In \emph{IEEE Conf. Comput. Vis. Pattern Recog.}, pages 2289--2297,
  2017.

\bibitem[Pawlikowska et~al.(2017)Pawlikowska, Halimi, Lamb, and
  Buller]{pawlikowskaSinglephotonThreedimensionalImaging2017}
Agata~M. Pawlikowska, Abderrahim Halimi, Robert~A. Lamb, and Gerald~S. Buller.
\newblock Single-photon three-dimensional imaging at up to 10 kilometers range.
\newblock \emph{Optics Express}, 25\penalty0 (10):\penalty0 11919--11931, 2017.

\bibitem[Pediredla et~al.(2018)Pediredla, Sankaranarayanan, Buttafava, Tosi,
  and Veeraraghavan]{pediredlaSignalProcessingBased2018}
Adithya~K. Pediredla, Aswin~C. Sankaranarayanan, Mauro Buttafava, Alberto Tosi,
  and Ashok Veeraraghavan.
\newblock Signal processing based pile-up compensation for gated single-photon
  avalanche diodes.
\newblock arXiv:1806.07437, 2018.

\bibitem[Pellegrini et~al.(2000)Pellegrini, Buller, Smith, Wallace, and
  Cova]{pellegrini2000laser}
Sara Pellegrini, Gerald~S Buller, Jason~M Smith, Andrew~M Wallace, and Sergio
  Cova.
\newblock Laser-based distance measurement using picosecond
  resolutiontime-correlated single-photon counting.
\newblock \emph{Measurement Science and Technology}, 11\penalty0 (6):\penalty0
  712, 2000.

\bibitem[Po et~al.(2022)Po, Pediredla, and Gkioulekas]{po2022adaptive}
Ryan Po, Adithya Pediredla, and Ioannis Gkioulekas.
\newblock Adaptive gating for single-photon {3D} imaging.
\newblock In \emph{IEEE Conf. Comput. Vis. Pattern Recog.}, pages 16354--16363,
  2022.

\bibitem[Rapp and Goyal(2017)]{rappFewPhotonsMany2017}
Joshua Rapp and Vivek~K Goyal.
\newblock A few photons among many: Unmixing signal and noise for
  photon-efficient active imaging.
\newblock \emph{IEEE Transactions on Computational Imaging}, 3\penalty0
  (3):\penalty0 445--459, 2017.

\bibitem[Rapp et~al.(2019)Rapp, Ma, Dawson, and
  Goyal]{rappDeadTimeCompensation2019}
Joshua Rapp, Yanting Ma, Robin M.~A. Dawson, and Vivek~K. Goyal.
\newblock Dead time compensation for high-flux ranging.
\newblock \emph{IEEE Transactions on Signal Processing}, 67\penalty0
  (13):\penalty0 3471--3486, 2019.

\bibitem[Rapp et~al.(2020{\natexlab{a}})Rapp, Ma, Dawson, and
  Goyal]{rappHighfluxSinglephotonLidar2020}
Joshua Rapp, Yanting Ma, Robin M.~A. Dawson, and Vivek~K Goyal.
\newblock High-flux single-photon lidar.
\newblock \emph{Optica}, 8\penalty0 (1):\penalty0 30--39, 2020{\natexlab{a}}.

\bibitem[Rapp et~al.(2020{\natexlab{b}})Rapp, Tachella, Altmann, McLaughlin,
  and Goyal]{rappAdvancesSinglePhotonLidar2020}
Joshua Rapp, Julian Tachella, Yoann Altmann, Stephen McLaughlin, and Vivek~K
  Goyal.
\newblock Advances in single-photon lidar for autonomous vehicles: Working
  principles, challenges, and recent advances.
\newblock \emph{IEEE Signal Processing Magazine}, 37\penalty0 (4):\penalty0
  62--71, 2020{\natexlab{b}}.

\bibitem[Shangguan et~al.(2023)Shangguan, Yang, Lin, Lee, Xia, and
  Weng]{shangguan2023compact}
Mingjia Shangguan, Zhifeng Yang, Zaifa Lin, Zhongping Lee, Haiyun Xia, and
  Zhenwu Weng.
\newblock Compact long-range single-photon underwater lidar with high
  spatial--temporal resolution.
\newblock \emph{IEEE Geoscience and Remote Sensing Letters}, 20:\penalty0 1--5,
  2023.

\bibitem[Shin et~al.(2016)Shin, Xu, Wong, Shapiro, and
  Goyal]{shinComputationalMultidepthSinglephoton2016}
Dongeek Shin, Feihu Xu, Franco N.~C. Wong, Jeffrey~H. Shapiro, and Vivek~K
  Goyal.
\newblock Computational multi-depth single-photon imaging.
\newblock \emph{Optics Express}, 24\penalty0 (3):\penalty0 1873--1888, 2016.

\bibitem[Snyder and Miller(2012)]{snyderRandomPointProcesses2012}
Donald~L. Snyder and Michael~I. Miller.
\newblock \emph{Random {{Point Processes}} in {{Time}} and {{Space}}}.
\newblock Springer Science \& Business Media, 2012.

\bibitem[Song et~al.(2022)Song, Vahdat, Mardani, and
  Kautz]{songPseudoinverseGuidedDiffusionModels2022}
Jiaming Song, Arash Vahdat, Morteza Mardani, and Jan Kautz.
\newblock Pseudoinverse-guided diffusion models for inverse problems.
\newblock In \emph{Int. Conf. Learn. Represent.}, 2022.

\bibitem[Sun et~al.(2024)Sun, Wu, Chen, Feng, and
  Bouman]{sunProvableProbabilisticImaging2023}
Yu Sun, Zihui Wu, Yifan Chen, Berthy~T. Feng, and Katherine~L. Bouman.
\newblock Provable probabilistic imaging using score-based generative priors.
\newblock \emph{IEEE Transactions on Computational Imaging}, 10:\penalty0
  1290--1305, 2024.

\bibitem[Swatantran et~al.(2016)Swatantran, Tang, Barrett, DeCola, and
  Dubayah]{swatantran2016rapid}
Anu Swatantran, Hao Tang, Terence Barrett, Phil DeCola, and Ralph Dubayah.
\newblock Rapid, high-resolution forest structure and terrain mapping over
  large areas using single photon lidar.
\newblock \emph{Scientific Reports}, 6\penalty0 (1):\penalty0 28277, 2016.

\bibitem[Tachella et~al.(2019{\natexlab{a}})Tachella, Altmann, Mellado,
  McCarthy, Tobin, Buller, Tourneret, and
  McLaughlin]{tachellaRealtime3DReconstruction2019}
Juli{\'a}n Tachella, Yoann Altmann, Nicolas Mellado, Aongus McCarthy, Rachael
  Tobin, Gerald~S. Buller, Jean-Yves Tourneret, and Stephen McLaughlin.
\newblock Real-time {{3D}} reconstruction from single-photon lidar data using
  plug-and-play point cloud denoisers.
\newblock \emph{Nature Communications}, 10\penalty0 (1):\penalty0 4984,
  2019{\natexlab{a}}.

\bibitem[Tachella et~al.(2019{\natexlab{b}})Tachella, Altmann, Ren, McCarthy,
  Buller, McLaughlin, and Tourneret]{tachellaBayesian3DReconstruction2019}
Juli{\'a}n Tachella, Yoann Altmann, Ximing Ren, Aongus McCarthy, Gerald~S.
  Buller, Stephen McLaughlin, and Jean-Yves Tourneret.
\newblock Bayesian {3D} reconstruction of complex scenes from single-photon
  lidar data.
\newblock \emph{SIAM Journal on Imaging Sciences}, 12\penalty0 (1):\penalty0
  521--550, 2019{\natexlab{b}}.

\bibitem[Tobin et~al.(2021)Tobin, Halimi, McCarthy, Soan, and
  Buller]{tobinRobustRealtime3D2021}
Rachael Tobin, Abderrahim Halimi, Aongus McCarthy, Philip~J. Soan, and
  Gerald~S. Buller.
\newblock Robust real-time {{3D}} imaging of moving scenes through atmospheric
  obscurant using single-photon {{LiDAR}}.
\newblock \emph{Scientific Reports}, 11\penalty0 (1):\penalty0 11236, 2021.

\bibitem[Wei et~al.(2023)Wei, Nousias, Gulve, Lindell, and
  Kutulakos]{weiPassiveUltraWidebandSinglePhoton2023}
Mian Wei, Sotiris Nousias, Rahul Gulve, David~B. Lindell, and Kiriakos~N.
  Kutulakos.
\newblock Passive ultra-wideband single-photon imaging.
\newblock In \emph{Int. Conf. Comput. Vis.}, pages 8135--8146, 2023.

\bibitem[Wu et~al.(2024)Wu, Sun, Chen, Zhang, Yue, and
  Bouman]{wuPrincipledProbabilisticImaging2024}
Zihui Wu, Yu Sun, Yifan Chen, Bingliang Zhang, Yisong Yue, and Katherine
  Bouman.
\newblock Principled probabilistic imaging using diffusion models as
  plug-and-play priors.
\newblock In \emph{Adv. Neural Inform. Process. Syst.}, pages 118389--118427.
  Curran Associates, Inc., 2024.

\bibitem[Yu et~al.(2020)Yu, Kukko, Kaartinen, Wang, Liang, Matikainen, and
  Hyypp{\"a}]{yu2020comparing}
Xiaowei Yu, Antero Kukko, Harri Kaartinen, Yunsheng Wang, Xinlian Liang, Leena
  Matikainen, and Juha Hyypp{\"a}.
\newblock Comparing features of single and multi-photon lidar in boreal
  forests.
\newblock \emph{ISPRS Journal of Photogrammetry and Remote Sensing},
  168:\penalty0 268--276, 2020.

\bibitem[Zhang et~al.(2024)Zhang, Weerasooriya, Chennuri, and
  Chan]{zhangParametricModelingEstimation2024}
Weijian Zhang, Hashan~K. Weerasooriya, Prateek Chennuri, and Stanley~H. Chan.
\newblock Parametric modeling and estimation of photon registrations for {3D}
  imaging.
\newblock In \emph{IEEE International Workshop on Multimedia Signal
  Processing}, pages 1--6, 2024.

\bibitem[Zhu et~al.(1997)Zhu, Byrd, Lu, and Nocedal]{zhu1997algorithm}
Ciyou Zhu, Richard~H Byrd, Peihuang Lu, and Jorge Nocedal.
\newblock Algorithm 778: {L-BFGS-B}: {Fortran} subroutines for large-scale
  bound-constrained optimization.
\newblock \emph{ACM Transactions on Mathematical Software (TOMS)}, 23\penalty0
  (4):\penalty0 550--560, 1997.

\end{thebibliography}
